\documentclass[12pt,a4paper]{article}
\pdfoutput=1

\usepackage{latexsym,amsmath,amssymb,color,mathrsfs,verbatim,epsfig, wasysym}
\usepackage{hyperref}
\usepackage{slashed}
\usepackage{multirow}
\usepackage{xspace}
\usepackage{color}
\usepackage{cite}
\usepackage{appendix}
\usepackage{graphicx,array}
\usepackage{mathtools}
\allowdisplaybreaks

\newcommand{\CC}{\mathcal{C}}
\newcommand{\LL}{\mathcal{L}}
\newcommand{\OO}{\mathcal{O}}

\newcommand{\Br}{\textrm{Br}}

\newcommand{\be}{\begin{equation}}
\newcommand{\ee}{\end{equation}}

\definecolor{darkblue}{cmyk}{1,0.3,0,0.2}
\definecolor{violet}{cmyk}{0,1,0,0.2}
\hypersetup{colorlinks, bookmarksnumbered, citecolor=darkblue, linkcolor=darkblue, pdfstartview=FitH, urlcolor=darkblue, linktocpage}

\begin{document}

\baselineskip=18pt


%
\thispagestyle{empty}
\vspace{20pt}

\begin{flushright}
\small 
\end{flushright}

\hfill
\vspace{20pt}

\begin{center}
{\Large \textbf
{
Bottom-Flavored Mono-Tau Tails at the LHC\\[20pt]
}}
\end{center}

\vspace{15pt}
\begin{center}
{\large David Marzocca$^{\, a}$, Ui~Min$^{\, b}$ and Minho~Son$^{\, b}$}
\vspace{30pt}

$^{a}$ {\small \it INFN, Sezione di Trieste, SISSA, Via Bonomea 265, 34136, Trieste, Italy }
\vskip 3pt
$^{b}$ {\small \it Department of Physics, Korea Advanced Institute of Science and Technology, \\ 291 Daehak-ro, Yuseong-gu, Daejeon 34141, Republic of Korea
}

\end{center}

\vspace{20pt}
\begin{center}
\textbf{Abstract}
\end{center}
\vspace{5pt} {\small
We study the effective field theory sensitivity of an LHC analysis for the $\tau \nu$  final state with an associated $b$-jet. To illustrate the improvement due to the $b$-tagging, we first recast the recent CMS analysis in the $\tau\nu$ channel, using an integrated luminosity of 35.9 fb$^{-1}$ at $\sqrt{s}=13$ TeV, and provide limits on all the dimension-six effective operators which contribute to the process.
The expected limits from the $b$-tagged analysis are then derived and compared. We find an improvement of approximately $\sim 30\%$ in the bounds for operators with a $b$ quark. We also discuss in detail possible angular observables to be used as a discriminator between dimension-six operators with different Lorentz structure. Finally, we study the impact of these limits on some simplified scenarios aimed at addressing the observed deviations from the Standard Model in lepton flavor universality ratios of semileptonic $B$-meson decays. In particular, we compare the collider limits on those scenarios set by our analysis either with or without the $b$-tagging, assuming an integrated luminosity of 300 fb$^{-1}$, with relevant low-energy flavor measurements.
}

\vfill\eject
\noindent

\newpage

\tableofcontents
%

\section{Introduction}
\label{sec:intro}

The high-energy tails of two-to-two scattering processes at the LHC are some of the most sensitive probes for New Physics (NP) at the collider. 
In absence of direct evidence for new physics, and assuming the mass scale of new particles lies above the energy reach of the collisions, these searches can provide very strong and model-independent limits on dimension-six operators. 
Scattering amplitudes involving such operators grow with the square of the energy, $ E^2$, compared to the corresponding Standard Model (SM) amplitudes. This enhancement of new physics effects at high energies can be leveraged to compensate the limited statistical and systematic precision of these processes, allowing the limits obtained in this way to be competitive with those derived from precision low-energy data.
For instance, it has already been shown that high-energy tails of 2 to 2 processes at LHC can provide complementary information to low-energy flavor physics on the flavor structure of New Physics \cite{Cirigliano:2012ab,deBlas:2013qqa,Gonzalez-Alonso:2016sip,Faroughy:2016osc,Greljo:2017vvb,Altmannshofer:2017poe,Greljo:2018tzh,Afik:2018nlr,Afik:2019htr,Angelescu:2020uug,Fuentes-Martin:2020lea} or even be competitive with LEP in putting constraints on electroweak precision tests \cite{Farina:2016rws,Alioli:2017nzr,Alioli:2017jdo,Franceschini:2017xkh,Liu:2018pkg}.

In case of the process at hands, $p p \to \tau \nu$, the relevant operators are semileptonic four-fermion operators.
In the formalism of the SM Effective Field Theory (SMEFT) and in the Warsaw basis \cite{Grzadkowski:2010es}, the ones which show a growth with energy of the scattering amplitude, compared to the SM, are
\begin{equation}\label{eq:basis:linear}
\begin{split}
	 {\mathcal L}^{\rm dim6}_{\rm SMEFT}  \supset -\frac{1}{v^2} &\Big [ [C^{(3)}_{lq}]_{ijkl} \left ( \bar{l}_i \gamma_\mu \sigma^I l_j \right ) \left ( \bar{q}_k \gamma^\mu \sigma^I q_l \right )
 \\[3pt]
	 &+ [C_{ledq}]_{ijkl} \left ( \bar{l}^\alpha_i e_j \right ) \left ( \bar{d}_k q^\alpha_l \right )
 + [C^{(1)}_{lequ}]_{ijkl} \left ( \bar{l}^\alpha_i e_j \right ) \epsilon_{\alpha\beta} \left ( \bar{q}^\beta_k u_l \right ) + \text{h.c.}
 \\[3pt]
	 &+ [C^{(3)}_{lequ}]_{ijkl} \left ( \bar{l}^\alpha_i \sigma_{\mu\nu} e_j \right ) \epsilon_{\alpha\beta}\left ( \bar{q}^\beta_k \sigma^{\mu\nu} u_l \right ) + \text{h.c.} \Big ]~,
\end{split}
\end{equation}
where $i,j,k,l$ are flavor indices, $\alpha, \beta$ are $SU(2)$ indices, and $1/v^2 = 2 G_F / \sqrt{2}$, $v = 246\, \text{GeV}$. 
Lepton and quark doublets are $l_i = (\nu^i_L, \ell^i_{L})$ and $q_i = (V^*_{ji} u_L^j, d_L^i)$, respectively, where $V$ is Cabbibo-Kobayashi-Maskawa (CKM) matrix.

This specific process is particularly interesting now due to the close connection with the measurements of lepton flavor universality (LFU) ratios of semileptonic $B$-meson decays $R(D^{(*)}) = \Br(B \to D^{(*)} \tau \nu) / \Br(B \to D^{(*)} \ell \nu)$ (with $\ell = e, \mu$) \cite{Lees:2012xj,Lees:2013uzd,Aaij:2015yra,Huschle:2015rga,Sato:2016svk,Hirose:2016wfn,Hirose:2017dxl,Aaij:2017uff,Aaij:2017deq,Siddi:2018avt,Belle:2019rba}, which in a combined fit of BaBar, Belle, and LHCb data, show a deviation from the SM prediction at the $\sim 3 \sigma$ level \cite{Amhis:2016xyh}, hinting for a possible presence of new physics in the $b \to c \tau \nu$ transition.\footnote{The light leptons channels are instead consistent with each other and with the SM expectation}
Since the mass scale of new resonances indicated by these deviations lies in the few-TeV range, testing this process in high-energy scattering at the LHC is clearly particularly motivated. 
CMS \cite{Sirunyan:2018lbg} and ATLAS \cite{Aaboud:2018vgh} searches in the $\tau\nu$ channel have been recasted to provide limits on EFT operators in~\cite{Greljo:2018tzh,Fuentes-Martin:2020lea}.

The main goal of this work is to design an LHC analysis of the $p p \to \tau \nu$ process, including also the requirement of a $b$-jet in the final state. This is expected to improve the sensitivity on operators involving a $b$ quark, such as those involved in the $R(D^{(*)})$ observables.
In order to quantify the gain in sensitivity due to the $b$-tagging, and to validate our background analysis, we also recast the CMS analysis of the $p p \to \tau \nu$ search \cite{Sirunyan:2018lbg}. We thus provide the present EFT limits from this search, as well as the future sensitivity of the searches for both cases with and without the $b$-tagging.

In Section~\ref{sec:EFT} we describe the EFT operators employed in the analysis, and the approach used to derive the EFT dependence of the cross section in each bin of the transverse mass.
In Section~\ref{sec:btag} we validate our analysis and simulation for $pp\rightarrow \tau\nu$ against the CMS analysis in~\cite{Sirunyan:2018lbg}.  After that, we perform a new analysis for $p p \to \tau \nu + b$ for further improvement. Also, we discuss the potential of some angular distributions for extracting more information on the tensor structure of four-fermion operators.  In Section~\ref{sec:EFTfit} we obtain the present limits and future sensitivity on the EFT coefficients from both $\tau\nu$ and $\tau\nu + b$ analyses.
In Section~\ref{sec:flavor} we discuss some implications of these constraints on some flavor structures, comparing with low-energy flavor measurements such as $R(D^{(*)})$, $B \to \tau \nu$, and $\tau$ decays. We conclude in Section \ref{sec:conclusions}. In Appendices we provide the cross section fit in terms of EFT coefficients and full differential cross section of 2 to 3 process as well as some simulation details.

\section{EFT contributions to high-energy tails}
\label{sec:EFT}

New physics effects in low-energy flavor observables are usually discussed in terms of an effective Hamiltonian defined at the low-energy scale with quarks in the mass basis. For the charged-current transitions at hand, the relevant effective Lagrangian is usually defined as
\begin{align}
	{\mathcal{L}}^{\rm CC}_{\rm eff}  = - {\mathcal{H}}^{\rm CC}_{\rm eff}  =
	 - \frac{4 G_f V_{ij}}{\sqrt{2}} \Big[
	&	C_{VLL}^{ij} (\bar u_i \gamma_\mu P_L d_j) (\bar \tau \gamma^\mu P_L \nu_\tau) +
		C_{VRL}^{ij} (\bar u_i \gamma_\mu P_R d_j) (\bar \tau \gamma^\mu P_L \nu_\tau) + \nonumber \\
	& 	C_{SL}^{ij} (\bar u_i P_L d_j) (\bar \tau P_L \nu_\tau) +
		C_{SR}^{ij} (\bar u_i P_R d_j) (\bar \tau P_L \nu_\tau) + \label{eq:eft:banomaly:mbscale} \\
	&	C_{T}^{ij} (\bar u_i \sigma_{\mu\nu} P_L d_j) (\bar \tau \sigma^{\mu\nu} P_L \nu_\tau)\Big] + h.c.~. \nonumber
\end{align}
These coefficients, evaluated at the matching scale, can be easily translated into those in the linear basis, Eq.~\eqref{eq:basis:linear}:
\be\begin{split}
	C_{VLL}^{ij} &= \frac{1}{V_{ij}} \sum_k V_{ik} [C_{lq}^{(3)}]_{33kj} , \\
	C_{SL}^{ij} &= \frac{1}{2V_{ij}}  [C_{lequ}^{(1)}]^*_{33ji} , \\
	C_{T}^{ij} &= \frac{1}{2V_{ij}}  [C_{lequ}^{(3)}]^*_{33ji} , \\
	C_{SR}^{ij} &= \frac{1}{2V_{ij}}  \sum_k V_{ik} [C_{ledq}]^*_{33jk} .
\label{eq:LEFTtoSMEFT}
\end{split}\ee
Going from the matching scale down to the low-energy scale relevant for flavor processes, the anomalous dimension induced by QCD interactions must be taken into account~\cite{Gonzalez-Alonso:2017iyc}. It can be noted that the $\mathcal{O}_{VLL}$ operator has no QCD anomalous dimension.
The $\mathcal{O}_{VRL}$ operator is generated at dimension-6 in the SMEFT only via anomalous $W$ boson couplings to right-handed quarks, and at energies above the electroweak scale is therefore resolved into a vertex correction for the $W$, so does not behave as a four-fermion operator (no growth with energy of the scattering amplitude). It can also be generated as a dimension-8 operator, thus receiving a further $v^2 / \Lambda^2$ suppression compared to dimension-6 operators. For this reason we keep it in the analysis done in the mass basis but drop it in the SMEFT analysis.

The parametrization in Eq.~(\ref{eq:eft:banomaly:mbscale}) is convenient for discussing low-energy flavor observables, but also for the high-energy tails studied here, as it features a non-interference among different EFT coefficients in the limit of negligible fermion masses.\footnote{Only $\OO_{SL}$ and $\OO_{T}$ with same flavor content have a non-vanishing interference among themselves.}
We thus implement in a {\sc FeynRules}~\cite{Alloul:2013bka} model the effective operators in Eq.~\eqref{eq:eft:banomaly:mbscale}.

Since these semileptonic operators contribute to the scattering amplitude with a single insertion, in general the cross section is quadratic in the EFT coefficients and be written as
\begin{equation}\label{eq:xsec:EFTcoeff}
 \sigma = \sigma_{SM} + C^{ij}_X\, \sigma^{ij,X}_{SM-EFT} +  (C^{ij}_X)^2\, \sigma^{ij,X}_{EFT^2}~,
\end{equation}
where $i,j$ are flavor indices and $X$ runs over all possible operators in Eq~(\ref{eq:eft:banomaly:mbscale}). Operators with the top quark do not contribute to this process. This leaves thirty EFT coefficients (six from each type of opeartor). In the limit of negligible fermion masses, the interference terms $\sigma^{ij,X}_{SM-EFT}$ vanish for all operators, except for the one associated with $C_{VLL}$.
We obtain the linear and quadratic terms by simulating them separately using \textsc{\sc MadGraph}5\_aMC$@$NLO~\cite{Alwall:2014hca}.
The complete cross section dependence on the EFT coefficients is provided in Appendix~\ref{sec:XsecEFT}.

Employing the EFT approach to discuss high-energy tails of scattering processes comes with important caveats regarding the validity of the EFT expansion.
By assumption, the energy scale of new states should be much above the typical energy of the process, $M_{\rm NP}^2 \gg \hat{s}$, where $\hat{s} \sim 1$~TeV in our case. Due to the growth with the energy of the EFT scattering amplitude, the cross section in the most sensitive bins is dominated by the EFT-squared contribution, rather than the SM-EFT interference.
Since quadratic terms are formally of order $\sim 1 / M_{\rm NP}^4$, like the interference of possible dimension-8 operators with the SM, the validity and generality of the approach could be questioned if their inclusion were to affect the results.
Nevertheless, in case of single tree-level mediators this is not an issue, since it turns out that the interference of dimension-8 operators with the SM is always smaller than the interference of dimension-6 terms with SM, if $M_{\rm NP}^2 > \hat{s}$, as shown in~\cite{Fuentes-Martin:2020lea}. A cancellation between dimension-six and eight contributions would require a specific multi-mediator scenario with tuned couplings.

Even if the mediator has a mass lower than the scattering energy, thus invalidating the EFT expansion, the limits obtained in the EFT approach can still be indicative of the true limits. In case of a mediator exchanged in the $s$-channel, the true signal includes a resonance and is always larger than the EFT prediction, implying that the bounds obtained in the EFT would be conservative \cite{Greljo:2017vvb}. In case of an exchange in the $t$ or $u$ channel, instead, the true signal can be smaller but, as shown in \cite{Greljo:2018tzh}, the EFT limits approximate well those obtained in the complete model.
We refer to \cite{Fuentes-Martin:2020lea} for a more detailed discussion of possible caveats due to the EFT expansion in the $p p \to \tau \nu (+b)$ process at the LHC.

\section{Boosting flavor precision at the LHC}
\label{sec:btag}

\subsection{Tagging bottom flavor}
\label{sec:heavyflavor}
Tagging a $b$-quark is beneficial in two aspects. First, while the dominant SM contribution to the $\tau\nu$ final state comes from the parton distribution function (PDF) of light quarks, the beyond the SM (BSM) contribution of interest are initiated by $cb$ and $ub$ initial state partons. Tagging a $b$-quark exclusively will suppress only the SM contribution and thus the sensitivity of the cross section on the EFT coefficients is enhanced. Secondly, by tagging a $b$-quark, one can restrict the analysis to the subset of four-fermion operators where one of the field is a $b$-quark, thus reducing the dimensionality of EFT parameter space entering the analysis. The dimension could be further reduced by an extra $c$-tagging.

The relevant collider search is $pp\rightarrow \tau + \slashed{\vec{E}}_T+b$. 
Inclusive $\tau\nu$ resonance searches without $b$-tagging using data at $\sqrt{s} = 13$ TeV have been performed in~\cite{Sirunyan:2018lbg,Aaboud:2018vgh}. To best of our knowledge, the experimental searches in $pp\rightarrow \tau\nu+b$ is not available.  Collider studies of the process $pp\rightarrow \tau\nu+b$ in the context of $W'$ and leptoquark searches have been performed in~\cite{Abdullah:2018ets,Brooijmans:2020yij}.

\subsection{Validation against CMS $\tau\nu$ analysis}
\label{sec:CMSvalidation}
We adopt the analysis of the CMS $\tau\nu$ resonance search at $\sqrt{s}=13$ TeV~\cite{Sirunyan:2018lbg} with an integrated luminosity of 35.9 fb$^{-1}$, recasting it to derive the sensitivity on the EFT coefficients. We collect all simulation details in Appendix~\ref{app:simulation}. Here, we focus on describing our main analysis procedure and results.

We first identify the isolated leptons according to the criteria $p_T(l)/(p_T(l)+p_T({\rm cone})) > 0.85$ where $p_T({\rm cone})$ is the surrounding transverse momentum within the isolation cone size of $R_{iso} = 0.3$. Any events with isolated leptons with $p_T(l) > 20$ GeV and $|\eta(l)| < 2.5$ are vetoed. All particles in the event are clustered by \textsc{Fastjet} 3.1.3~\cite{Cacciari:2005hq} using the anti-$k_T$ algorithm~\cite{Cacciari:2008gp} with a jet size of $R=0.5$. Events with at least one jet that satisfies $p_T(j) > 20$ GeV and $|\eta(j)| < 2.5$ are selected~\footnote{The jet definition is not provided in~\cite{Sirunyan:2018lbg}, we believe that what we have adopted here is close to the commonly used selection cuts for jets in literature.}. Jets are classified into four categories depending on whether they match to either heavy flavors or truth-level tau-lepton, namely $b,\, c,\, \tau$-jets and light jets. Jets are first iterated to identify $\tau$-jet candidates. While the CMS analysis in~\cite{Sirunyan:2018lbg} uses the sophisticated multivariant-based (MVA-based) $\tau$-jet identification, we classify a jet as a $\tau$-jet candidate if a truth-level tau lepton in the hard process is found inside a jet within a distance of $R=0.25$ from a jet vector. Events with more than one $\tau$-jet candidate are vetoed. The remaining jets are further iteratively searched for $b$-hadrons or $c$-hadrons inside them to identify $b,\, c$-jets candidates. If a $b$-hadron ($c$-hadron) is found inside a jet, it is declared to be a $b$-jet candidate ($c$-jet candidate). The leftover jets are classified as light jets. The missing transverse momentum $\vec{p}^{\ miss}_T$ is defined as the negative vectorial sum of all visible reconstructed objects such as $\tau$-jet and QCD-jets. 

\begin{figure}[t]
\begin{center}
\includegraphics[width=0.55\textwidth]{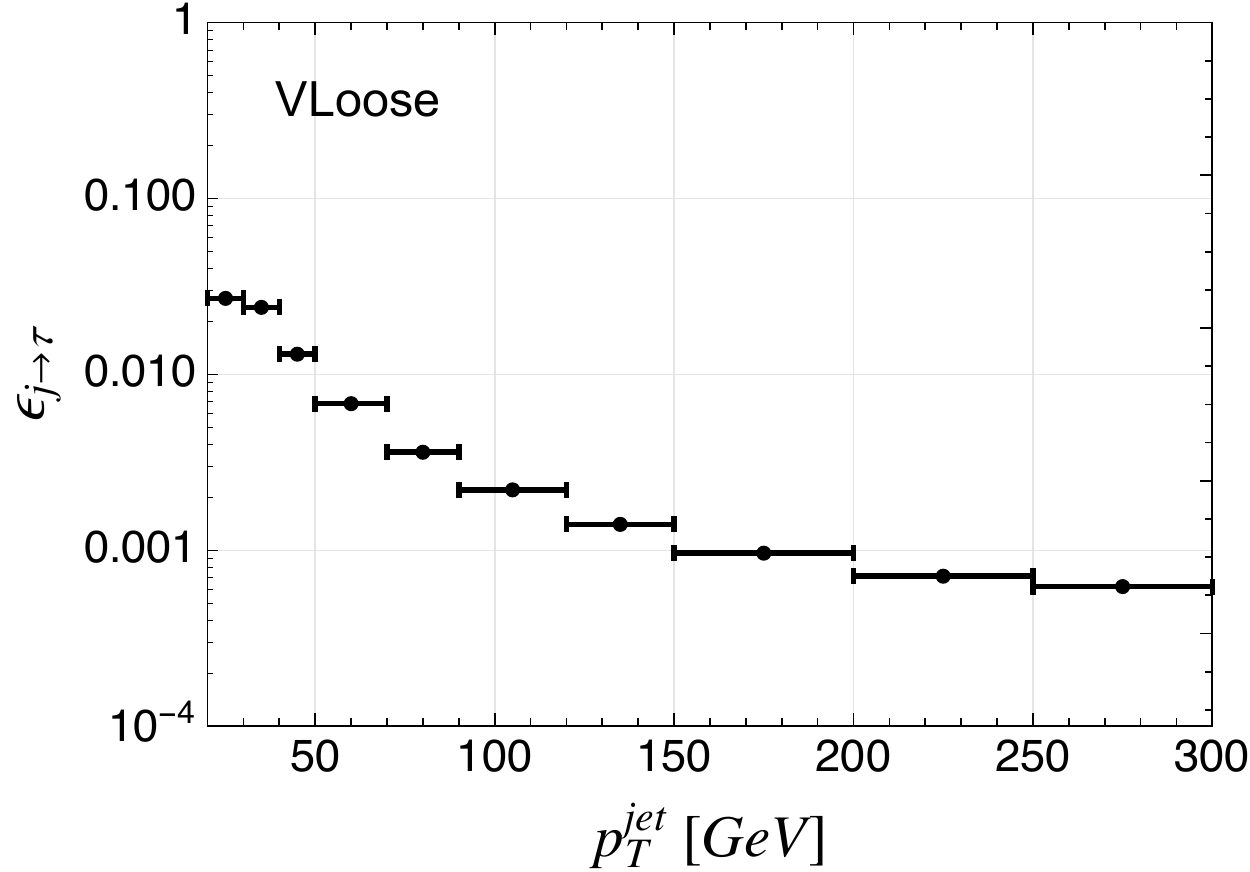}
\caption{\small The misidentification rate of $j\rightarrow\tau$ for the VLoose working point taken from the CMS performance of reconstruction and identification of tau leptons using data at $\sqrt{s} = 13$ TeV~\cite{Sirunyan:2018pgf}.}
\label{fig:mistag:tau:CMS}
\end{center}
\end{figure}

Similarly to the analysis in~\cite{Sirunyan:2018lbg}, we adopt the very loose (VLoose) working point for tag and mistag rates of the MVA-based $\tau$-jet identification taken from~\cite{Sirunyan:2018pgf} (see Fig.4 of~\cite{Sirunyan:2018pgf}).  The tag rate in VLoose working point is roughly 70\%, $\epsilon_{\tau\rightarrow \tau} = 0.7$, whereas the mistag rate $\epsilon_{j\rightarrow \tau}$ is shown in Fig.~\ref{fig:mistag:tau:CMS}. 
The mistag rate decreases with an increasing $p_T(\tau)$ and its value is smaller than 0.4\% for $p_T(\tau) \gtrsim 80$ GeV. In applying the mistag rate in Fig.~\ref{fig:mistag:tau:CMS} to QCD-jets in the $\tau\nu$ analysis, we do not distinguish the heavy flavor jets from the light jets. In our analysis, we assume that the mistag rate is saturated to the smallest value in Fig.~\ref{fig:mistag:tau:CMS} for the transverse momentum $p_T(\tau) > 300$ GeV as it is not available in~\cite{Sirunyan:2018pgf}.
 
The analysis cuts imposed in the CMS analysis~\cite{Sirunyan:2018lbg} are
\begin{equation}\label{eq:cut:CMS:pt}
p_T(\tau) > 80\ {\rm GeV}~,\quad |\eta(\tau)| < 2.1~, \quad p^{\ miss}_T > 200\ {\rm GeV}~,
\end{equation}
and, to reflect the back-to-back configuration of $\tau\nu$ system,
\begin{equation}\label{eq:cut:CMS:backtoback}
 0.7 <  p^\tau_T/p^{\ miss}_T < 1.3~, \quad \triangle\phi (\vec{p}^{\ \tau}_T,\, \vec{p}^{\ miss}_T) > 2.4~,
\end{equation}
where $\vec{p}^{\ \tau}_T$ is the transverse momentum of the $\tau$-jet, while its magnitude is denoted by $p^\tau_T$ (similarly for the missing transverse momentum). The variable $\Delta\phi$ in Eq.~(\ref{eq:cut:CMS:backtoback}) is an azimuthal angle.
Finally, events that passed the cuts in Eqs.~(\ref{eq:cut:CMS:pt}) and~(\ref{eq:cut:CMS:backtoback}) are binned in the transverse mass, $m_T$, defined as
\begin{equation}
  m_T = \sqrt{ 2 p^\tau_T p^{\ miss}_T [ 1- \cos\Delta\phi (\vec{p}^\tau_T,\, \vec{p}^{\ miss}_T)] }~.
\end{equation}
Following the description above, we validate our background simulation against the CMS analysis. They are illustrated in Table~\ref{tab:CMS:bkg}. 
\begin{table}[t]
\centering
\scalebox{1.0}{
\begin{tabular}{c|ccc}  
\hline
$m_T$ [TeV]  & \quad $m_T < 0.5 $TeV  & \quad  $0.5 < m_T < 1 $TeV & \quad $m_T > 1$ TeV
\\
\hline \hline
$W$+jets   & 653 (786$\pm$110) & 366 (355$\pm$68)  & 18 (22$\pm$6.2)  \\[2.5pt]
$Z\rightarrow \nu\nu$+jets  & 181 (236$\pm$120)  & 96 (68$\pm$35) & 5.2 (0.9$\pm$0.5)  \\[2.5pt]
$t\bar{t}$ & 112 (68$\pm$15) & 41 (14.5$\pm$4.5) & 0.44 ($<$0.1) \\[2.5pt]
$Z/\gamma^*\rightarrow ll$+jets & 34.5 (36$\pm$8.7) & 13.2 (10$\pm$5.1)  &  0.0025 ($<$ 0.1) \\[2.5pt]
$VV$  & 22.4(24.9$\pm$6.4) & 16.5(9.6$\pm$3.5) & 1.7(0.7$\pm$0.1) \\[2.5pt] 
single-$t$ & 15.6 (21.5$\pm$6.5)  & 4.3 (7.0$\pm$2.9) & 0.1 ($<$0.1) \\[2.5pt]
\hline
Total & 1018.5 ($1243\pm 160$) & 537 ($485 \pm 77$) & 25.4 ($23.4 \pm 7.2$) \\[2.5pt]
\hline
\end{tabular}
}
\caption{\small Expected number of events in the SM from our simulation, for $\sqrt{s}$ =13 TeV and an integrated luminosity of 35.9 fb$^{-1}$. The numbers in parenthesis are the CMS result in~\cite{Sirunyan:2018lbg} with associated total systematic uncertainties.}
\label{tab:CMS:bkg}
\end{table}
While the first two bins of $m_T$ variable in Table~\ref{tab:CMS:bkg} are in a good agreement with CMS result (values in parenthesis) except for $t\bar{t}$ background which differs more than twice (in a conservative way), our estimate of the last bin turns out to be more conservative except for the dominant one, $W+$jets~\footnote{One possibility is that the transverse momentum of $\tau$-jet (either tagged one or fake) in the last bin, $m_T > 1$ TeV, is likely above 300 GeV for which we assumed a conservative saturated mistag rate instead of taking $p_T$ dependent values. We have also tried a few different definitions of missing transverse momenta and we found that it caused minor effect. For top backgrounds, other than CMS~\cite{Sirunyan:2018lbg} using {\tt POWEG}, no further simulation information such as the matching or $k$-factor is available. (see Appendix~\ref{app:simulation} for our simulation).  We decided to leave our estimate of $t\bar{t}$ as is as it is more conservative.}.  Although we decided not to further investigate to resolve the discrepancy in Table~\ref{tab:CMS:bkg}, due to limited available information from Ref.~\cite{Sirunyan:2018lbg}, we point out that the dominant background $W+$jets agrees well with the CMS analysis and thus sensitivities on the EFT coefficients derived either from our estimate or the CMS one will be similar. 

The same set of cuts in Eqs.~(\ref{eq:cut:CMS:pt}) and (\ref{eq:cut:CMS:backtoback}) are imposed on the signal event samples for thirty EFT coefficients in Eq.~(\ref{eq:eft:banomaly:mbscale}). While those signal samples for the inclusive $\tau\nu$ analysis were not matched due to limited computing resources, we apply the nominal unit  $k$-factor to all EFT signal samples, based on our numerical comparison between unmatched samples and some selected matched ones up to one jet (see Appendix~\ref{app:signal} for details).

\begin{figure}[p]
\begin{center}
\includegraphics[width=0.48\textwidth]{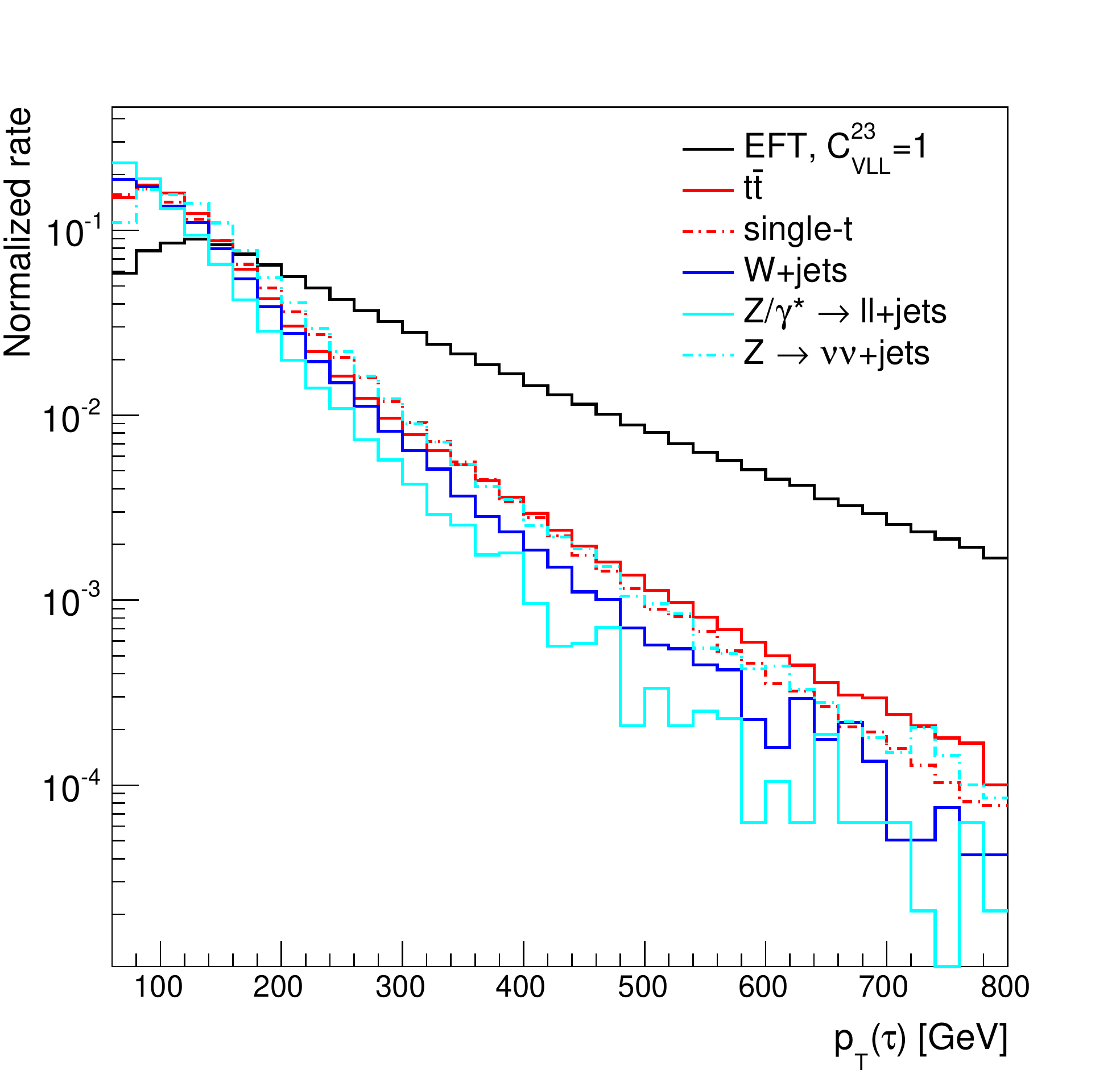}\quad
\includegraphics[width=0.48\textwidth]{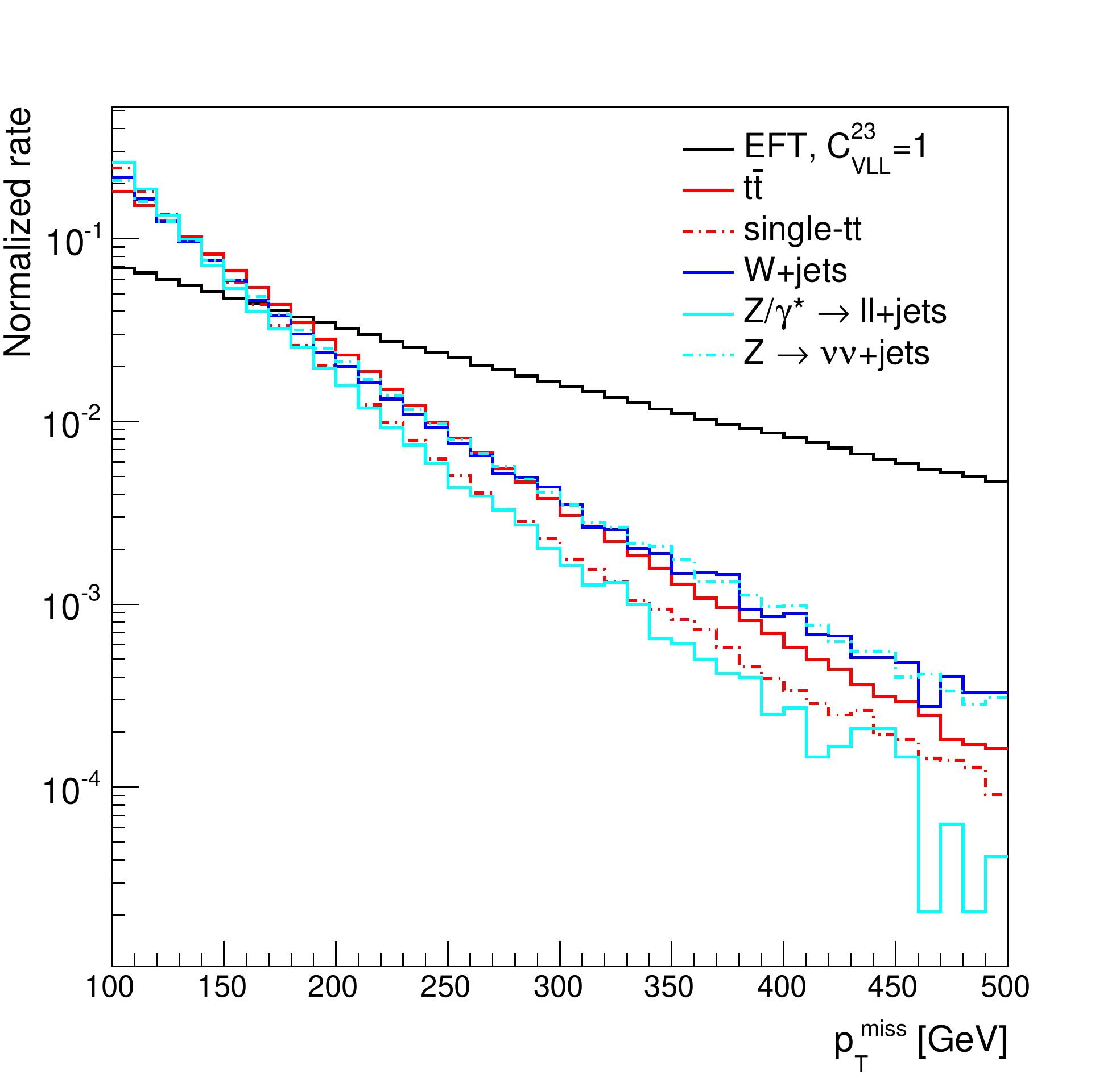}\\[3pt]
\includegraphics[width=0.48\textwidth]{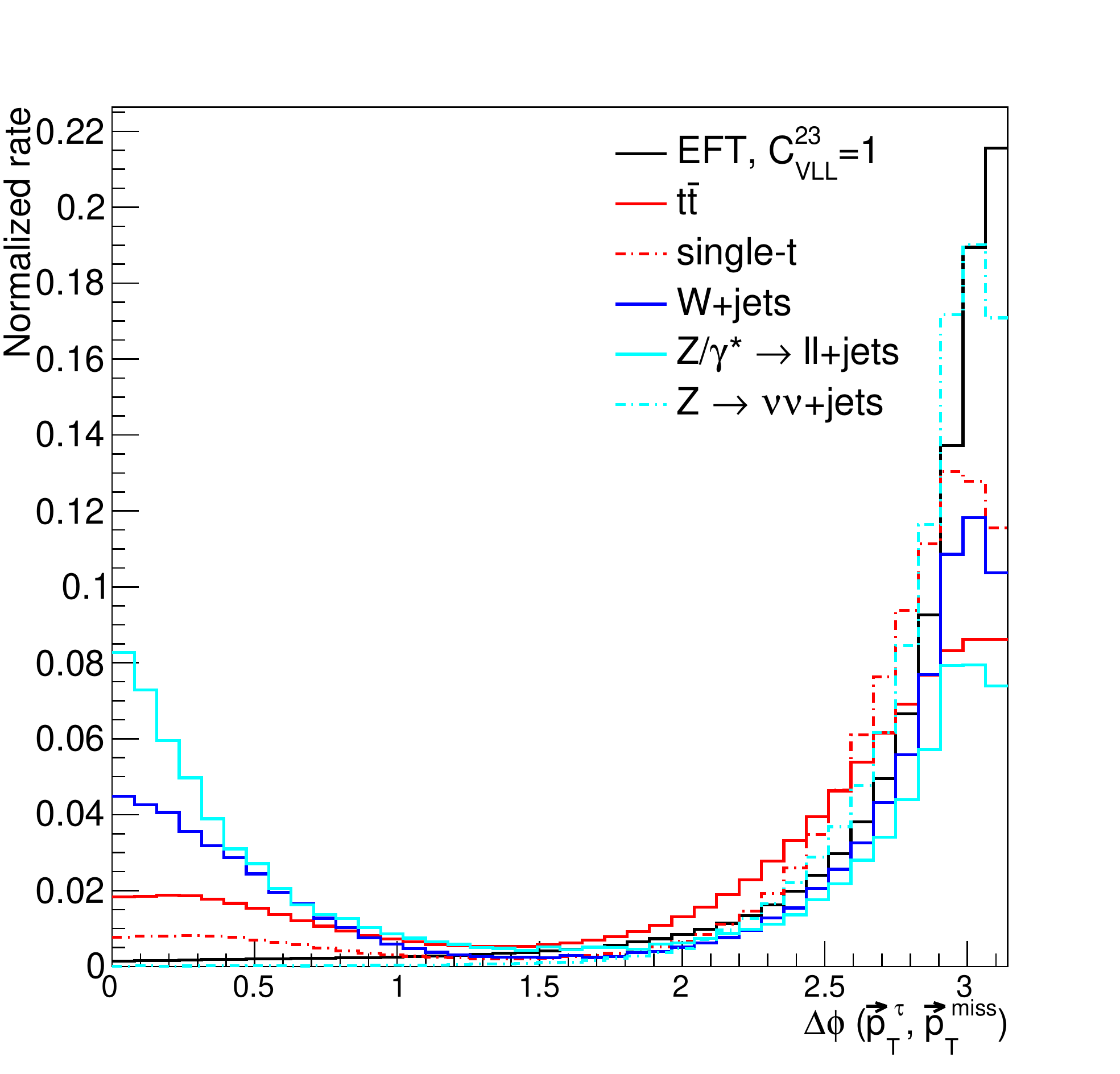}\quad
\includegraphics[width=0.48\textwidth]{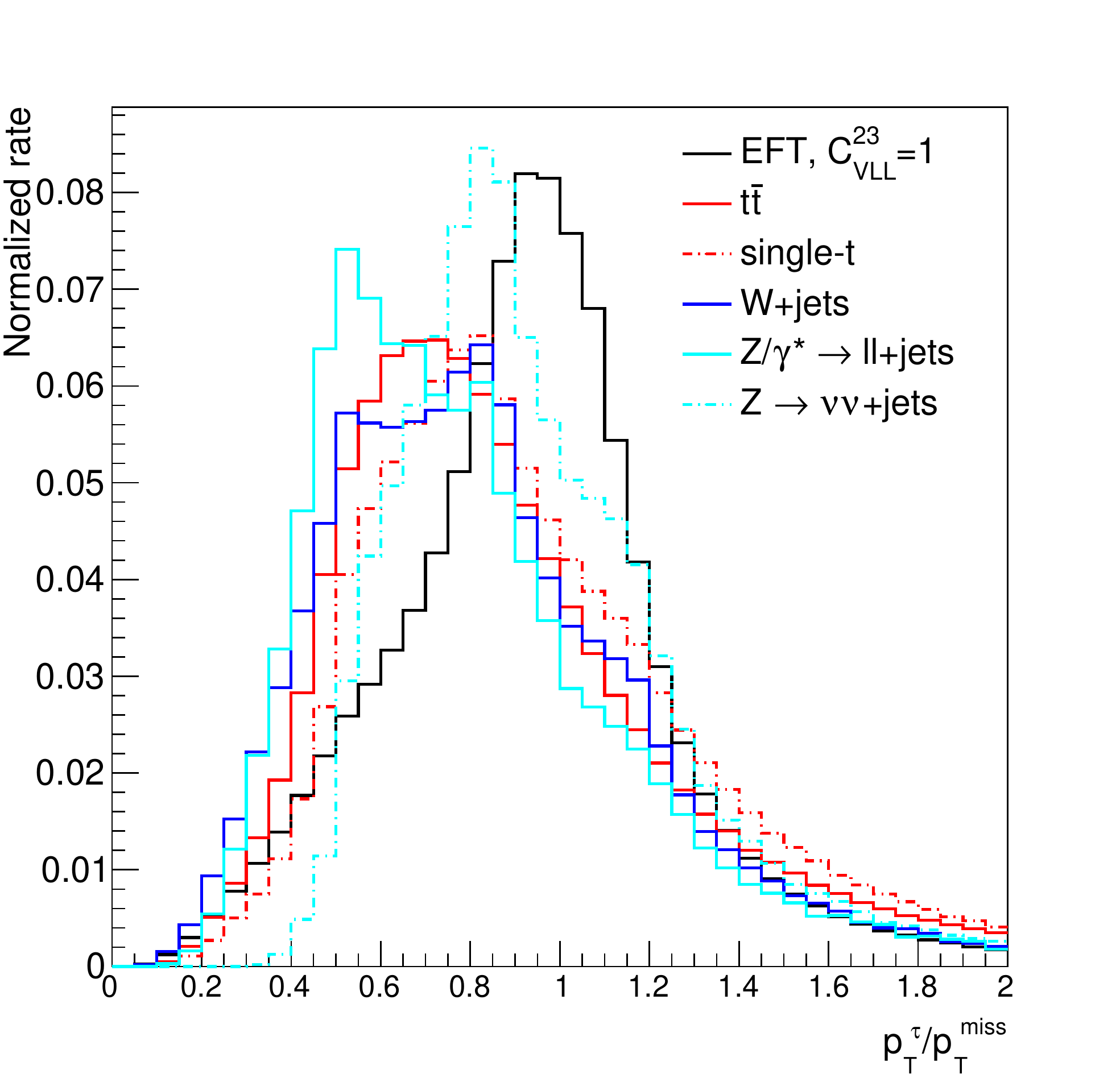}
\caption{\small The distributions of $p_T(\tau)$ (top left), the missing transverse momentum, $p_T^{\ miss}$, (top right), $\Delta\phi (\vec{p}^{\ \tau}_T,\, \vec{p}^{\ miss}_T)$ (bottom left), and $p_T(\tau)/p_T^{\ miss}$ (bottom right) for the signal with $C^{23}_{VLL}=1$ and backgrounds. Events in all plots are restricted to include at least two jets, $N_j \geq 2$, and satisfy $p_T(\tau) > 50$ GeV (for the leading jet if no $\tau$-jet is found) and $p^{\ miss}_T > 100$ GeV.}
\label{fig:taunuj:CMSvar}
\end{center}
\end{figure}

\subsection{Analysis of $\tau\nu$ with an associated $b$-jet}
\label{sec:taunub}

\begin{figure}[t]
\begin{center}
\includegraphics[width=0.6\textwidth]{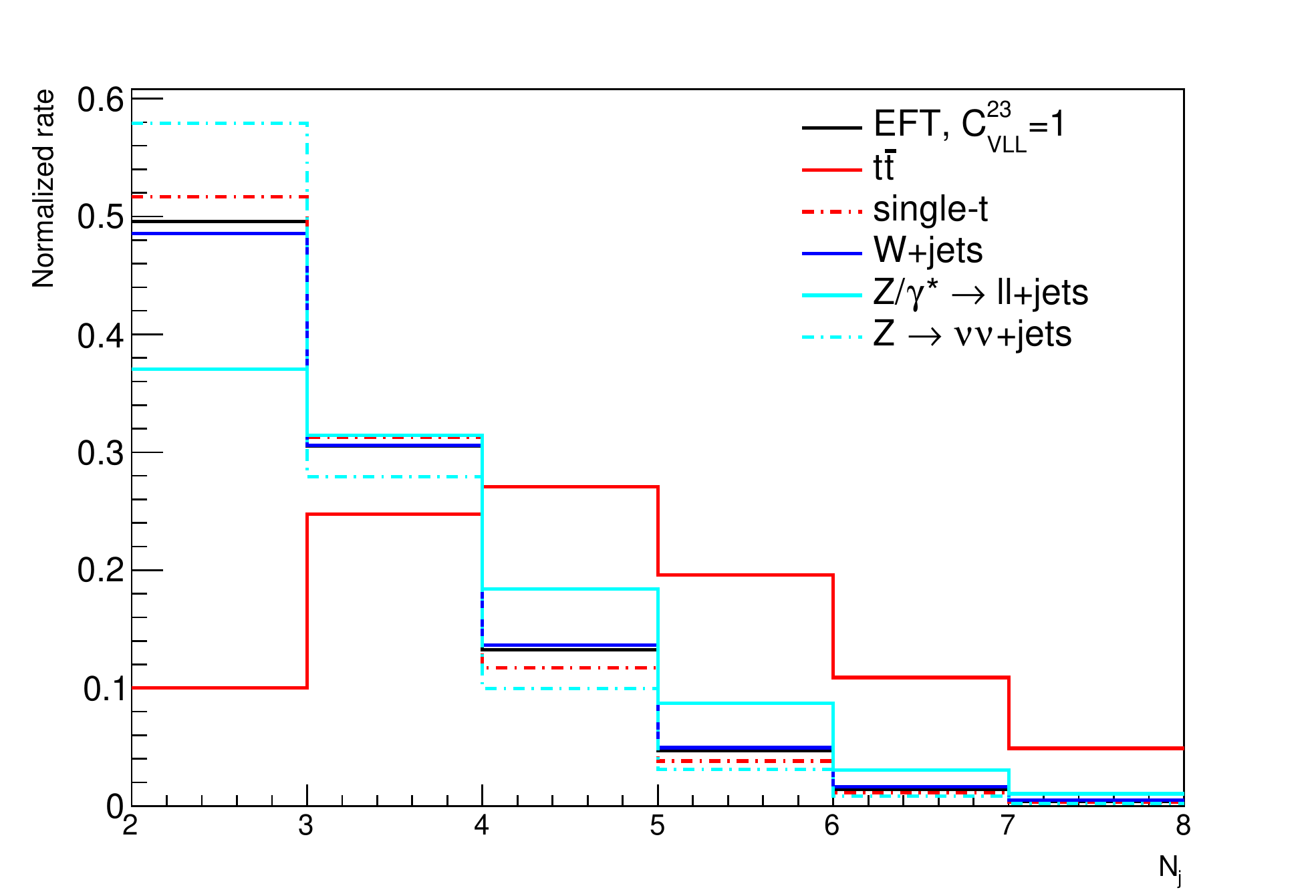}
\caption{\small The $N_{j}$ distribution of the signal with $C^{23}_{VLL}=1$ and backgrounds. Events are restricted to include at least two jets, $N_j \geq 2$, and less than two $b$-jets, $N_b <2$, and satisfy $p_T(\tau) > 50$ GeV (for the leading jet if no $\tau$-jet is found) and $p^{\ miss}_T > 100$ GeV.}
\label{fig:dist:nj}
\end{center}
\end{figure}

For the analysis with a $b$-jet, the event selection is the same as in Section~\ref{sec:CMSvalidation}, except that events with at least two jets are considered. The extra jets, in addition to the $\tau$-jet, in signal samples is likely to include a $b$-jet, whereas those in the background samples are likely light jets faking $b$-jets. Events with more than one $\tau$-jet or $b$-jet are vetoed.
We adopt the following tag and mistag rates for $b$-jet identification, along with the VLoose working point for the $\tau$-identification explained in Section~\ref{sec:CMSvalidation}, 
\begin{equation}\label{eq:bc:mistagrate}
  \epsilon_{b\rightarrow b} = 0.7~, \quad \epsilon_{c\rightarrow b} = 0.3~, \quad \epsilon_{j\rightarrow b} = 0.015~.
\end{equation}

$W+$jets is the irreducible SM contribution to the $\tau\nu (+b)$ channel, and will interfere with the contribution from the EFT operators with the same helicity structure. In order to develop our analysis, we choose $C_{VLL}^{cb} = 1$ as benchmark point for signal events. To avoid double counting the SM contribution, we take only BSM event samples from the interference and quadratic terms in Eq.~(\ref{eq:xsec:EFTcoeff}).
The benchmark signal events were generated through the process $pp\rightarrow \tau\nu$ matched up to an extra-jet (using $k_T$-jet MLM matching~\cite{Alwall:2007fs}) in the 5-flavor scheme. The distributions of the same variables used in the CMS analysis described in Section~\ref{sec:CMSvalidation} are illustrated in Fig.~\ref{fig:taunuj:CMSvar}, where $\tau$ refers to the $\tau$-jet (or the leading jet if not found). As is evident in Fig.~\ref{fig:taunuj:CMSvar}, they continue to be efficient discriminators for the $\tau\nu$ process with the associated $b$-jet.

We impose the following cuts on the events,
\begin{equation}\label{eq:ourcut:pt}
\begin{split}
&p_T(\tau) > 70\ {\rm GeV}~,\quad |\eta(\tau)| < 2.1~, \quad p^{\ miss}_T > 150\ {\rm GeV}~,\\[3.5pt]
&\hspace{1cm} p_T(b) > 20\ {\rm GeV}~,\quad |\eta(b)| < 2.5~,
\end{split}
\end{equation}
and, similarly to reflect the back-to-back configuration of $\tau\nu$ system,
\begin{equation}\label{eq:cut:ours:backtoback}
 0.7 <  p^\tau_T/p^{\ miss}_T < 1.3~, \quad \triangle\phi (\vec{p}^{\ \tau}_T,\, \vec{p}^{\ miss}_T) > 2.4~.
\end{equation}
The cuts on $p_T(\tau)$ and $p^{\ miss}_T$ in Eq.~(\ref{eq:ourcut:pt}) were relaxed to retain more events, compared to those in $\tau\nu$ analysis in Section~\ref{sec:CMSvalidation}. Additionally, we impose a cut on jet multiplicity whose definition includes $\tau$-jet as well,
\begin{equation}\label{eq:cut:Nj}
 N_j \leq 4~,
\end{equation}
that is efficient in reducing $t\bar{t}$ background as is evident in Fig.~\ref{fig:dist:nj}.
The cuts in Eqs.~(\ref{eq:ourcut:pt}),\ref{eq:cut:Nj}) were not optimized (similar cuts are also found in~\cite{Abdullah:2018ets}). We leave optimizing the cuts using multivariate method or machine learning for future work.

\begin{figure}[t]
\begin{center}
\includegraphics[width=0.48\textwidth]{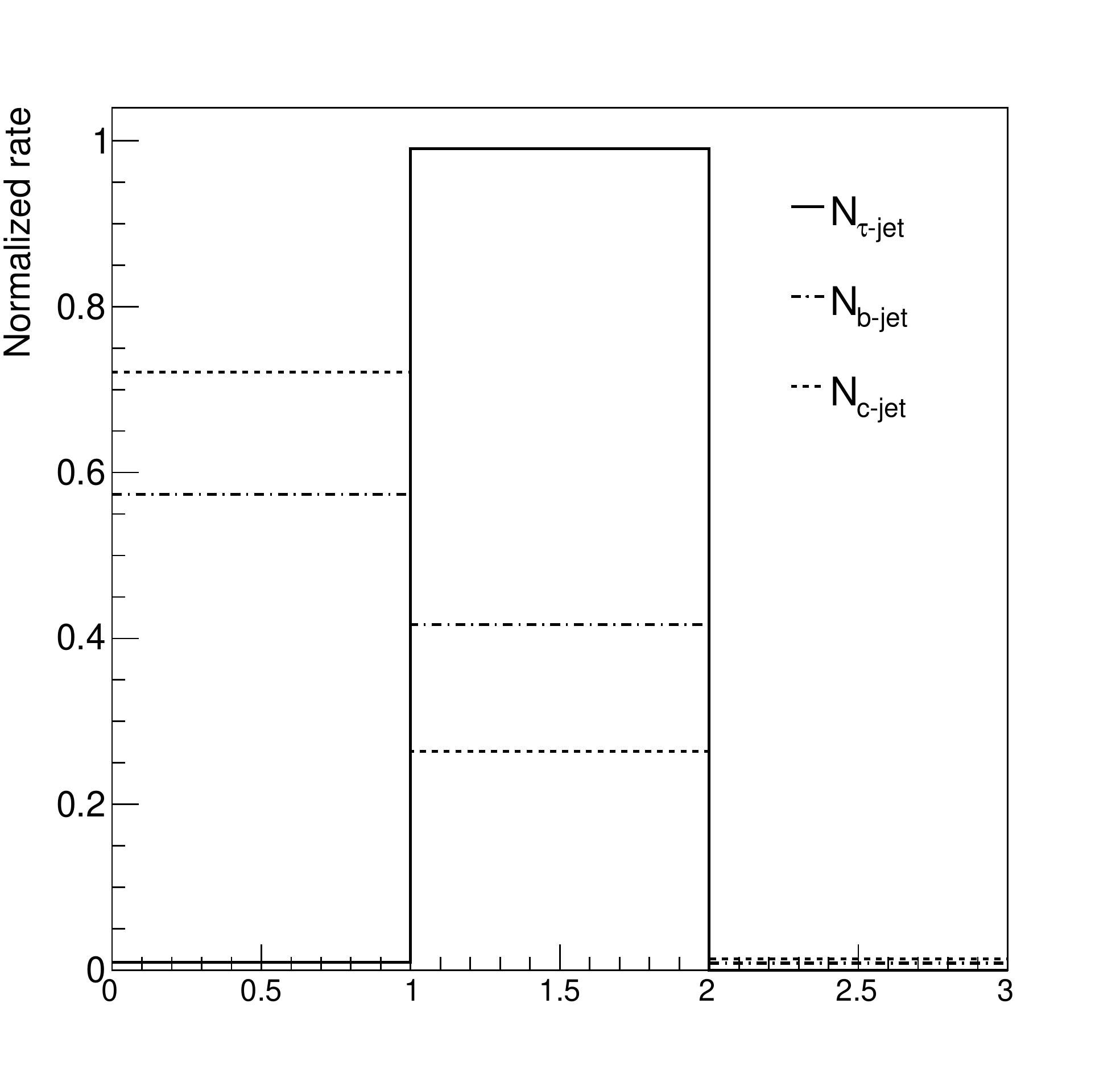} \quad
\includegraphics[width=0.48\textwidth]{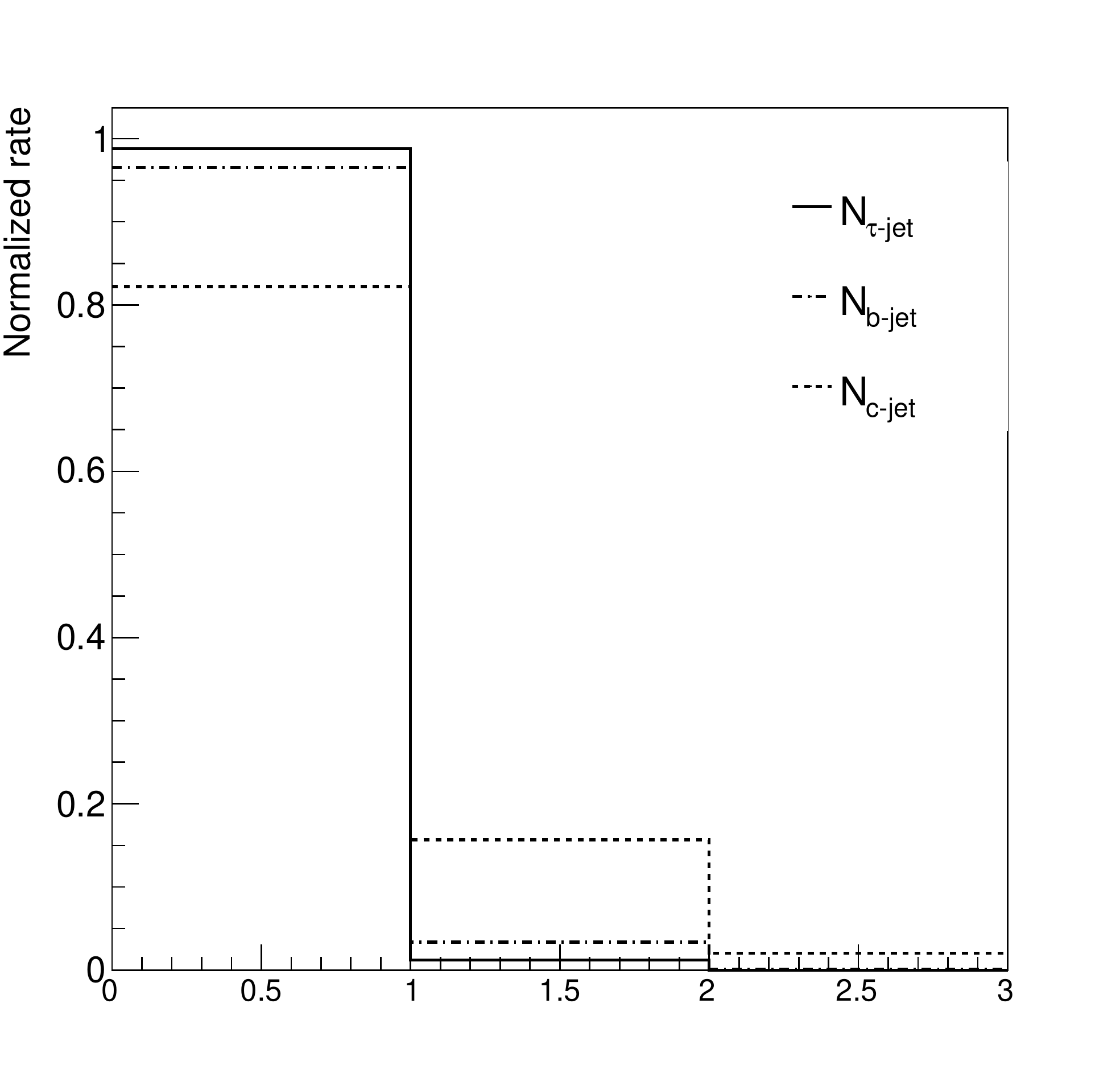}
\caption{\small The $N_{\tau,b,c\text{-}jet}$ distributions of the benchmark signal with $C^{23}_{VLL}=1$ (left) and $W+$jets background (right) in the signal region. Events in both plots are restricted to include at least two jets, $N_j \geq 2$ and satisfy $p_T(\tau) > 70$ GeV (for the leading jet if no $\tau$-jet is found), $p^{\ miss}_T > 150$ GeV, $\triangle\phi (\vec{p}^{\ \tau}_T,\, \vec{p}^{\ miss}_T) > 2.4$, $0.7 <  p^\tau_T/p^{\ miss}_T < 1.3$ and $m_T > 500$ GeV.}
\label{fig:njnbnc:sig:wjets}
\end{center}
\end{figure}

Interestingly, we find that the dominant contribution of $W+$jets to the signal region comes from fakes as is illustrated in Fig.~\ref{fig:njnbnc:sig:wjets}. To be specific, most $\tau$-tagged jets in $W+$jets are found not to be in a back-to-back configuration with the missing transverse momentum, and what mimics the signal topology are fakes. Therefore, the estimation of $W+$jets background becomes sensitive to the $p_T$-dependent tau mistag rate. Whereas the signal region for the signal events is enriched by $\tau$-tagged jets as is evident in Fig.~\ref{fig:njnbnc:sig:wjets}. Although $W+$jets is an irreducible background in terms of Feynman diagrams, this property makes it a kinematically reducible background to the signal, which implies further suppression of the interference between the signal and background. Assuming this property remains true even at the level of dimension-8 operators, it will help in establishing the better EFT expansion, namely $\sigma_{{\rm dim6}^2} \gg \sigma_{SM-{\rm dim8}}$. 

According to the jet flavor distribution in Fig.~\ref{fig:njnbnc:sig:wjets}, the $c$-jet population in $W+$jets is close to 16\% followed by a few \% of $b$-tagged jets. Given the mistag rates in Eq.~(\ref{eq:bc:mistagrate}), we find that the dominant contribution to $W+$jets comes from $c$-jet faking $b$-jet followed by $b$-jet and light jets faking $b$-jet (last two have similar sizes). While we used rather conservative mistag rate for $c$-jet, any improvement will further reduce $W+$jet background. However, note that the signal from the $bc\tau\nu$ type operator has a benefit from the higher mistag rate for $c$-jet as the extra-jet can be easily $c$-flavored as is seen in left panel of Fig.~\ref{fig:njnbnc:sig:wjets}. 

\begin{table}[t]
\centering
\scalebox{1.0}{
\begin{tabular}{c|ccc}  
\hline
$m_T$ [TeV]  & \quad $m_T < 0.5 $TeV  & \quad  $0.5 < m_T < 1 $TeV & \quad $m_T > 1$ TeV
\\
\hline \hline
$W$+jets   				& 181$\pm$ 25 	& 19.4$\pm$ 3.7  	& 0.18$\pm$ 0.05  \\[2.5pt]
$Z\rightarrow \nu\nu$+jets  	& 26.3$\pm$ 13   	& 3.44$\pm$ 1.8 	& 0.21$\pm$ 0.12 \\[2.5pt]
$t\bar{t}$ 					& 173$\pm$ 38  	& 15.8$\pm$ 4.9 	& 0.29$\pm$ 0.03 \\[2.5pt]
$Z/\gamma^*\rightarrow ll$+jets& 17.9$\pm$ 4.3  	& 0.49$\pm$ 0.25  	&  $(4.2 \pm 0.4) \times 10^{-5}$ \\[2.5pt]
$VV$  					& 10.5$\pm$ 2.7 	& 2.91$\pm$ 1.1  	& 0.35$\pm$ 0.05 \\[2.5pt] 
single-$t$ 					& 39.4$\pm$ 12  	& 1.80$\pm$ 0.75  	& 0.067$\pm$ 0.007  \\[2.5pt]
\hline
Total						& 448$\pm$ 49		& 43.8$\pm$ 6.5		& 1.10$\pm$ 0.14 \\[2.5pt]
\hline
\end{tabular}
}
\caption{\small Our estimate for SM background number of events in $p p \to \tau \nu + b$ at $\sqrt{s}$ =13 TeV and an integrated luminosity of 35.9 fb$^{-1}$. The systematic uncertainty in table was obtained by rescaling each uncertainty in the CMS analysis (see Table~\ref{tab:CMS:bkg}) with the ratio of events between the two analyses.}
\label{tab:bctaunub:bkg}
\end{table}

As an estimate of the systematic uncertainty for the backgrounds, we rescaled each uncertainty in the CMS analysis in Table~\ref{tab:CMS:bkg} with the ratio of events between the two analyses. These were summed in quadrature for the total number of background events. Our final background estimates for $pp \rightarrow \tau\nu +b$ are reported in Table~\ref{tab:bctaunub:bkg}. 

\begin{table}[t]
\centering
\scalebox{1.0}{
\begin{tabular}{c|ccc}  
\hline
$m_T$ [TeV]  & \quad $m_T < 0.5 $TeV  & \quad  $0.5 < m_T < 1 $TeV & \quad $m_T > 1$ TeV \\[2.5pt]
\hline \hline
$\tau\nu$ & 143 & 272 & 83.3 \\[3.5pt]
$\tau\nu$ with $b$-tagging & 100 & 83 & 25.6 \\
\hline
\end{tabular}
}
\caption{\small Our estimate of signal events for the benchmark model with $C_{VLL}^{23} = 1$, with an integrated luminosity of 35.9 fb$^{-1}$ at $\sqrt{s}$ =13 TeV, for the two analyses without and with $b$-tagging. }
\label{tab:sig:bornob}
\end{table}

As was mentioned in Section~\ref{sec:heavyflavor}, the $b$-tagging is beneficial as it suppresses mainly the SM contribution, $W+$jets for instance, while retaining most BSM signals from the operators with $b$-quark. Indeed, we can see by comparing two Tables~\ref{tab:CMS:bkg} and~\ref{tab:bctaunub:bkg} that the size of $W+$jets is significantly reduced by simply demanding $b$-tagged jet.  On the contrary, our benchmark signal with $C_{VLL}^{cb} = 1$ is reduced at most by a factor of three in presence of the $b$-tagging, as is illustrated in Table~\ref{tab:sig:bornob}.

\subsection{Studying angular distributions}
\label{sec:angular}

The heavy flavor tagging can improve the sensitivity on operators involving a $b$-quark but has little or no impact on the different tensor structures. In order to increase the sensitivity on these, the natural candidate are angular observables. Furthermore, in case of an observation of a deviation from the SM, studying angular distributions can help to address the degeneracy in operator space that would otherwise be present.

\begin{figure}[t]
\begin{center}
\includegraphics[width=0.50\textwidth]{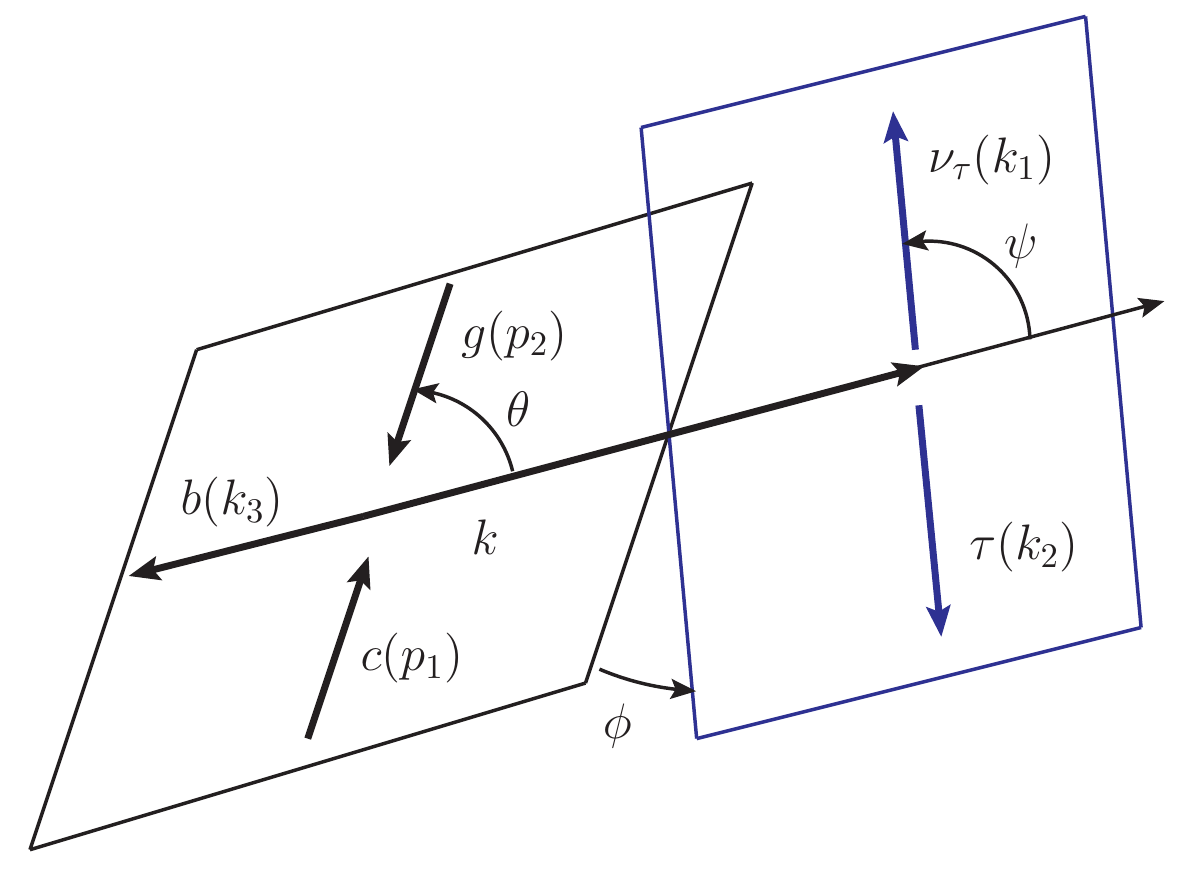}
\caption{\small The definition of three angles in our coordinate system for the process $cg \rightarrow \tau \nu b$. We factorized 2 to 3 process effectively as the product of 2 to 2 process and 1 to 2 process. The artificially introduced intermediate momentum $k$ corresponds to the momentum of the $\tau\nu$ system (whether or not it is associated with a resonance).}
\label{fig:2to3:coordinate}
\end{center}
\end{figure}
For better understanding of the angular dependence, we evaluate analytically the partonic differential cross sections with respect to various angles defining our coordinate system of 2 to 3 process, consisting of five variables, namely $\sqrt{\hat{s}},\, z,\, \theta,\, \psi,\, \phi$. The three angles are illustrated in Fig.~\ref{fig:2to3:coordinate}. When the process is thought of as 2 to 2 process like, for instance, $cg \rightarrow (\tau\nu) +b$ by treating $\tau\nu$ effectively as one particle (whether or not it is associated with the resonance), we use $\theta$ to refer to the polar angle in the rest frame of this effective 2 to 2 process. On the other hand, $\psi$ refers to the polar angle of the $\tau\nu$ system in its rest frame. The remaining angle $\phi$ denotes the relative angle between two planes of the $\tau\nu$ system and the aforementioned effective 2 to 2 process. The variable $z$ is the fraction of the partonic energy $\sqrt{\hat{s}}$ flowing into the $\tau\nu$ system. More detailed description is given in Appendix~\ref{app:sec:helicity:diff}.

\begin{figure}[t]
\begin{center}
\includegraphics[width=0.45\textwidth]{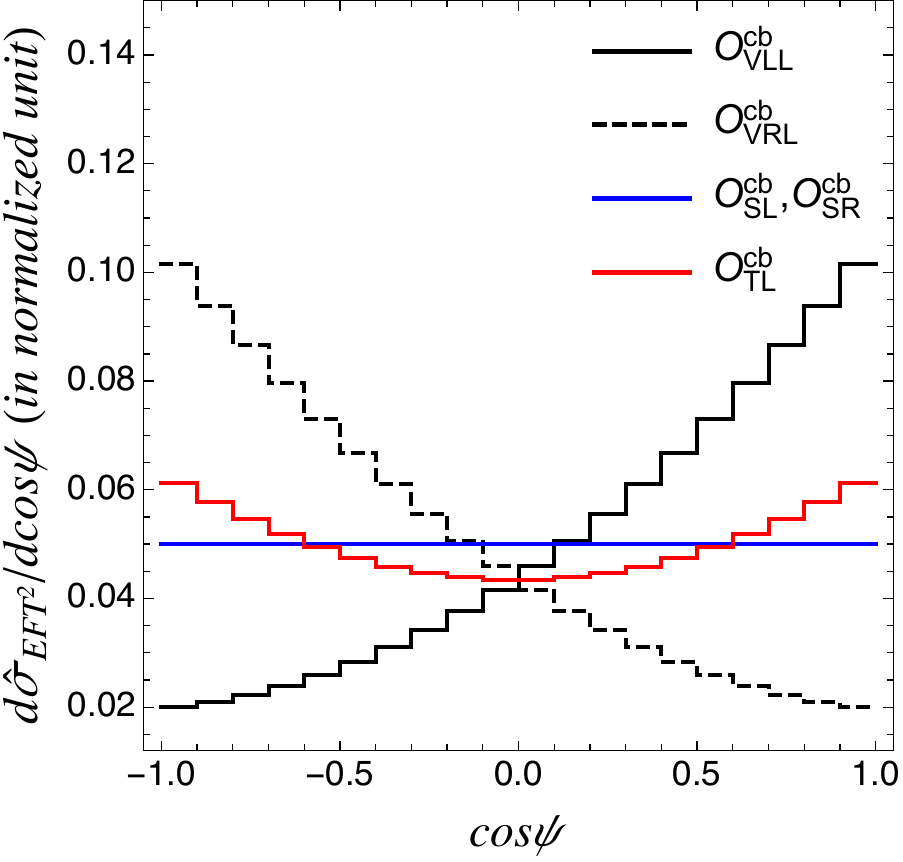} \quad
\includegraphics[width=0.45\textwidth]{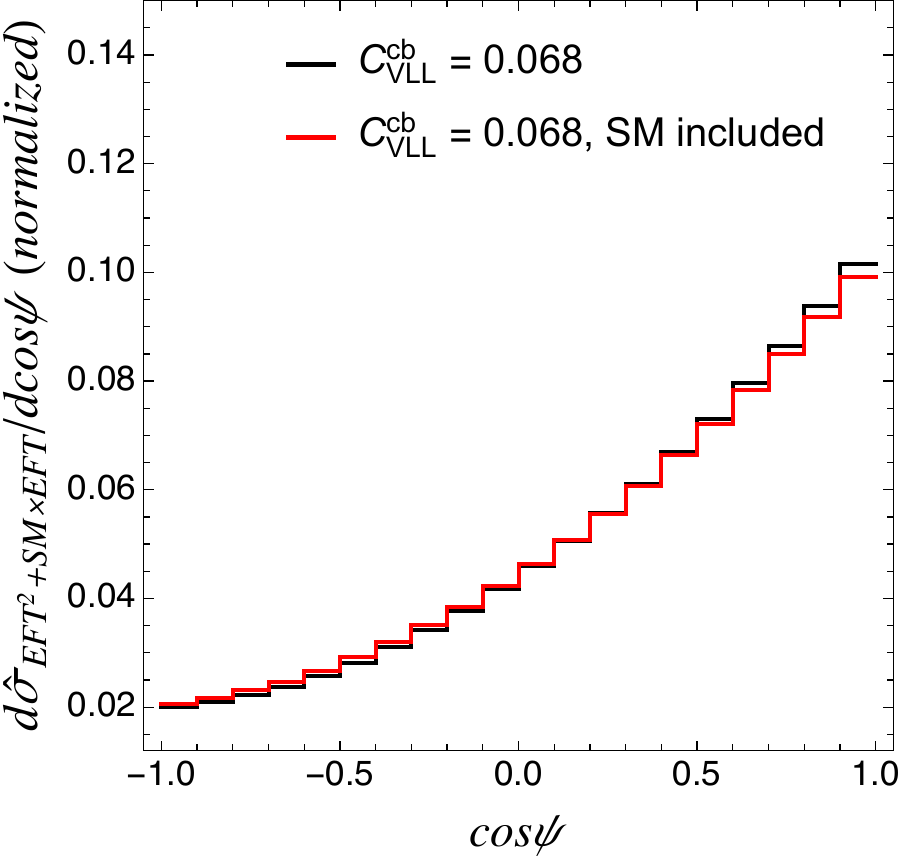}
\caption{\small Left: the normalized partonic differential cross section, $d{\hat\sigma}_{EFT^2}/d\cos\psi$, for an individual EFT coefficient by switching on each coefficient at a time. Right: similarly for the $\mathcal{O}_{VLL}^{cb}$ operator with the best-fit value $C^{cb}_{VLL}|_{\rm best\text{-}fit}=0.068$ including the SM contribution. In both plots, the energy of the system is fixed to be $\sqrt{\hat{s}} = 1$ TeV and $p_T(b) \geq 20$ GeV was imposed.}
\label{fig:angular:distr}
\end{center}
\end{figure}

Assuming, for simplicity, that all particles in the processes are massless and that all EFT coefficients are real, the partonic differential cross section from the BSM is evaluated to be
\begin{equation}\label{eq:ddsigma}
\begin{split}
  &\frac{d^2\hat\sigma_{EFT^2}(cg\rightarrow \tau\nu b)}{d\cos\theta d\cos\psi} 
  = \frac{\alpha_s   }{36864 \pi^2 }\frac{C_{VLL}^{cb\ 2} V^2_{cb}}{v^4} \frac{\hat{s}}{1-\cos\theta} \left  [ 
   144 \cos\theta  \left ( \cos\psi - \frac{1}{12}\cos 2\psi -\frac{1}{4} \right ) \right .
  \\[2.5pt]
 &\hspace{2.5cm} \left . \quad  -12\cos 2\theta \left ( \cos\psi - \frac{25}{12} \cos2 \psi - \frac{11}{12} \right )-4 \cos\psi +19 \cos 2 \psi +121 \right ]
 \\[2.5pt]
 &\hspace{2.0cm} + \frac{\alpha_s   }{36864 \pi^2 }\frac{C_{VRL}^{cb\ 2} V_{cb}^2}{v^4} \frac{\hat{s}}{1-\cos\theta} \left  [ 
   144 \cos\theta  \left ( - \cos\psi - \frac{1}{12}\cos 2\psi -\frac{1}{4} \right ) \right .
  \\[2.5pt]
 &\hspace{2.5cm} \left . \quad -12\cos 2\theta \left ( -\cos\psi - \frac{25}{12} \cos2 \psi - \frac{11}{12} \right ) + 4 \cos\psi +19 \cos 2 \psi +121 \right ]
 \\[2.5pt]
 &\hspace{2.0cm} +\frac{\alpha_s}{18432 \pi ^2} \frac{\left ( C_{SL}^{cb\ 2} + C_{SR}^{cb\ 2} \right ) V_{cb}^2}{v^4} \frac{\hat{s}}{1-\cos\theta} \left (4 \cos\theta +\cos 2\theta+27 \right )
 \\[2.5pt]
 &\hspace{2.0cm} + \frac{\alpha_s}{1152 \pi ^2} \frac{C^{cb\ 2}_{T}V^2_{cb}}{v^4} \frac{\hat{s}}{1-\cos\theta} 
 \left [ -14  \cos\theta \left ( \cos 2\psi + 1 \right ) + \frac{45}{2} \cos 2\theta \left ( \cos 2\psi + \frac{1}{9} \right )  \right .
 \\[2.5pt]
 &\hspace{2.5cm} \left . \quad   + \frac{15}{2} \left ( \cos 2\psi + \frac{11}{3} \right ) \right ]
 \\[2.5pt]
 &\hspace{2.0cm} +\frac{\alpha _s}{2304 \pi ^2} \frac{C_{SL}^{cb} C_{T}^{cb}V_{cb}^2}{v^4} \frac{\hat{s}}{1-\cos \theta}   \left (-36 \cos \theta -\cos 2 \theta + 5 \right )  \cos \psi~, 
\end{split}
\end{equation}
where the integration over $\phi$ and $z$ has been performed (see Appendix~\ref{app:sec:helicity:diff} for the full differential cross section before the integration). While $\mathcal{O}_{VLL}$ operator interferes with the SM contribution from the $W$ boson exchange, we have not generalized the differential cross section in Eq.~(\ref{eq:ddsigma}) to include it for a technical reason~\footnote{The propagator of the intermediate $W$ boson in the SM diagram carries the momentum squared of $(2z-1)\hat{s}$ and it becomes challenging to get any simple analytic expression out of the integration over $z$.}. Instead, we included the SM contribution numerically (see Fig.\ref{fig:angular:distr}). Although the SM contribution makes a visible effect for a lower energy, $\sqrt{\hat{s}} \ll \mathcal{O}({\rm TeV})$, its effect is found to be negligible around TeV scale for the chosen EFT coefficient in Fig.\ref{fig:angular:distr}. One notes that the interference term between $\mathcal{O}_{SL}$ and $\mathcal{O}_{T}$ in Eq.~(\ref{eq:ddsigma}) disappears in our massless limit upon integrating over the polar angle $\psi$~\footnote{The interference can survive through the imperfect cancellation in the $\phi$ integration when kinematic cuts are imposed. In our analysis, we have checked numerically that non-vanishing interference terms are small enough to be ignored.}. 

The distinction between operators with different Lorentz structures will be pronounced in the differential distribution of the polar angle $\psi$ of the $\tau\nu$ system, namely $d\hat\sigma/d\cos\psi$.
We can integrate the partonic cross section in Eq.~(\ref{eq:ddsigma}) over $\theta$. However, to avoid the singularity in the forward region, namely near $\theta \sim 0$, (from $t$-channel diagram of the process) as is evident in Eq.~(\ref{eq:ddsigma}), we need to impose a cut on the $p_T$ of the $b$-quark. The transverse momentum of the $b$-quark in our coordinate is $p_T(b) = \sqrt{\hat{s}}(1-z) \sin\theta$. For the given cut on $p_T(b) \geq p_{T\, min}$ and fixed energy $\sqrt{\hat{s}}$, the differential cross section is obtained by integrating over $\theta$ and $z$,
\begin{equation}
\begin{split}
 \frac{d\hat\sigma_{EFT^2}(cg\rightarrow \tau\nu b)}{d\cos\psi} 
=\int_{1/2}^{1-p_{T\, min}/\sqrt{\hat{s}}} dz \int_{\cos\theta_{min}(z)}^{\cos\theta_{max}(z)} d\cos\theta\ \frac{d^3\hat\sigma_{EFT^2}(cg\rightarrow \tau\nu b)}{d\cos\psi\, dz\, d\cos\theta}~,
\end{split}
\end{equation}
where the boundary values of $\cos\theta$ are given by
\begin{equation}
  \cos\theta_{max/min}(z) = \pm \sqrt{ 1 - \frac{p^2_{T\, min}}{\hat{s}(1-z)^2}} ~.
\end{equation}
We performed the integration numerically for a fixed partonic energy $\sqrt{\hat{s}} = 1$ TeV with $p_{T\, min} = 20$ GeV. The resulting differential angular distribution is shown in Fig.~\ref{fig:angular:distr}. As is evident in Fig.~\ref{fig:angular:distr}, the distribution of $d\hat\sigma/d\cos\psi$ looks promising as a discriminant for different Lorentz structure of four-fermion operators. However, the distributions  in Fig.~\ref{fig:angular:distr} could be far from the reality as they are affected by kinematic cuts. 

\begin{figure}[tph]
\begin{center}
\includegraphics[width=0.40\textwidth]{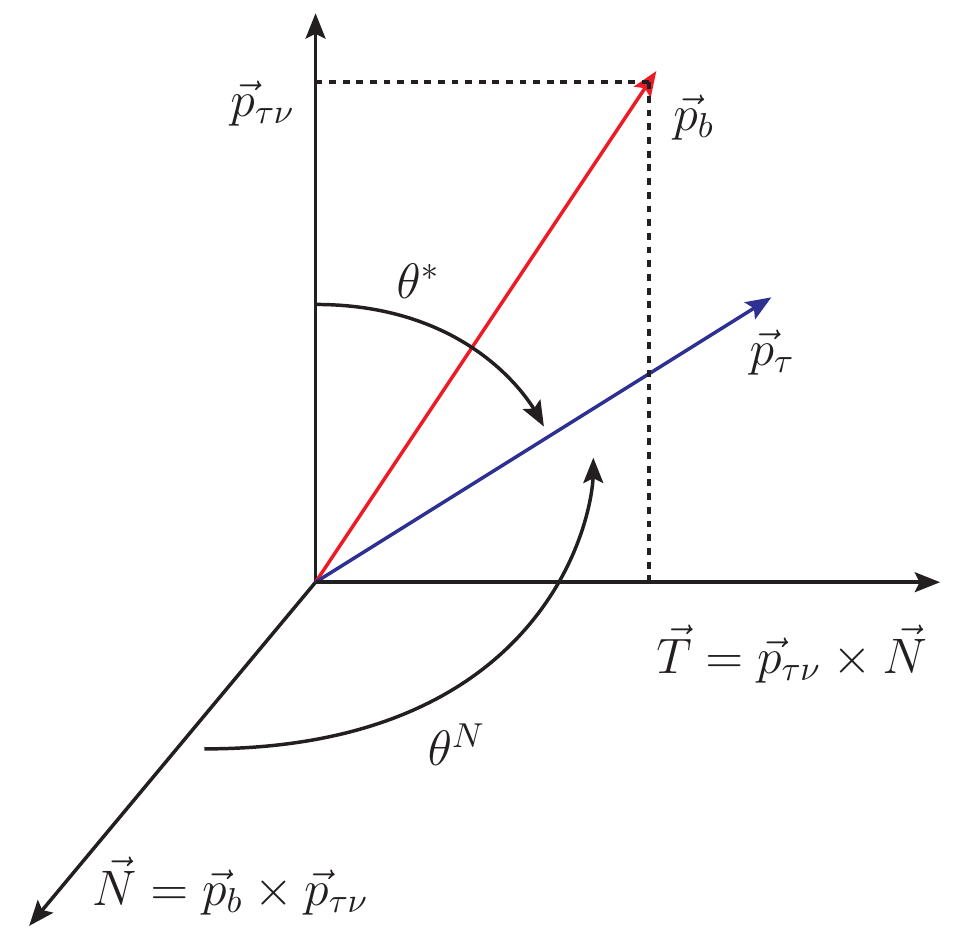}
\caption{\small Another choice for angular variables for the $p p \to \tau\nu b$ process. $\vec{p}_{\tau\nu}$ corresponds to the 3-vector of $\tau\nu$ system. $\vec{p}_{\tau}$ is the 3-vector in the $\tau\nu$ rest frame. $\vec{N}$ ($\vec{T}$) is the normal (tangential) 3-vector to the plane made by $\vec{p}_{\tau\nu}$ and $\vec{p}_b$.}
\label{fig:thetaN:thetaS}
\end{center}
\end{figure}
\begin{figure}[tph]
\begin{center}
\includegraphics[width=0.48\textwidth]{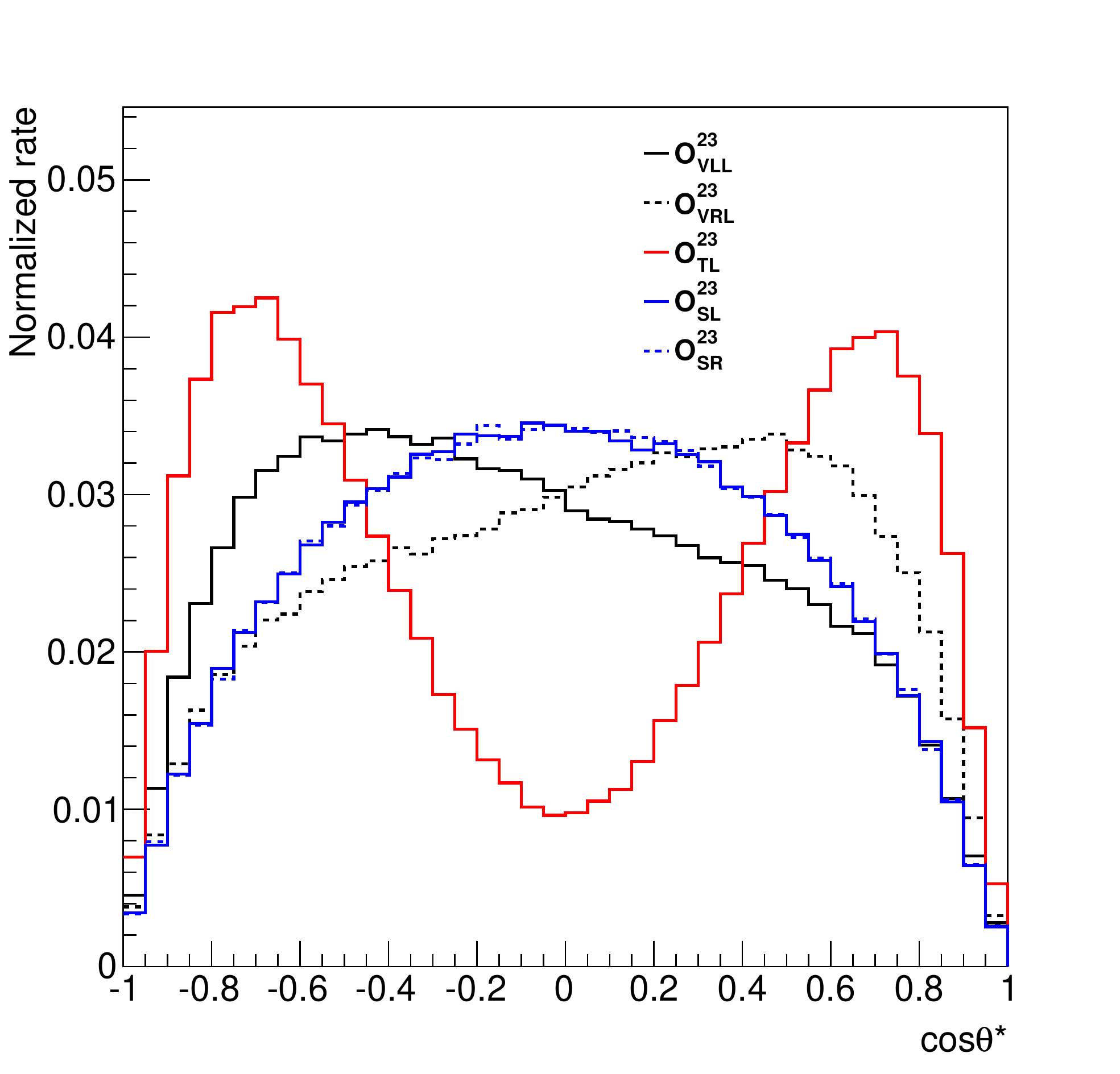}\quad
\includegraphics[width=0.48\textwidth]{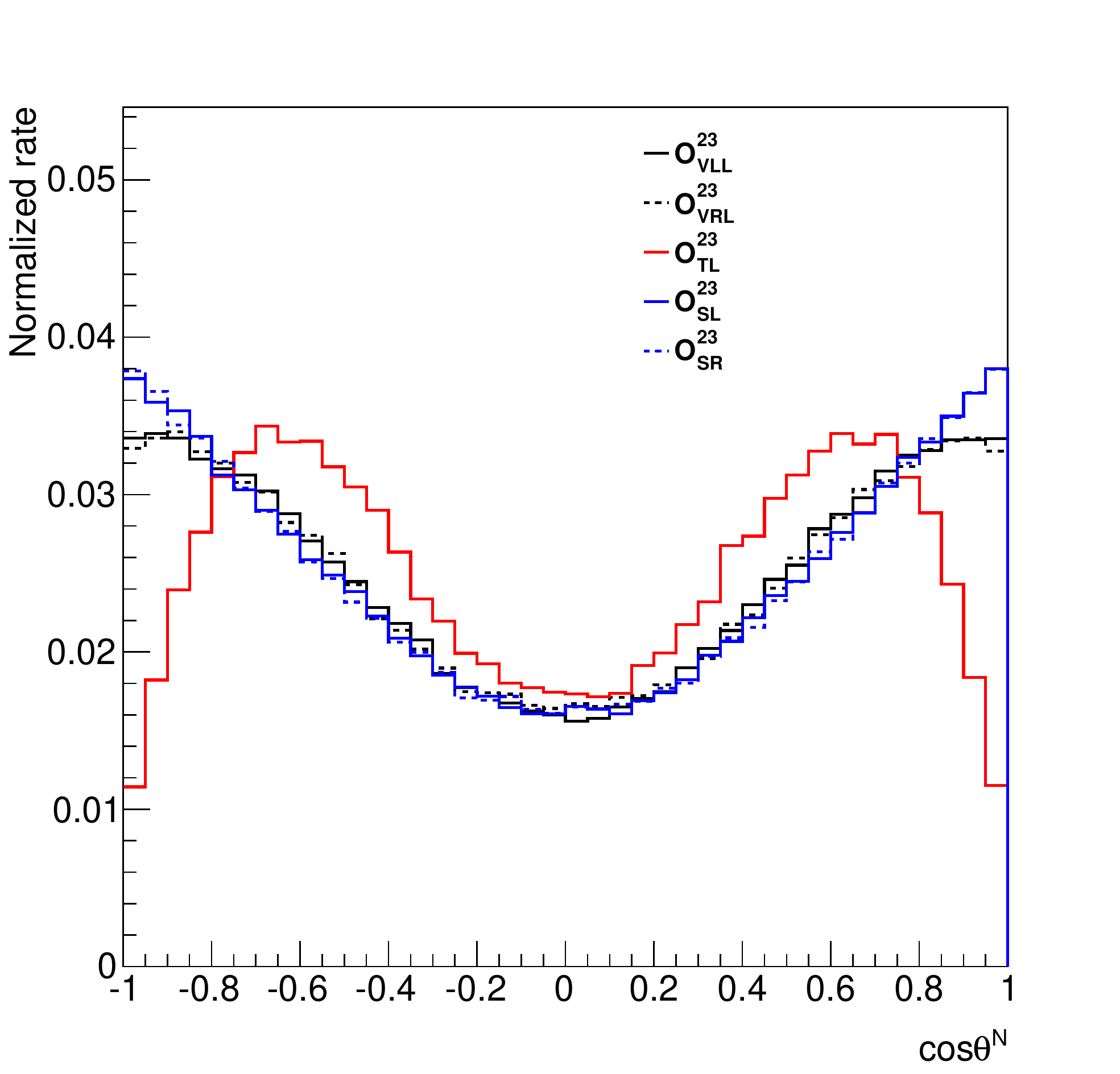}
\caption{\small Normalized differential cross sections, $d\sigma/d\cos\theta^*$ and $d\sigma/d\cos\theta^N$, for individual EFT coefficients (set equal to 1). The angles, $\theta^*$ and $\psi$, are related through $\theta^* = \pi - \psi$. Both plots are made using the partonic MC events of $pp\rightarrow \tau\nu b$ process generated by {\sc MadGraph5}. Events in both plots are required to satisfy $p_T(\tau) > 80$ GeV, $p^{\ miss}_T > 200$ GeV, $\triangle\phi (\vec{p}^{\ \tau}_T,\, \vec{p}^{\ miss}_T) > 2.4$, $0.7 <  p^\tau_T/p^{\ miss}_T < 1.3$ and $m_T > 500$ GeV.}
\label{fig:angular:distr:CMS}
\end{center}
\end{figure}

We investigate the implication of the kinematic cuts on the angular distributions using the partonic MC events of $pp\rightarrow \tau\nu b$ process in terms of the angular variables shown in Fig.~\ref{fig:thetaN:thetaS}, motivated by what has been explored in the single top process~\cite{AguilarSaavedra:2010nx}. The angular variables in Fig.~\ref{fig:thetaN:thetaS} are more suited for experimental measurements, whereas those in Fig.~\ref{fig:2to3:coordinate} were more convenient for the analytic evaluation. $\theta^N$ in Fig.~\ref{fig:thetaN:thetaS} is the angle of $\vec{p}_{\tau}$ with respect to the normal vector $\vec{N} = \vec{p}_b\times \vec{p}_{\tau\nu}$ whereas the variable $\theta^*$ is the polar angle of $\tau$ vector (denoted by $\vec{p}_{\tau}$) with respect to the $\tau\nu$ vector ($\vec{p}_{\tau\nu}$) in $\tau\nu$ rest frame. $\theta^*$ is related to $\psi$ in our coordinate through the relation, $\theta^* = \pi - \psi$. One could also define angle between $\vec{p}_\tau$ and $\vec{p}_b$ (3-vector of $b$) in Fig.~\ref{fig:thetaN:thetaS}. We found that its differential distribution is more pronounced, while having similar shapes, than that of $\cos\theta^*$.

After imposing the CMS type cuts in Section~\ref{sec:CMSvalidation} on $\tau$-lepton and missing transverse momentum, the resulting angular distributions are illustrated in Fig.~\ref{fig:angular:distr:CMS}. Comparing two plots in the left panels of Figs.~\ref{fig:angular:distr} and~\ref{fig:angular:distr:CMS}, we observe that both edges of the distributions in Fig.~\ref{fig:angular:distr:CMS} are depleted due to kinematic cuts~\footnote{For instance, the forward/backward region along the collider will be excluded due to kinematic cuts. When $\theta^* \sim 0,\, \pi$, the events will be similarly restricted by the same kinematic cuts.}. Interestingly, the distribution from the tensor operator becomes more pronounced. On the other hand, the distribution of $d\sigma/d\cos\theta^N$ in presence of kinematic cuts does not look promising.

We have not implemented the angular observables described in this section to our analysis as it requires more detailed study at the hadron level including a realistic reconstruction of neutrinos. We leave more comprehensive study on them for future work.

\section{Sensitivity on EFT coefficients}
\label{sec:EFTfit}

In this Section we present the limits on the EFT coefficients in Eq.~\eqref{eq:eft:banomaly:mbscale} obtained by recasting the CSM $\tau\nu$ analysis \cite{Sirunyan:2018lbg}.
We then compare the prospects for an integrated luminosity of 300~fb$^{-1}$ using the same analysis, with those derived from our analysis with a $b$-tagged jet.

For each of the three $m_T$ bins the total cross section is the sum of the SM background cross section, as detailed in the previous sections, and the EFT contribution consisting in the interference and quadratic terms of Eq.~\eqref{eq:xsec:EFTcoeff}.
From this cross section we build a log-likelihood by assuming the number of events in each bin follows a Gaussian distribution. Given the sufficiently large number of expected events in each bin, the central limit theorem assures us that using a Gaussian distribution instead of a Poisson one is a good approximation.
We thus have
\be
	\chi^2 \equiv - 2 \log \mathcal{L} = \sum_{\rm bin} \frac{1}{\sigma_{\rm bin}^2} \left[ \mathcal{L} (\sigma_{\rm SM, \, bin} + \sigma_{\rm EFT, \, bin}(C^{ij}_X) ) - N_{\rm ev, \, bin}^{\rm obs} \right]^2~,
	\label{eq:chiSQ}
\ee
where $\mathcal{L}$ indicates the luminosity, $\sigma_{\rm SM, \, bin}$ is the SM prediction for the cross section in each bin,  $\sigma_{\rm EFT, \, bin}(C^{ij}_X)$ is the EFT-dependent cross section, and $N_{\rm ev, \, bin}^{\rm obs}$ is either the observed number of events in that bin (for recasting the CMS analysis) or is fixed to the expected number of events in the SM for the prospects. The variance $\sigma_{\rm bin}^2$ is obtained, for each bin, by combining in quadrature the statistical and systematic uncertainty. Correlations between different bins are neglected since they are not reported by the experiment.

\begin{table}[p]
\centering
\scalebox{0.85}{
\begin{tabular}{|c|c|c|c|} 
\hline
	EFT coeff.		& CMS ($\mathcal{L}$=35.9 fb$^{-1}$)  & $\tau \nu$ - $\mathcal{L}$=300 fb$^{-1}$ & $\tau \nu b$ -  $\mathcal{L}$=300 fb$^{-1}$ \\[5pt]
\hline \hline
$|C^{11}_{SL}|$   & $1.5 \times 10^{-3}$ 	& $1.1 \times 10^{-3}$ & -- \\[3.5pt]
$|C^{12}_{SL}|$   & $9.8 \times 10^{-3}$ 	& $7.5 \times 10^{-3}$ & -- \\[3.5pt]
$|C^{13}_{SL}|$   & 2.2 		& 1.7 & 1.1 \\[3.5pt]
$|C^{21}_{SL}|$   & $1.6 \times 10^{-2}$ &	 $1.2 \times 10^{-2}$ & -- \\[3.5pt]
$|C^{22}_{SL}|$   & $9.8 \times 10^{-3}$ & $7.5 \times 10^{-3}$ & -- \\[3.5pt]
$|C^{23}_{SL}|$   & 0.33 		& 0.26 & 0.18 \\[3.5pt]
\hline
$|C^{23}_{SL}| = 4 |C^{23}_{T}|$ & 0.31 & 0.24 & 0.17 \\[3.5pt]
\hline
$|C^{11}_{SR}|$   & $1.5 \times 10^{-3}$ 	& $1.1 \times 10^{-3}$  & -- \\[3.5pt]
$|C^{12}_{SR}|$   & $9.9 \times 10^{-3}$ 	& $7.5 \times 10^{-3}$  & -- \\[3.5pt]
$|C^{13}_{SR}|$   & 2.2 		& 1.7 & 1.1 \\[3.5pt]
$|C^{21}_{SR}|$   & $1.6 \times 10^{-2}$ 	& $1.2 \times 10^{-2}$  & -- \\[3.5pt]
$|C^{22}_{SR}|$   & $9.7 \times 10^{-3}$	& $7.5 \times 10^{-3}$  & -- \\[3.5pt]
$|C^{23}_{SR}|$   & 0.33 		& 0.26 &  0.19\\[3.5pt]
\hline
$|C^{11}_{T}|$   & $8.5 \times 10^{-4}$	&  $6.5 \times 10^{-4}$ & -- \\[3.5pt]
$|C^{12}_{T}|$   & $5.5 \times 10^{-3}$	&  $4.2 \times 10^{-3}$ & -- \\[3.5pt]
$|C^{13}_{T}|$   & 1.3		& 0.97 & 0.57 \\[3.5pt]
$|C^{21}_{T}|$   & $9.4 \times 10^{-3}$	& $7.2 \times 10^{-3}$  & -- \\[3.5pt]
$|C^{22}_{T}|$   & $5.8 \times 10^{-3}$	& $4.5 \times 10^{-3}$  & -- \\[3.5pt]
$|C^{23}_{T}|$   & 0.20		& 0.16 & 0.099 \\[3.5pt]
\hline
$C^{11}_{VLL}$   & $[-0.40 , 3.2] \times 10^{-3}$	& $3.1 \times 10^{-4}$  & -- \\[3.5pt]
$C^{12}_{VLL}$   & $[-0.78, 1.1]  \times10^{-2}$	& $9.0 \times 10^{-3}$  & -- \\[3.5pt]
$C^{13}_{VLL}$   & $[-2.1, 2.1]$				& 1.6 & 0.93 \\[3.5pt]
$C^{21}_{VLL}$   & $[-1.4, 1.8] \times 10^{-2}$		& $1.4 \times 10^{-2}$  & -- \\[3.5pt]
$C^{22}_{VLL}$   & $[-0.73, 1.2]  \times 10^{-2}$	& $1.5 \times 10^{-3}$ & -- \\[3.5pt]
$C^{23}_{VLL}$   & $[-0.33, 0.34]$				& $[-0.25, 0.26]$ & $[-0.14, 0.15]$ \\[3.5pt]
\hline
$|C^{11}_{VRL}|$   & $1.5 \times 10^{-3}$	& $1.1 \times 10^{-3}$  & -- \\[3.5pt]
$|C^{12}_{VRL}|$   & $9.6 \times 10^{-3}$	& $7.3 \times 10^{-3}$  & -- \\[3.5pt]
$|C^{13}_{VRL}|$   & 2.1				& 1.6 & 0.94 \\[3.5pt]
$|C^{21}_{VRL}|$   & $1.6 \times 10^{-2}$	& $1.2 \times 10^{-2}$  & -- \\[3.5pt]
$|C^{22}_{VRL}|$   & $9.6 \times 10^{-3}$	& $7.4 \times 10^{-3}$  & -- \\[3.5pt]
$|C^{23}_{VRL}|$   & 0.33		& 0.26 & 0.15 \\[3.5pt]
\hline
\end{tabular}
}
\caption{\small In the second column we show the recasted 95\% CL intervals for the EFT coefficients defined in Eq.~(\ref{eq:eft:banomaly:mbscale}), evaluated at the 1TeV scale and switched on one at a time, using the CMS $\tau\nu$ analysis at $\sqrt{s}=13$ TeV and an integrated luminosity of 35.9~fb$^{-1}$. In the third and fourth column we show the prospects with a luminosity of 300~fb$^{-1}$ for the same $\tau \nu$ analysis and $\tau \nu + b$-jet analysis we propose, respectively.}
\label{tab:result}
\end{table}

\begin{table}[p]
\centering
\scalebox{0.85}{
\begin{tabular}{|c|c|c|c|} 
\hline
	SMEFT coeff.		& CMS ($\mathcal{L}$=35.9 fb$^{-1}$)  & $\tau \nu$ - $\mathcal{L}$=300 fb$^{-1}$ & $\tau \nu b$ -  $\mathcal{L}$=300 fb$^{-1}$ \\[5pt]
\hline \hline
$[C^{(3)}_{lq}]_{3311}$   & $[-0.39 , 3.2] \times 10^{-3}$	& $3.1 \times 10^{-4}$  & -- \\[3.5pt]
$[C^{(3)}_{lq}]_{3312}$   & $[-1.1, 2.6]  \times10^{-3}$	& $[-0.85, 2.2] \times 10^{-3}$  & -- \\[3.5pt]
$[C^{(3)}_{lq}]_{3313}$  & $[- 7.9, 7.9] \times 10^{-3}$				& $[-6.1, 6.0] \times 10^{-3}$ & $ 3.5 \times 10^{-3}$ \\[3.5pt]
$[C^{(3)}_{lq}]_{3322}$  & $[-4.8, 8.8] \times 10^{-3}$		& $[-3.5, 7.1] \times 10^{-3}$  & -- \\[3.5pt]
$[C^{(3)}_{lq}]_{3323}$  & $[-1.3, 1.4]  \times 10^{-2}$	& $[-1.0, 1.1] \times 10^{-2}$ & $5.8 \times 10^{-3}$ \\[3.5pt]
$[C^{(3)}_{lq}]_{3333}$  & $[-0.33, 0.33]$				& $[-0.25, 0.26] $ & $[-0.14, 0.15]$ \\[3.5pt]
\hline
$|[C^{(1)}_{lequ}]_{3311}|$   & $ 2.9 \times 10^{-3}$	& $2.2 \times 10^{-3}$  & -- \\[3.5pt]
$|[C^{(1)}_{lequ}]_{3312}|$   & $ 7.2 \times 10^{-3}$	& $5.5 \times 10^{-3}$  & -- \\[3.5pt]
$|[C^{(1)}_{lequ}]_{3321}|$   & $ 4.4 \times 10^{-3}$	& $3.4 \times 10^{-3}$  & -- \\[3.5pt]
$|[C^{(1)}_{lequ}]_{3322}|$   & $ 1.9 \times 10^{-2}$	& $1.5 \times 10^{-2}$  & -- \\[3.5pt]
$|[C^{(1)}_{lequ}]_{3331}|$  & $ 1.6 \times 10^{-2}$	& $1.2 \times 10^{-2}$  & $0.80 \times 10^{-2}$ \\[3.5pt]
$|[C^{(1)}_{lequ}]_{3332}|$  & $ 2.8 \times 10^{-2}$	& $2.2 \times 10^{-2}$  & $1.5 \times 10^{-2}$ \\[3.5pt]
\hline
$|[C^{(3)}_{lequ}]_{3311}|$   & $ 1.7 \times 10^{-3}$	& $1.3 \times 10^{-3}$  & -- \\[3.5pt]
$|[C^{(3)}_{lequ}]_{3312}|$   & $ 4.2 \times 10^{-3}$	& $3.2 \times 10^{-3}$  & -- \\[3.5pt]
$|[C^{(3)}_{lequ}]_{3321}|$   & $ 2.5 \times 10^{-3}$	& $1.9 \times 10^{-3}$  & -- \\[3.5pt]
$|[C^{(3)}_{lequ}]_{3322}|$   & $ 1.1 \times 10^{-2}$	& $0.87 \times 10^{-2}$  & -- \\[3.5pt]
$|[C^{(3)}_{lequ}]_{3331}|$  & $ 0.93 \times 10^{-2}$	& $0.71 \times 10^{-2}$  & $0.42 \times 10^{-2}$ \\[3.5pt]
$|[C^{(3)}_{lequ}]_{3332}|$  & $ 1.7 \times 10^{-2}$	& $1.3 \times 10^{-2}$  & $0.83 \times 10^{-2}$ \\[3.5pt]
\hline
$|[C_{ledq}]_{3311}|$   & $ 3.0 \times 10^{-3}$	& $2.3 \times 10^{-3}$  & -- \\[3.5pt]
$|[C_{ledq}]_{3312}|$   & $ 6.5 \times 10^{-3}$	& $5.0 \times 10^{-3}$  & -- \\[3.5pt]
$|[C_{ledq}]_{3313}|$   &	 	$ 0.17$		& $0.13$  & -- \\[3.5pt]
$|[C_{ledq}]_{3321}|$   & $ 4.5 \times 10^{-3}$	& $3.5 \times 10^{-3}$  & -- \\[3.5pt]
$|[C_{ledq}]_{3322}|$   & $ 1.4 \times 10^{-2}$	& $1.1 \times 10^{-2}$  & -- \\[3.5pt]
$|[C_{ledq}]_{3323}|$   & $ 0.42 \times 10^{-3}$	& $0.32$  & -- \\[3.5pt]
$|[C_{ledq}]_{3331}|$   & $ 1.6 \times 10^{-2}$	& $1.2 \times 10^{-2}$  & $0.81 \times 10^{-2}$ \\[3.5pt]
$|[C_{ledq}]_{3332}|$   & $ 2.7 \times 10^{-2}$	& $2.0 \times 10^{-2}$  & $1.5 \times 10^{-2}$ \\[3.5pt]
$|[C_{ledq}]_{3333}|$   & $ 0.66 \times 10^{-3}$	& $0.51$  & 0.37 \\[3.5pt]
\hline
\end{tabular}
}
\caption{\small In the second column we show the recasted 95\% CL intervals for the SMEFT coefficients defined in Eq.~(\ref{eq:basis:linear}), evaluated at the 1TeV scale and switched on one at a time, using the CMS $\tau\nu$ analysis at $\sqrt{s}=13$ TeV and an integrated luminosity of 35.9~fb$^{-1}$. In the third and fourth column we show the prospects for a  luminosity of 300~fb$^{-1}$, for the same $\tau \nu$ analysis and the $\tau \nu + b$-jet analysis we propose, respectively.}
\label{tab:result_SMEFT}
\end{table}

\subsection{Sensitivity from CMS $\tau \nu$ analysis and future prospects}

In order to extract the present EFT limits from the CMS measurements in the $\tau \nu$ channel, we fix the integrated luminosity to 35.9 fb$^{-1}$ and employ the CMS prediction for SM background events, see Table~\ref{tab:CMS:bkg}. We also use their estimate for the systematic uncertainty in each bin and combine it in quadrature with the statistical uncertainty.
We checked that using the CMS prediction for the SM backgrounds or our results doesn't affect in a sizeable way the results of the fit.

By setting the integrated luminosity to 300~fb$^{-1}$ and the number of events to the expected number in the SM, we obtain the future prospects for the EFT limits. We scale both statistical and systematic uncertainties as $\sqrt{\mathcal{L}}$, assuming that also systematic uncertainties will decrease with time thanks to improved SM computations and understanding of the detector performance. 
We avoid extrapolating to the full HL-LHC luminosity since it is expected that the analysis will qualitatively improve with more data, for example thanks to finer binning in the transverse mass that will be allowed when more events are collected, as well as improved experimental techniques.

The present limits and future prospects on all the EFT coefficients, switched on one at a time, are collected in Table~\ref{tab:result} (second and third column, respectively). We also derived 2D limits in all pairs of mass-basis EFT coefficients and checked that no relevant correlations are present, as expected from the fact that coefficients with different fermion flavor or chirality do not interfere with each other. The present limits obtained from the CMS analysis are in agreement with those derived in~\cite{Greljo:2018tzh}, comparing 2D limits with those reported in~\cite{Shi:2019gxi} we also find a good agreement.

Using the relations in Eq.~\eqref{eq:LEFTtoSMEFT} we translate the $\chi^2$ of Eq.~\eqref{eq:chiSQ} as function of the SMEFT coefficients in Eq.~\eqref{eq:basis:linear}. The corresponding single-coefficient limits are shown in Table~\ref{tab:result_SMEFT}.
In this scenario the only large correlation between coefficients is between the $[C^{(3)}_{lq}]_{3333}$ and $[C^{(3)}_{lq}]_{3323}$ coefficients, since for both the leading contribution to $p p \to \tau \nu$ is mainly due to the same $b c \to \tau \nu$ partonic process, as will be discussed in more details below.

In the supplementary material \texttt{chSQ\_LEFT\_CMS36fb.m} and \texttt{chSQ\_SMEFT\_CMS36fb.m} we provide the complete $\chi^2$ functions for the CMS recast in the two EFT bases, so that limits can be easily derived in any specific direction in the EFT coefficient space.

\subsection{Sensitivity from the $\tau \nu + b$ analysis}

In a completely analogous manner we obtain the future prospects for the proposed $\tau \nu$ analysis with an associated $b$-jet, discussed in Section~\ref{sec:taunub}.
We use the estimate for the SM background contributions, and their systematic uncertainty, reported in Table~\ref{tab:bctaunub:bkg} and the cross-section dependence on EFT operators with a $b$-quark, obtained with the same analysis.

The expected 95\% CL intervals for each coefficient, taken one at a time, with an integrated luminosity of 300~fb$^{-1}$ are collected in the fourth column of Tables~\ref{tab:result} and \ref{tab:result_SMEFT}. Comparing with the expected bounds obtained for the same integrated luminosity from the analysis without the $b$-jet requirement (third column) we observe a 30$\div$35\% improvement on the sensitivity on those EFT coefficients.
This improvement, with same luminosity, is larger than the one obtained when increasing the luminosity from 36 to 300~fb$^{-1}$ with the standard analysis.
This improvement from the $b$-tagging is consistent with what has been found in the $p p \to \mu \mu (+ b\text{-jet})$ channel in~\cite{Afik:2018nlr}, where the limit, for a luminosity of $36$ fb$^{-1}$, improved by $\sim 33\%$ when compared with the analysis without the $b$-tag done in~\cite{Greljo:2017vvb}.

\section{Flavor physics from collider tails}
\label{sec:flavor}

In this section we discuss what information on the flavor structure of New Physics can be extracted from high-$p_T$ tails of $p p \to \tau \nu (+b)$ at LHC, and how this compares with limits from low-energy flavor processes.
This topic has already been the focus of several works in recent years, see \cite{Greljo:2018tzh,Fuentes-Martin:2020lea} for $\tau \nu$ searches and \cite{Cirigliano:2012ab,deBlas:2013qqa,Gonzalez-Alonso:2016sip,Faroughy:2016osc,Farina:2016rws,Greljo:2017vvb,Afik:2018nlr,Afik:2019htr,Angelescu:2020uug,Fuentes-Martin:2020lea} for other leptonic final states.

Since the main focus of our work is in the high-energy tails with a $b$-tagged jet, we concentrate on operators involving a $b$ quark.
For the purpose of illustration, among the operators in Eq.~(\ref{eq:eft:banomaly:mbscale}) we focus for the moment on the left-handed vector operator $\mathcal{O}_{VLL}$.
The two charged-current contact interactions involving a $b$ quark are $ c b \to \tau \nu$ and $u b \to \tau \nu$, generated by the $\CC_{cb}$ and $\CC_{ub}$ coefficients, respectively.
Since, by assumption, the new physics mediators should be above the energy scale of collisions, these coefficients should be matched to the SMEFT operators in Eq.~(\ref{eq:basis:linear}) (see Eq.~\eqref{eq:LEFTtoSMEFT} for the relations):
\be \begin{split} \label{eq:HpTcoeff:cb}
	(c b  \to \tau \nu) \qquad \CC_{cb} \equiv V_{cb} C_{VLL}^{cb} &=  [C^{(3)}_{l q}]_{33  13} V_{cd} +  [C^{(3)}_{l q}]_{33  23} V_{cs}  + [C^{(3)}_{l q}]_{33 33} V_{cb}~, \\
	(u b  \to \tau \nu) \qquad	\CC_{ub} \equiv V_{ub} C_{VLL}^{ub} &=  [C^{(3)}_{l q}]_{33  13} V_{ud} +  [C^{(3)}_{l q}]_{33  23} V_{us}  + [C^{(3)}_{l q}]_{33 33} V_{ub} ~.
\end{split}\ee
The three $[C^{(3)}_{l q}]_{33 i3}$ coefficients involved in these partonic transitions also generate, via CKM misalignment, contributions to other transitions involved in $p p \to \tau \nu$:
\be \begin{split}
	(u_i s \to \tau \nu) \qquad	\CC_{u_i s} \equiv V_{is} C_{VLL}^{is} &=  [C^{(3)}_{l q}]_{33  32} V_{ib}  =  [C^{(3)}_{l q}]_{33  23}^* V_{ib} ~, \\[1.5pt]
	(u_i d \to \tau \nu) \qquad	\CC_{u_i d} \equiv V_{id} C_{VLL}^{id} &=  [C^{(3)}_{l q}]_{33  31} V_{ib}  =  [C^{(3)}_{l q}]_{33  13}^* V_{ib} ~.
	\label{eq:HpTcoeffs}
\end{split}\ee
Note however that these transitions do not contribute to $ p p \to \tau \nu b$.
Depending on the specific direction in UV flavor space of the SMEFT coefficients $[C_{lq}^{(3)}]_{33 i3}$ the collider signal rate can be enhanced with respect to the contribution arising only from $\CC_{cb}$, thanks to the different parton luminosities and CKM factors.

Let us consider $\mathcal{C}_{cb}$ and $\mathcal{C}_{ub}$, that contribute also to $p p \to \tau \nu b$.
The naive estimate of the interference term between the SM and BSM amplitudes, taking into account the PDF luminosity, is
\be\label{eq:SM:BSM:inter}
 \sigma_{\rm INT}(\hat{s}) \sim \CC_{cb} \LL_{cb} \hat{\sigma}_{\rm SM-EFT}^{cb} + \CC_{ub} \LL_{ub} \hat{\sigma}_{\rm SM-EFT}^{ub}  \approx
 \CC_{cb} \LL_{cb} \hat{\sigma}_{\rm SM-EFT}^{cb}\left ( 1 + \frac{V_{ub}}{V_{cb}}\  \frac{\mathcal{L}_{ub}}{\mathcal{L}_{cb}} \ \frac{\mathcal{C}_{ub}}{\mathcal{C}_{cb}} \right )~,
\ee
where $\hat{\sigma}^{ij}$ denotes the partonic cross section and $\mathcal{C}_{cb}$, $\mathcal{C}_{ub}$ were pulled out of the partonic cross sections for clear comparison, whereas the quadratic terms are
\be\label{eq:SM:BSM:quad}
 \sigma_{\rm QUAD}(\hat{s}) \sim |\CC_{cb}|^2 \LL_{cb} \hat{\sigma}_{\rm EFT^2}^{cb} + |\CC_{ub}|^2 \LL_{ub} \hat{\sigma}_{\rm EFT^2}^{ub}  \approx
 |\CC_{cb}|^2 \LL_{cb} \hat{\sigma}_{\rm EFT^2}^{cb}\left ( 1 +  \frac{\mathcal{L}_{ub}}{\mathcal{L}_{cb}} \ \frac{\left |\mathcal{C}_{ub} \right |^2}{\left | \mathcal{C}_{cb} \right |^2} \right )~.
\ee
Switching on a single $[C_{lq}^{(3)}]_{33i3}$ coefficient at a time, the interference and quadratic terms in Eqs.~(\ref{eq:SM:BSM:inter}) and (\ref{eq:SM:BSM:quad}) become
\begin{equation}\label{eq:SMvsBSM:eft}
  \CC_{cb} \LL_{cb} \hat{\sigma}_{\rm SM-EFT}^{cb}
  \left ( 1 + \frac{V_{ub}}{V_{cb}}\ \frac{\mathcal{L}_{ub}}{\mathcal{L}_{cb}} \ \kappa_{i3} \right )~ \quad {\rm and}\quad
  |\CC_{cb}|^2 \LL_{cb} \hat{\sigma}_{\rm EFT^2}^{cb}
  \left ( 1 +  \frac{\mathcal{L}_{ub}}{\mathcal{L}_{cb}}  \ \kappa^2_{i3} \right )~,
\end{equation}
where $\kappa_{i3} = V_{ui}/V_{ci}$  ($ = 4.22 ,\, 0.24 ,\, 0.09$ for $i=1,2,3$, respectively) and the PDF luminosity ratio $\mathcal{L}_{ub} / \mathcal{L}_{cb} \approx (13, 24, 50)$ for partonic scattering energy of $\sqrt{q^2} = (0.5, 1, 2)$ TeV, respectively, given collision energy $\sqrt{s} = 13$ TeV. Also, the quadratic terms in the EFT are larger than the interference with the SM for the parameter space and energy range relevant for the bounds.

\begin{figure}[p]
\begin{center}
\includegraphics[height=6.84cm]{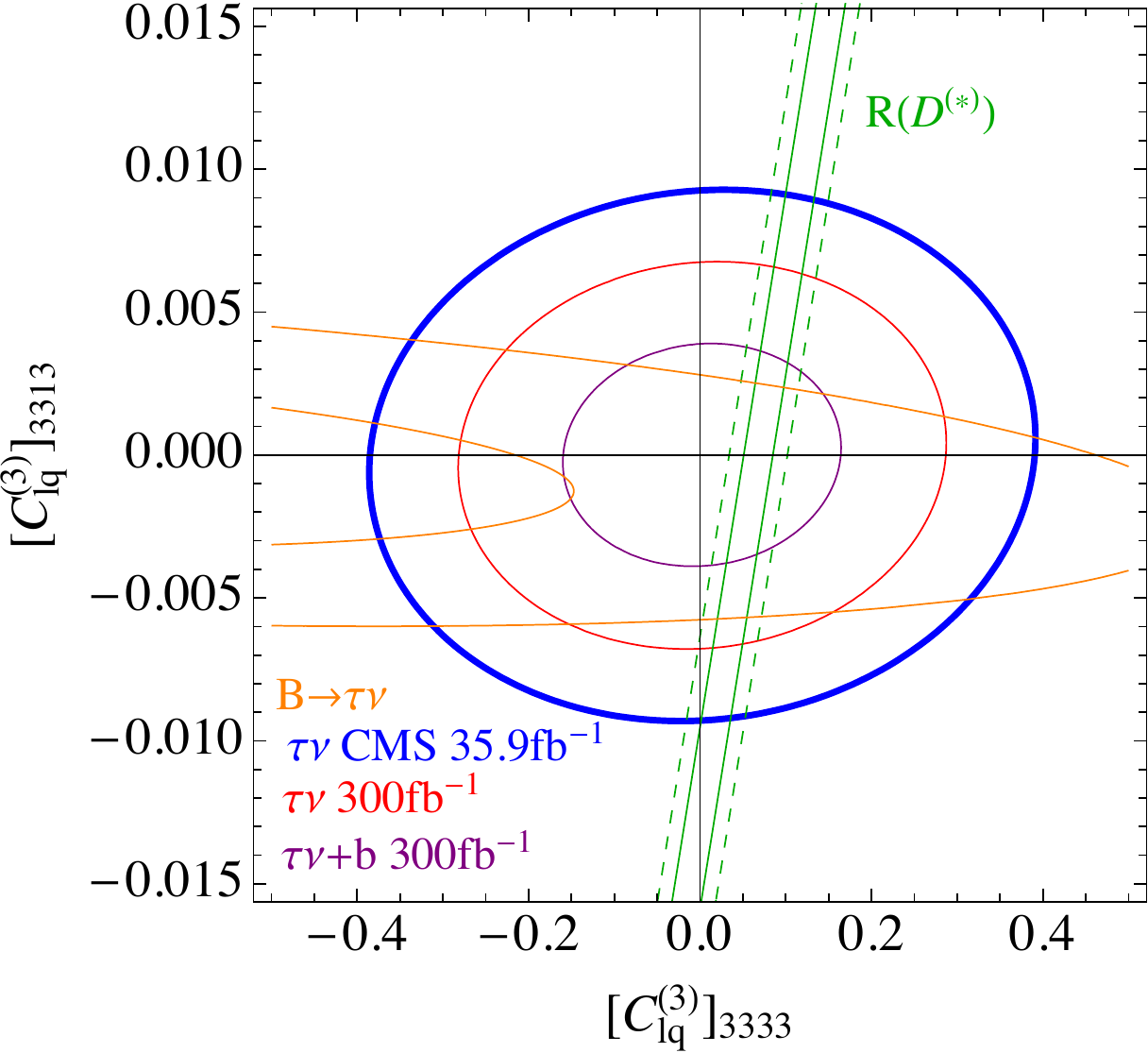} ~
\includegraphics[height=6.8cm]{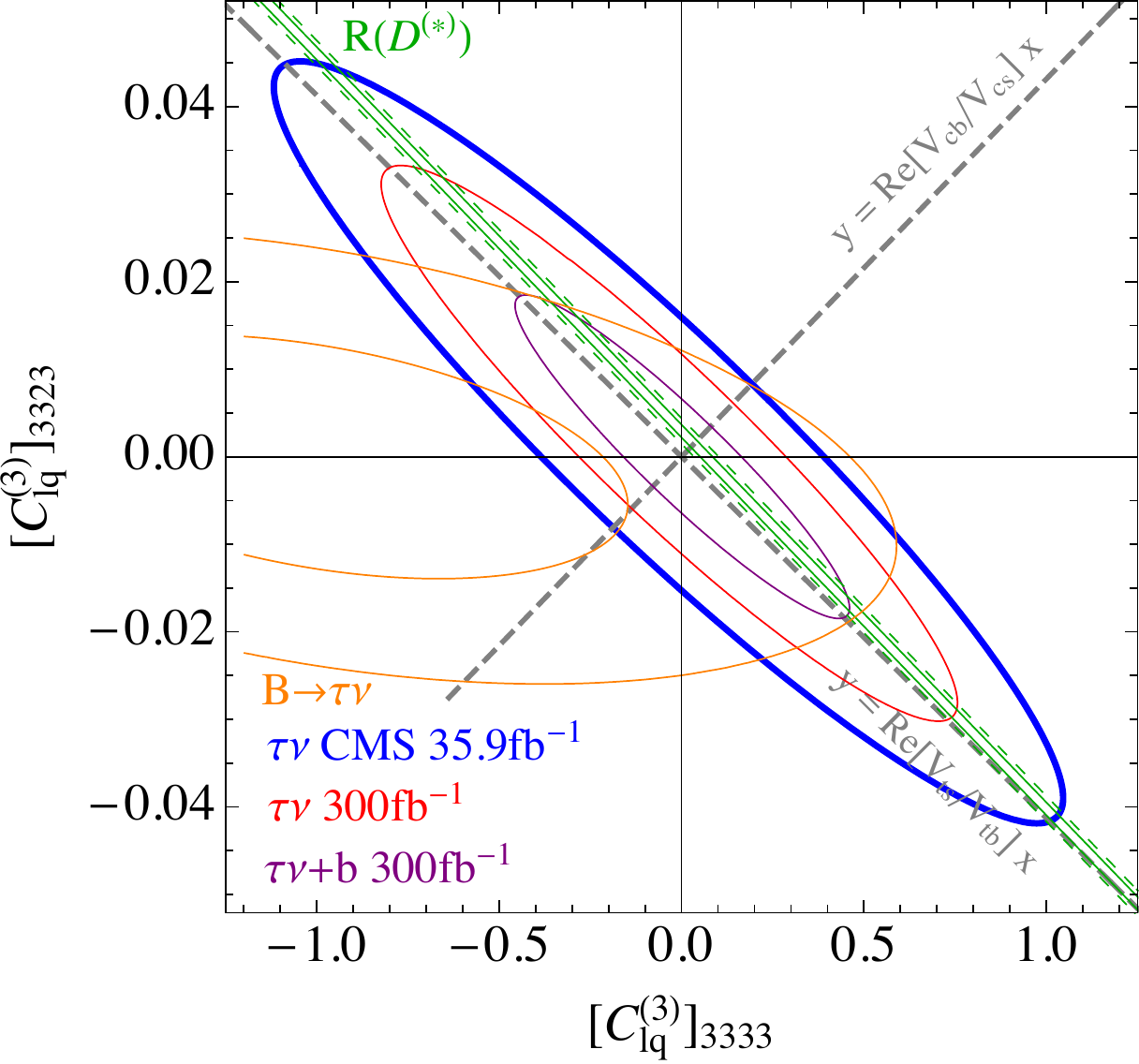}\\[0.2cm]
\includegraphics[height=6.8cm]{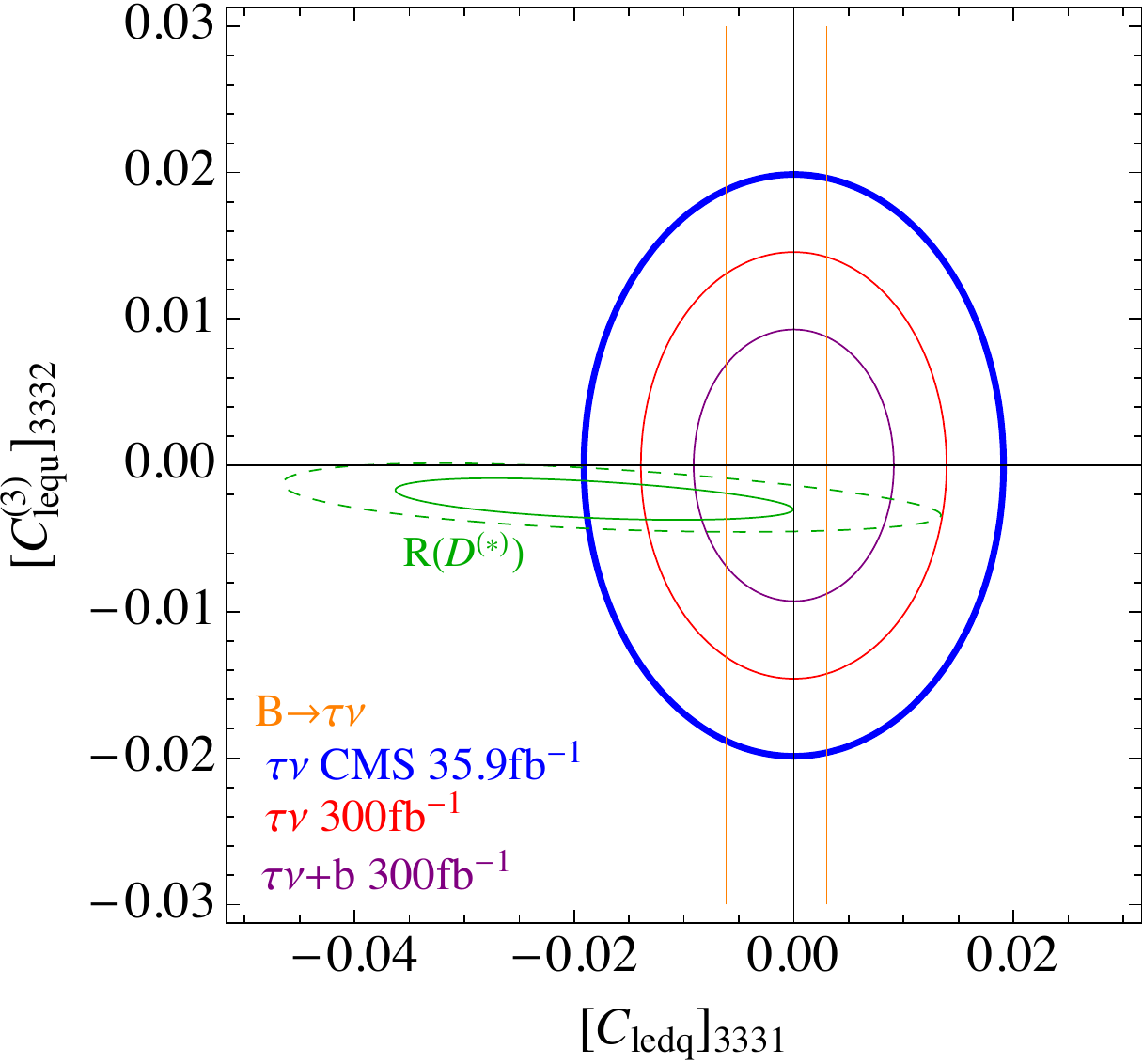}
\includegraphics[height=6.8cm]{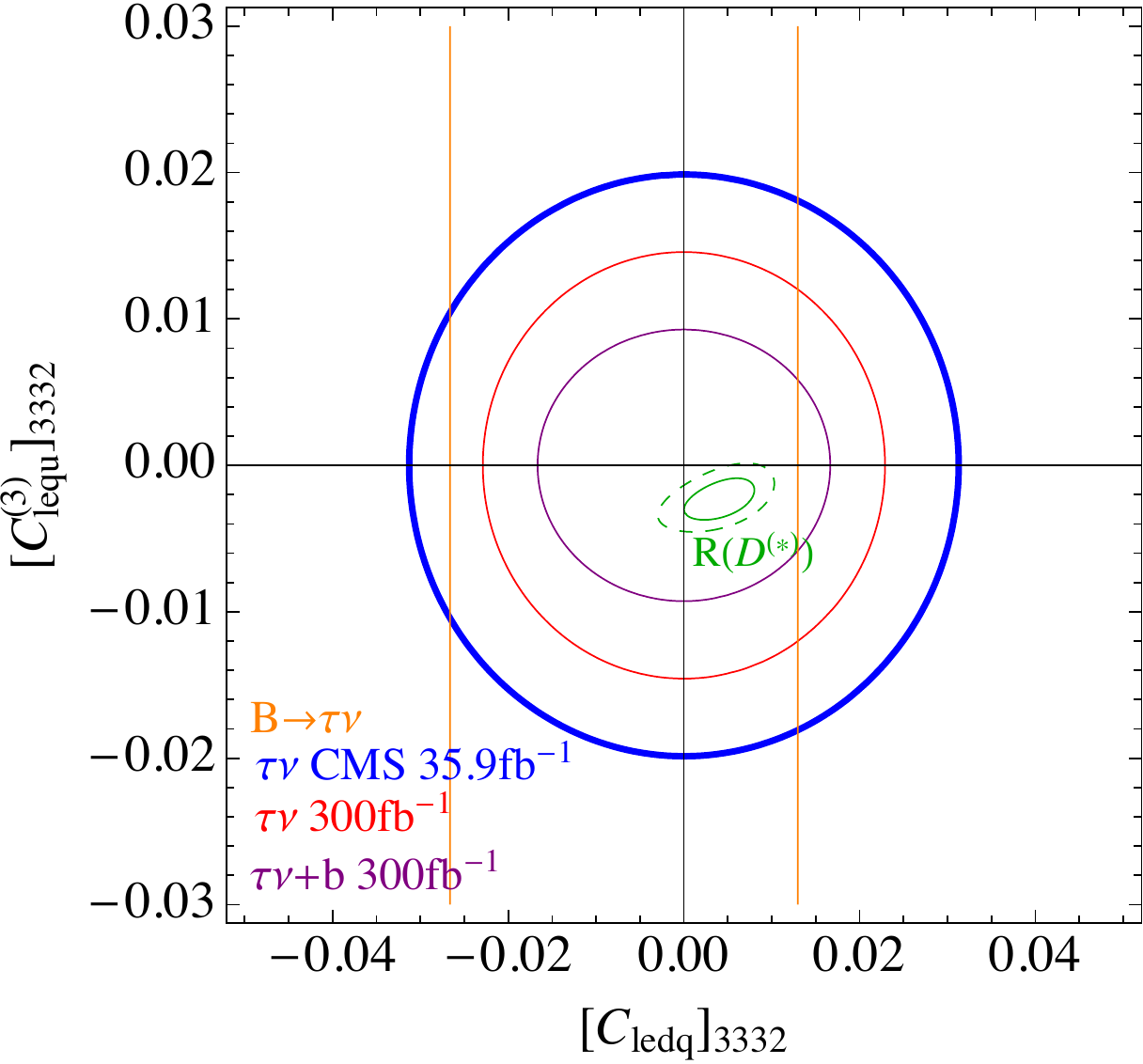}\\[0.2cm]
\includegraphics[height=6.8cm]{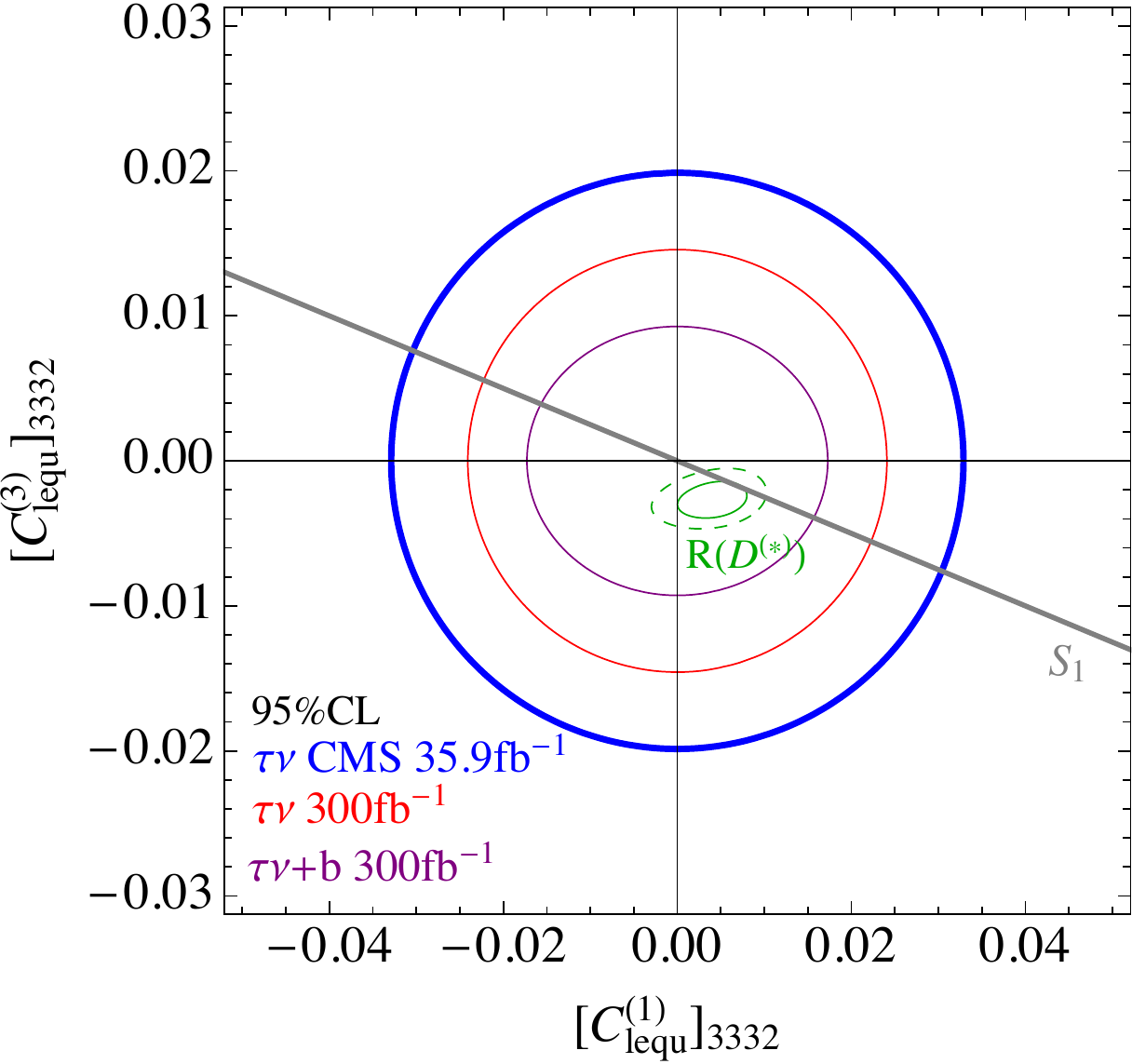}
\caption{\small 95\% CL limits and prospects in several pairs of SMEFT coefficients, while other operators are set to zero. 
The solid (dashed) green lines are $1 (2) \sigma$ contours from the $R(D^{(*)})$ fit \cite{Amhis:2016xyh}, while orange lines are 95\% CL limits from $B \to \tau \nu$.}
\label{fig:2Dsmeft_bound}
\end{center}
\end{figure}

In case of the $[C_{lq}^{(3)}]_{33 13}$ coefficient, the contribution from the $ub$ initial state to the quadratic term of the cross section is enhanced by a large factor $\sim (4.22)^2\ (\mathcal{L}_{ub}/\mathcal{L}_{cb})$ compared to the $cb$ initial state and thus dominates.
For the $[C_{lq}^{(3)}]_{33 23}$ coefficient, on the other hand, the contributions from $ub$ and $cb$ initial states are of the same order at 1 TeV, with $ub$ ($cb$) becoming more important at higher (lower) energies.
For $[C_{lq}^{(3)}]_{33 33}$, the suppression due to the small numerical coefficient $|V_{ub}/V_{cb}|^2 = \kappa^2_{33} = (0.09)^2$ is not compensated by the enhancement due to the up-quark PDF even up to 2 TeV of scattering energy, therefore the contribution from $ub$ initial state will be subdominant with respect to the one from $cb$ (see~\cite{Greljo:2018tzh} for a related discussion).

Another potentially interesting case would be the contribution from $tb$ initial states for the operator with $[C_{lq}^{(3)}]_{33 33}$. The relative contribution from $tb$ initial states will roughly scale like $|V_{tb}/V_{cb}|^2\ (\mathcal{L}_{tb}/\mathcal{L}_{cb})$ (with $|V_{tb}/V_{cb}| \sim 24.9$) up to the different phase space contribution. Although top PDF in the proton is negligibly small, including a top quark in the final state, $p p \to \tau \nu + t b$, would modify completely the set of backgrounds and it may be worth investigating.

To illustrate quantitatively this discussion, we show in Fig.~\ref{fig:2Dsmeft_bound} the 95\%CL limits and 
prospects in the planes ($[C^{(3)}_{lq}]_{3333}$, $[C^{(3)}_{lq}]_{3313}$) [top-left panel], ($[C^{(3)}_{lq}]_{3333}$, $[C^{(3)}_{lq}]_{3323}$)[top-right panel], including those from the $\tau \nu + b$ analysis described in this work.
The sizeable correlation in the top-right panel of Fig.~\ref{fig:2Dsmeft_bound} is due to the fact that both limits arise mostly from the same partonic process $c b \to \tau \nu$, that is, the $C_{VLL}^{cb}$ coefficient, while the weaker limit in the perpendicular direction arises from $u b \to \tau \nu$, due to the CKM suppression.
In the central panels of Fig.~\ref{fig:2Dsmeft_bound} we show the limits in the planes ($[C_{ledq}]_{3331}$, $[C^{(3)}_{lequ}]_{3332}$) [left] and  ($[C_{ledq}]_{3332}$, $[C^{(3)}_{lequ}]_{3332}$) [right]. For the $[C_{ledq}]_{333 k}$ coefficients, a reasoning analogous to the one illustrated above for $[C^{(3)}_{lq}]_{33i3}$ applies.
Finally, in the bottom panel of Fig.~\ref{fig:2Dsmeft_bound} we illustrate the constraints in the plane of the scalar and tensor operators $[C^{(1)}_{lequ}]_{3332}$ and $[C^{(3)}_{lequ}]_{3332}$. The gray line represents the relation predicted by the single-mediator exchange of the $S_1 \sim ({\bf \bar 3}, {\bf 1}, -1/3)$ leptoquark.

The constraints from high-$p_T$ tails should be compared with those derived from low-energy flavor processes. In particular, the most sensitive observables to the $\CC_{cb}$ and $\CC_{ub}$ coefficients are the LFU ratios $R(D^{(*)}) = \Br(B \to D^{(*)} \tau \nu)/\Br(B \to D^{(*)} \ell \nu)$ and the leptonic decay $B \to \tau \nu$, respectively.

Taking the latest global fit results on $R(D^{(*)})$, updated in the Spring of 2019, \cite{Amhis:2016xyh}, the anomalous measurements can be reproduced, for example, for $C_{VLL}^{cb}({\rm TeV}) = 0.068 \pm 0.017$ or $C_{SL}^{cb}({\rm TeV}) = - 4 C_{T}^{cb}({\rm TeV}) \in [0.062 \div 0.093]_{1\sigma}$~\footnote{The combination $C_{SL}^{cb}({\rm TeV}) = - 4 C_{T}^{cb}({\rm TeV})$ is generated at the UV matching scale by integrating out the leptoquark $S_1 \sim ({\bf \bar 3}, {\bf 1}, -1/3)$. The QCD RG evolution down to $m_b$ modifies it to $C_{SL}^{cb}(mb) \approx - 8 C_{T}^{cb}(mb) \in [0.113 \div 0.170]_{1\sigma}$, which is the value quoted in \cite{Bardhan:2019ljo}.}, as well as for other combinations of coefficients. See for example Refs.~\cite{Bardhan:2019ljo,Shi:2019gxi,Murgui:2019czp,Blanke:2019qrx,Alok:2019uqc} for updated EFT fits of $R(D^{(*)})$ and related observables. 

The branching ratio $\Br(B \to \tau \nu)$ is given by
\be
	\text{Br}(B^- \to \tau^- \bar\nu) = \text{Br}(B^- \to \tau^- \bar\nu)_{\rm SM} \left| 1 + C_{VLL}^{ub} + \frac{m_B^2}{2 m_\tau (m_b + m_u)} (C_{SR}^{ub} - C_{SL}^{ub}) \right|^2~,
	\label{eq:Btaunu}
\ee
where $\text{Br}(B^- \to \tau^- \bar\nu)_{\rm SM} = (7.92 \pm 0.55) \times 10^{-5}$ \cite{Banelli:2018fnx} and the combination of experimental measurements is $\text{Br}(B^- \to \tau^- \bar\nu)_{\rm exp} = (1.09 \pm 0.24) \times 10^{-4}$ \cite{Tanabashi:2018oca}.
Taking one coefficient at a time, the $2\sigma$ limits are:
\be
	C_{VLL}^{ub}(m_b) \in [-0.13, 0.41]~, \quad
	C_{SR}^{ub}(m_b) - C_{SL}^{ub}(m_b) \in [-0.07, 0.22]~.
\ee

For all the 2D planes in Fig.~\ref{fig:2Dsmeft_bound} we also show with solid (dashed) green lines the $1 (2) \sigma$ contour from the $R(D^{(*)})$ fit (the RG evolution from $m_b$ up to 1 TeV is included, which is relevant for scalar and tensor operators \cite{Gonzalez-Alonso:2017iyc}) and with orange lines the 95\% CL limit from $B \to \tau \nu$.
Comparing the low-energy limits with those from high-$p_T$ tails, we see that the expected sensitivity at 300 fb$^{-1}$ with our analysis with the $b$-tagging starts to probe regions not already excluded by flavor measurements (see, for example, the upper two panels).
Furthermore, while the leptonic decay $B \to \tau \nu$ only tests the specific combination of EFT coefficients in Eq.~\eqref{eq:Btaunu}, LHC searches put independent limits on all of them, thanks to the vanishing interference between different coefficients.
One could also expect that the limits from high-$p_T$ tails will improve substantially with HL-LHC, thanks to larger number of events, finer binning, and possibly the addition of angular distributions.

Going beyond operators with a $b$ quark, let us consider low-energy observables constraining other $u_i d_j \tau \nu$ contact interactions. The other quark pairs are $ud$, $us$, $cd$, and $cs$.
The most sensitive observable to the first two are $\tau^- \to \nu \pi^- (K^-)$ decays, while charm transitions are tested in (semi-)tauonic tau decays. In order to compare the sensitivity reach on EFT operators let us focus for simplicity on left-handed operators $C_{VLL}^{ij}$.
Tau decays to pions and Kaons are tested at the per-mille level, and the limits can be written as \cite{Pich:2013lsa}:
\be\begin{split}
	\Gamma_{\tau \to \pi} / \Gamma_{\pi \to \mu} &\to |1 + C_{VLL}^{ud}| = 0.9962 \pm 0.0027~, \\
	\Gamma_{\tau \to K} / \Gamma_{K \to \mu} &\to |1 + C_{VLL}^{us}| = 0.9858 \pm 0.0070~, 
\end{split}\ee
providing the following $2\sigma$ intervals: $C_{VLL}^{ud} \in [-9.2, 1.6] \times 10^{-3}$, $C_{VLL}^{us} \in [-2.8, -0.02] \times 10^{-2}$. Note that the latter does not include zero due to some tension with the SM. Comparing these limits with those from $p p \to \tau \nu$ in Table~\ref{tab:result}, we observe that present LHC constraints are comparable, while future limits will be stronger.
For what regards the comparison with $D$ meson decays, a detailed analysis was done recently in~\cite{Fuentes-Martin:2020lea}, to which we refer for details. The limits obtained from (semi-)leptonic decays are $C_{VLL}^{cd} \in [-0.21, 0.27] $ and $C_{VLL}^{cs} \in [-1.4, 7.0] \times 10^{-2}$. Also in this case the high-$p_T$ limits are stronger.

\subsection{Collider limits for Rank-One-Flavor-Violation}

In several new physics scenarios, the UV physics responsible for the contributions in $R(D^{(*)})$ and $p p \to \tau \nu$ couples only to a specific combination of left-handed quarks. For example, the vector-leptoquark $U_1^\mu \sim ({\bf 3}, {\bf 1}, 2/3)$, which is one of the favourite scenarios for addressing the $B$-anomalies, couples to left-handed fermions as
\be
	\LL_{U_1} \supset g_{i 3} ( \bar q_i \gamma_\mu l_3)  U_1^\mu + h.c.~,
\ee
where we selected only the coupling to the third generation leptons as it is the one contributing to $p p \to \tau \nu$. The coupling to left-handed quarks and third generation leptons is thus parametrized by the vector in ${\rm U}(3)_q$ flavor space $g_{i3}$. As a consequence, the structure of SMEFT coefficients is of rank-one: $[C^{(3)}_{l q}]_{33 ij} \propto g_{i3} \, g_{j3}^*$. The same rank-one structure is generated for other single-leptoquark scenarios and in all cases where the new physics flavor structure is induced via the mixing of SM quark doublets with a single vector-like fermion.
The generalisation of this flavor structure has been dubbed Rank-One-Flavor-Violation (ROFV) in \cite{Gherardi:2019zil}.

\begin{figure}[t]
\begin{center}
\includegraphics[width=0.34\textwidth]{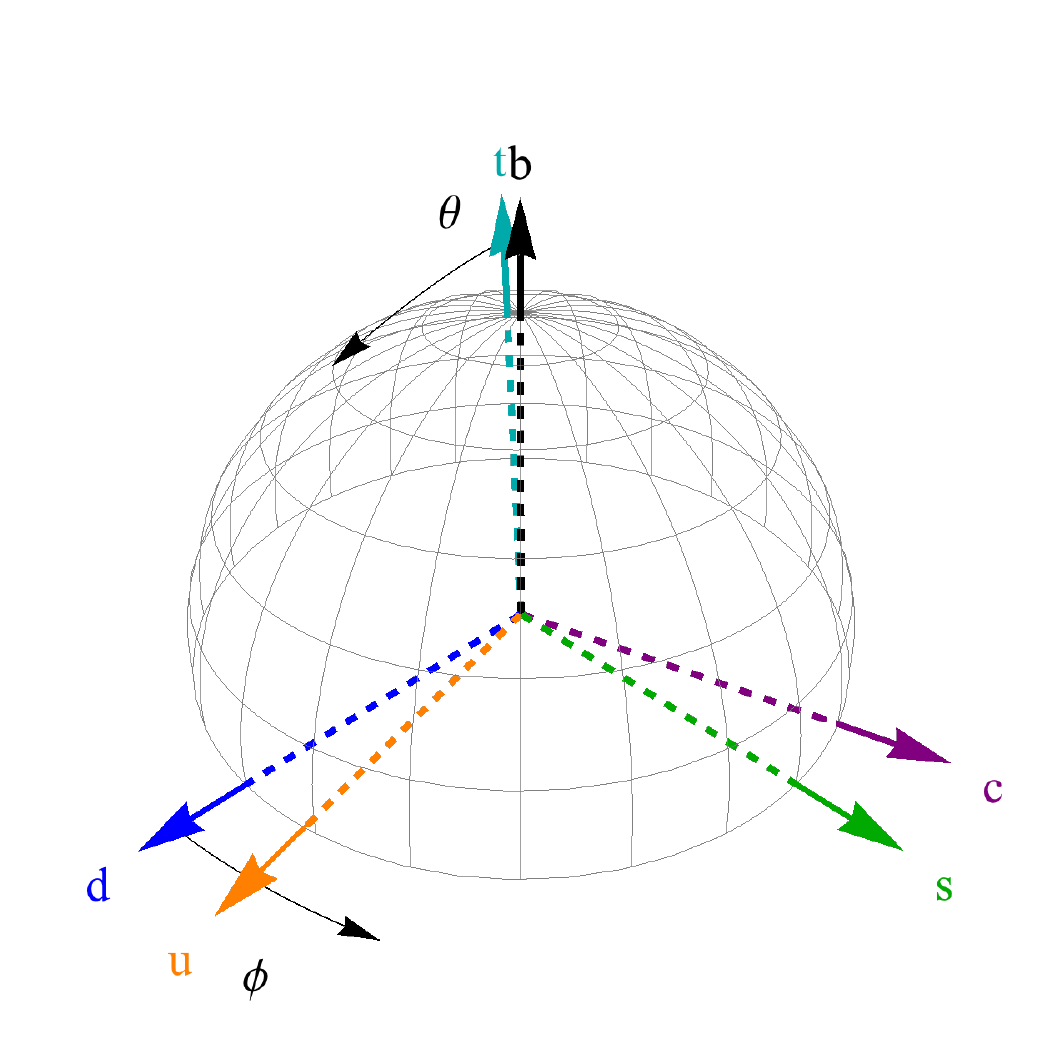}
\includegraphics[width=0.65\textwidth]{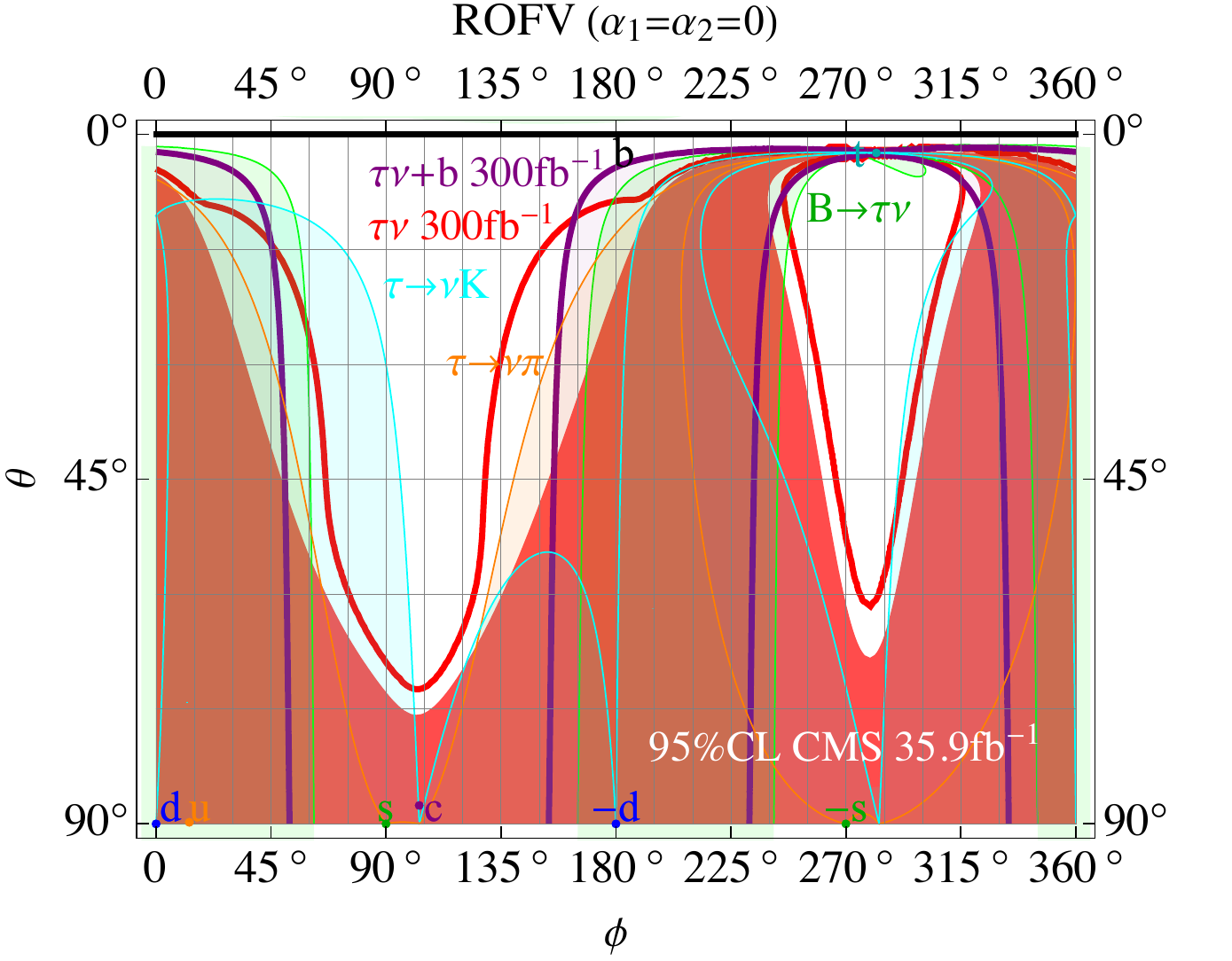}
\caption{\small Under the ROFV assumption  (for $\alpha_1=\alpha_2 =0$), we show with a red-coloured region the 95\% CL exclusion from the CMS $p p \to \tau \nu$ search, assuming that the best-fit value of $R(D^{(*)})$ is reproduced. The red and purple lines correspond to the expected 95\% CL limits with 300~fb$^{-1}$ from $pp \to \tau \nu$ and $pp \to \tau \nu b$, respectively. We also report the 95\% CL limits from $B \to \tau \nu$ (green), $\tau \to \nu K$ (cyan), and $\tau \to \nu \pi$ (orange).}
\label{fig:ROFVbounds}
\end{center}
\end{figure}

Following this hypothesis, we can parametrize the SMEFT coefficients as
\be
	[C^{(3)}_{l q}]_{33 i j} = C_L \hat{n}_i \hat{n}_j^*~,
	\label{eq:ROFVSMEFT}
\ee
where $C_L \in \mathbb{R}$ and $\hat{n}_i$ is a unitary vector in ${\rm U}(3)_q$ flavor space:
\be
	\hat{n} = \left( \begin{array}{c}	\sin \theta \cos \phi \, e^{i \alpha_1} \\
							\sin \theta \sin \phi \, e^{i \alpha_2} \\
							\cos \theta							\end{array} \right) ~,
\ee
with $\theta \in [0, \pi/2]$, $\phi \in [0,2\pi)$, $\alpha_{1,2} \in [- \pi/2, \pi/2]$. The directions aligned with down quarks form the chosen orthonormal basis in this space. Another possible choice of basis is the one aligned with up quarks, and the rotation between the two basis is given by the CKM matrix. In the left panel of Fig.~\ref{fig:ROFVbounds}, we draw the directions in $(\theta, \phi)$ associated with each SM quark direction (the corresponding $\alpha_{1,2}$ phases are not shown), and the corresponding $(\theta, \phi)$ are also shown as dots in the right panel.

With this parametrization, the combination of coefficients contributing to $R(D^{(*)})$ is given by
\be\begin{split}
	C_{VLL}^{cb} &= \frac{ \sum_{i} [C^{(3)}_{l q}]_{33 i3} V_{ci}}{V_{cb}} = 
		\frac{C_L \cos\theta}{V_{cb}} \left( \cos \theta V_{cb} + \sin \theta \sin \phi e^{i \alpha_2} V_{cs} + \sin \theta \cos \phi e^{i \alpha_1} V_{cd}\right)~.
		\label{eq:CVLL23_ROFV}
\end{split}\ee
By imposing that the \emph{measurement} of $C_{VLL}^{cb} = 0.068 \pm 0.017$ from $R(D^{(*)})$ is reproduced (for example at the best-fit point), we can fix the overall coefficient $C_L$ in Eq.~\eqref{eq:CVLL23_ROFV} as function of $C_{VLL}^{cb}$ (i.e. of $R(D^{(*)})$) and of the other parameters:
\be
	C_L = \frac{V_{cb} (C_{VLL}^{cb} )_{\rm best-fit}}{\cos\theta (\cos \theta V_{cb} + \sin \theta \sin \phi e^{i \alpha_2} V_{cs} + \sin \theta \cos \phi e^{i \alpha_1} V_{cd}) }~.
\ee
By plugging this in the definition of the ROFV structure of the SMEFT coefficients in Eq.~\eqref{eq:ROFVSMEFT}, all of the $[C^{(3)}_{l q}]_{33 i j}$ will depend only on $\theta$, $\phi$, and the two phases $\alpha_{1,2}$.
Fixing the phases, for instance to zero, we can study the collider limits (and prospects) from $p p \to \tau \nu \; (+b)$ in the plane of $\theta$ and $\phi$, see Fig.~\ref{fig:ROFVbounds}. 
In the same figure we also report the 95\% CL limits from low-energy processes sensitive to the $[C_{lq}^{(3)}]_{33ij}$ coefficients, specifically $B \to \tau \nu$ (green), $\tau \to \nu K$ (cyan), and $\tau \to \nu \pi$ (orange), as discussed in the previous section. The limits from $D$ meson decays, instead, are too weak for any value of $\theta$ and $\phi$.

\section{Conclusions}
\label{sec:conclusions}

In this work, we derived the sensitivity on the EFT coefficients of four-fermion operators from the collider study of both $\tau\nu$ and $\tau\nu+b$ channels at the LHC. The former has been extensively considered in literature, including in the context of the anomalous $R(D^{(*)})$ measurements \cite{Greljo:2018tzh} and comparing the sensitivity against $D$-meson decays \cite{Fuentes-Martin:2020lea}.
Using the existing CMS $\tau\nu$ analysis with an integrated luminosity of 35.9 fb$^{-1}$ at $\sqrt{s}=13$ TeV, we obtained the constraints for all EFT coefficients of four-fermion operators contributing to the process, both in the mass-eigenvalue basis and in the gauge-invariant Warsaw basis. The likelihood function for all coefficients is provided alongside this work in supplementary material, allowing the reader to study limits in any direction in the EFT space.

Using the $\tau \nu$ analysis to validate our procedure and estimates of all the background channels by comparing with CMS results, we studied the possibility of including bottom flavor tagging by devising a dedicated analysis.
The impact of $b$-jet tagging on the EFT is mainly two-folds. First, it allows to focus only on the subset of EFT operators involving a $b$-quark. Secondly, demanding a $b$-tagged jet suppresses the SM backgrounds while retaining most of the $b$-enriched signal events, thus improving the sensitivity on that subset of EFT coefficients. Comparing the sensitivity with the analysis without a $b$-jet, we estimate the improvement in the EFT limits to be approximately 30\%, for the same luminosity. 

We also discussed possible strategies for distinguish the operators with different Lorentz tensor structures using the angular observables. To isolate the pure angular properties from the impact of a realistic neutrino reconstruction, we worked at the parton level assuming  perfect neutrino reconstruction in both our analytic evaluation and the MC simulation. The differential distribution of the polar angle $\theta^*$ (equivalent to $\psi$ in our analytic evaluation) in $\tau\nu$ rest frame shows promising discrimination power. While the major limitation might be caused by a realistic neutrino reconstruction, it certainly deserves further detailed investigation. We provided full analytic differential cross section of 2 to 3 process in Appendix~\ref{app:sec:helicity:diff}. This allows to study analytically other sets of angular observables, as well as transforming easily to other coordinates.

Comparing the limits, and prospects, on pair of coefficients derived from mono-$\tau$ tails with those from low-energy flavor measurements, specifically $R(D^{(*)})$ and \mbox{$B \to \tau \nu$}, we find that in some cases the LHC prospects with a luminosity of 300 fb$^{-1}$ and the $b$-tagging requirement start to be competitive.
Furthermore, the higher luminosity reachable at HL-LHC is expected to further improve the picture by reducing the statistical uncertainty, allowing more $m_T$ bins at high energy, and possibly studying angular distributions.
A dedicated analysis is left for future work.

In several ultraviolet completions of the semi-tauonic operators $[\mathcal{O}_{lq}^{(3)}]_{33ij}$, the mediators couple to a single direction in quark-flavor space. For instance, this is automatic for single-leptoquark exchange \cite{Gherardi:2019zil}. In this case the EFT coefficient matrix is a rank-one tensor: $[C_{lq}^{(3)}]_{33ij} = C_L \hat{n}_i \hat{n}_j^*$, where $\hat{n}_i$ is the unitary vector in the quark flavor space. In our analysis, we have shown that, once the overall $C_L$ coefficient is fixed by the $R(D^{(*)})$ measurement and for a simplifying assumption for the phases, the collider limits on $[C_{lq}^{(3)}]_{33ij}$ can be recasted in terms of two angles that nicely visualize the collider probes of the flavor (mis)alignment. In the plane of these two angles, the collider limits from mono-tau tails are competitive with the constraint from $B \to \tau \nu$ and $\tau$ decays to $\nu \pi$ and $\nu K$.

The collider strategy we presented aims to improve the sensitivity to semileptonic four-fermion operators in the SMEFT containing a $b$-quark. This is part of a larger effort by the community, aimed at extracting the largest possible amount of information on EFT extensions of the Standard Model from LHC data, that will help us understanding the nature of NP better.

\section*{Acknowledgments}
We thank KyeongPil Lee and Hwidong Yoo for useful discussions regarding the experimental details and Monte-Carlo simulation. MS thanks Myeonghun Park for useful discussion. MS and MU were supported by the Samsung Science and Technology Foundation under Project Number SSTF-BA1602-04. 
DM acknowledges support by the INFN grant SESAMO, MIUR grant PRIN\_2017L5W2PT, and partial support by the European Research Council (ERC) under the European Union Horizon 2020 research and innovation programme, grant agreement 833280 (FLAY).


\appendix
\section{Cross section in terms of EFT coefficients}
\label{sec:XsecEFT}

In Table~\ref{tab:result}, we have presented one-dimensional sensitivity by switching on only one operator at a time. Here, we present our result for the EFT cross section, keeping all operators. Along with the background in Table~\ref{tab:CMS:bkg}, one should be able to construct the complete likelihood function in the space of EFT coefficients, c.f. Eq.~\eqref{eq:chiSQ}.
Following the description in Section~\ref{sec:CMSvalidation}, the BSM cross section, in fb, after imposing the same cuts as those in CMS analysis using 35.9$^{-1}$fb at $\sqrt{s}=13$ TeV  takes the form,
\begin{equation}\label{app:eq:xsec:EFTcoeff}
 \left [ \sigma (pp\rightarrow \tau\nu) - \sigma_{SM} \right ]_{\text{with CMS cuts}} = C^{ij}_X\, \sigma^{ij,X}_{SM-EFT} +  (C^{ij}_X)^2\, \sigma^{ij,X}_{EFT^2}~,
\end{equation}
where the interference terms for three $m_T$ bins are given by
\begin{equation}
\begin{split}
\left . C^{ij}_X\, \sigma^{ij,X}_{SM-EFT} [{\rm fb}] \right |_{\rm Bin 1}=&\ 
-1448\  C^{11}_{VLL} - 36.55\ C^{12}_{VLL} - 0.0008855\ C^{13}_{VLL}
 \\
 &-18.16\ C^{21}_{VLL} - 93.30\ C^{22}_{VLL} - 0.09312\ C^{23}_{VLL}~,
 \\[3pt]
\left . C^{ij}_X\, \sigma^{ij,X}_{SM-EFT} [{\rm fb}] \right |_{\rm Bin 2} =&\ 
-2056\ C^{11}_{VLL} - 50.01\ C^{12}_{VLL} - 0.001164\ C^{13}_{VLL}
 \\
 &-22.78\ C^{21}_{VLL} - 97.94\ C^{22}_{VLL} - 0.09520\ C^{23}_{VLL}~,
 \\[3pt]
\left . C^{ij}_X\, \sigma^{ij,X}_{SM-EFT} [{\rm fb}] \right |_{\rm Bin 3}=&\ 
-430.3\ C^{11}_{VLL} - 9.722\ C^{12}_{VLL} - 0.0002062\ C^{13}_{VLL}
 \\
 &-3.866\ C^{21}_{VLL} - 11.18\ C^{22}_{VLL} - 0.01043\ C^{23}_{VLL}~,
\end{split}
\end{equation}
and quadratic terms for three $m_T$ bins are
\begin{equation}
\begin{split}
 \left . (C^{ij}_X)^2\, \sigma^{ij,X}_{EFT^2} [{\rm fb}]\right |_{\rm Bin 1}
 =&\ 
55620 (C^{11}_{SL})^2+1345 (C^{12}_{SL})^2+0.03106 (C^{13}_{SL})^2 
\\
&+ 694.2 (C^{21}_{SL})^2+3146 (C^{22}_{SL})^2+3.091 (C^{23}_{SL})^2
\\
&+ 55540 (C^{11}_{SR})^2+1340 (C^{12}_{SR})^2+0.03109 (C^{13}_{SR})^2
\\
&+ 686.9 (C^{21}_{SR})^2+3151 (C^{22}_{SR})^2+3.093 (C^{23}_{SR})^2
 \\
 &+ 245200 (C^{11}_{T})^2+5627 (C^{12}_{T})^2+0.1262 (C^{13}_{T})^2 
 \\
 &+ 2814 (C^{21}_{T})^2+11770 (C^{22}_{T})^2+11.38 (C^{23}_{T})^2
\\
 &+ 70510 (C^{11}_{VLL})^2+1739 (C^{12}_{VLL})^2+0.03995 (C^{13}_{VLL})^2
 \\
 &+ 762.3 (C^{21}_{VLL})^2+3526 (C^{22}_{VLL})^2+3.415 (C^{23}_{VLL})^2
 \\
 &+ 64480 (C^{11}_{VRL})^2+1455 (C^{12}_{VRL})^2+0.03352 (C^{13}_{VRL})^2
 \\
 &+ 854.7 (C^{21}_{VRL})^2+3543 (C^{22}_{VRL})^2+3.449 (C^{23}_{VRL})^2~,
\end{split}
\end{equation}
\begin{equation}
\begin{split}
 \left . (C^{ij}_X)^2\, \sigma^{ij,X}_{EFT^2} [{\rm fb}]\right |_{\rm Bin 2}
 =&\ 
170400 (C^{11}_{SL})^2+3942 (C^{12}_{SL})^2+0.08904 (C^{13}_{SL})^2
\\
&+1846 (C^{21}_{SL})^2+6909 (C^{22}_{SL})^2+6.539 (C^{23}_{SL})^2
\\
&+169500 (C^{11}_{SR})^2+3938 (C^{12}_{SR})^2+0.08971 (C^{13}_{SR})^2
\\
&+1834 (C^{21}_{SR})^2+6938 (C^{22}_{SR})^2+6.565 (C^{23}_{SR})^2
 \\
 &+668700 (C^{11}_{T})^2+15270 (C^{12}_{T})^2+0.3320 (C^{13}_{T})^2
 \\
 &+6836 (C^{21}_{T})^2+23780 (C^{22}_{T})^2+21.99 (C^{23}_{T})^2
 \\
 &+201200 (C^{11}_{VLL})^2+4725 (C^{12}_{VLL})^2+0.1054 (C^{13}_{VLL})^2
 \\
 &+2004 (C^{21}_{VLL})^2+7421 (C^{22}_{VLL})^2+7.001 (C^{23}_{VLL})^2
 \\
 &+189800 (C^{11}_{VRL})^2+4283 (C^{12}_{VRL})^2+0.09593 (C^{13}_{VRL})^2
 \\
 &+2095 (C^{21}_{VRL})^2+7481 (C^{22}_{VRL})^2+7.016 (C^{23}_{VRL})^2~,
\end{split}
\end{equation}
\begin{equation}
\begin{split}
 \left . (C^{ij}_X)^2\, \sigma^{ij,X}_{EFT^2} [{\rm fb}]\right |_{\rm Bin 3}
 =&\ 
125900 (C^{11}_{SL})^2+2928 (C^{12}_{SL})^2 + 0.05954 (C^{13}_{SL})^2
\\
&+1081 (C^{21}_{SL})^2+2867 (C^{22}_{SL})^2+2.428 (C^{23}_{SL})^2
\\
&+126300 (C^{11}_{SR})^2+2924 (C^{12}_{SR})^2+0.05939 (C^{13}_{SR})^2
\\
&+1073 (C^{21}_{SR})^2+2882 (C^{22}_{SR})^2+2.437 (C^{23}_{SR})^2
 \\
 &+385200 (C^{11}_{T})^2+9126 (C^{12}_{T})^2+0.1753 (C^{13}_{T})^2
 \\
 &+3130 (C^{21}_{T})^2+8003 (C^{22}_{T})^2+6.469 (C^{23}_{T})^2
 \\
 &+133700 (C^{11}_{VLL})^2+3135 (C^{12}_{VLL})^2+0.06213 (C^{13}_{VLL})^2
 \\
 &+1105 (C^{21}_{VLL})^2+2928 (C^{22}_{VLL})^2 + 2.395 (C^{23}_{VLL})^2
 \\
 &+133600 (C^{11}_{VRL})^2+3104 (C^{12}_{VRL})^2+0.06213 (C^{13}_{VRL})^2
 \\
 &+1101 (C^{21}_{VRL})^2+2937 (C^{22}_{VRL})^2+2.436 (C^{23}_{VRL})^2~.
\end{split}
\end{equation}
Note that the large numerical factors in front of many EFT coefficients are artifacts of our definition of EFT coefficients. They do not invalidate the EFT expansion.
Similarly, the BSM cross section for the $pp\rightarrow \tau\nu b$ process, following the description in Section~\ref{sec:taunub}, can be written as
\begin{equation}\label{app:eq:xsec:EFTcoeff:taunub}
 \left [ \sigma(pp\rightarrow \tau\nu b) - \sigma_{SM} \right ]_{\text{with our cuts}} = C^{ij}_X\, \sigma^{ij,X}_{SM-EFT} +  (C^{ij}_X)^2\, \sigma^{ij,X}_{EFT^2}~,
\end{equation}
where the interference terms for three $m_T$ bins are given by 
\begin{equation}
\begin{split}
\left . C^{ij}_X\, \sigma^{ij,X}_{SM-EFT} [{\rm fb}] \right |_{\rm Bin 1}=&\ -0.001064\ C^{13}_{VLL} - 0.1236\ C^{23}_{VLL}~,
\\[3pt]
\left . C^{ij}_X\, \sigma^{ij,X}_{SM-EFT} [{\rm fb}] \right |_{\rm Bin 2}=&\ -0.0004469\ C^{13}_{VLL} - 0.03474\ C^{23}_{VLL}~,
\\[3pt]
\left . C^{ij}_X\, \sigma^{ij,X}_{SM-EFT} [{\rm fb}] \right |_{\rm Bin 3}=&\ -0.00006014\ C^{13}_{VLL} - 0.002814\ C^{23}_{VLL}~,
\end{split}
\end{equation}
and quadratic terms for three $m_T$ bins are given by
\begin{equation}
\begin{split}
\left . (C^{ij}_X)^2\, \sigma^{ij,X}_{EFT^2} [{\rm fb}] \right |_{\rm Bin 1}
 =&\  0.02059 (C^{13}_{SL})^2+1.814 (C^{23}_{SL})^2 
\\
&+ 0.02065 (C^{13}_{SR})^2+1.737 (C^{23}_{SR})^2 
\\
& + 0.1125 (C^{13}_{T})^2+9.318 (C^{23}_{T})^2 
\\
& + 0.03486 (C^{13}_{VLL})^2+3.201 (C^{23}_{VLL})^2  
\\
&+ 0.02533 (C^{13}_{VRL})^2+3.108 (C^{23}_{VRL})^2~,
\end{split}
\end{equation}
\begin{equation}
\begin{split}
\left . (C^{ij}_X)^2\, \sigma^{ij,X}_{EFT^2} [{\rm fb}] \right |_{\rm Bin 2}
 =&\  0.02730 (C^{13}_{SL})^2+1.643 (C^{23}_{SL})^2
\\
&+ 0.02757 (C^{13}_{SR})^2+1.573 (C^{23}_{SR})^2 
\\
& + 0.1268 (C^{13}_{T})^2+7.119 (C^{23}_{T})^2 
\\
& + 0.04131 (C^{13}_{VLL})^2+2.736 (C^{23}_{VLL})^2  
\\
&+ 0.03575 (C^{13}_{VRL})^2+2.723 (C^{23}_{VRL})^2~,
\end{split}
\end{equation}
\begin{equation}
\begin{split}
\left . (C^{ij}_X)^2\, \sigma^{ij,X}_{EFT^2} [{\rm fb}] \right |_{\rm Bin 3}
 =&\ 0.01605 (C^{13}_{SL})^2+0.5428 (C^{23}_{SL})^2
\\
&+ 0.01617 (C^{13}_{SR})^2 + 0.5083 (C^{23}_{SR})^2 
\\
& + 0.05797 (C^{13}_{T})^2+1.817 (C^{23}_{T})^2 
\\
& + 0.02199 (C^{13}_{VLL})^2+0.8496 (C^{23}_{VLL})^2  
\\
&+ 0.02152 (C^{13}_{VRL})^2+0.8569 (C^{23}_{VRL})^2~.
\end{split}
\end{equation}
As we briefly mentioned in Section~\ref{sec:angular}, the interference terms between $\mathcal{O}_{T}$ and $\mathcal{O}_{SL}$ operators exist due to kinematic cuts. We found that they are small enough to be ignored.

\section{Calculation of differential cross section}
\label{app:sec:helicity:diff}
The analytic evaluation of the $2\rightarrow 3$ amplitude should help us with the exact understanding of the $E$-growing behavior of the amplitude and various angular distributions. In this section, we calculate the helicity amplitude and differential cross section of the process $c g \rightarrow \tau \nu + b$  which is relevant for the $R({D^{(*)}})$ anomaly. For the helicity amplitude, we do it only for the $\mathcal{O}^{cb}_{VLL}$ operator as an example (see Section~\ref{sec:EFT} for the definition). For the differential cross section,  we include all operators with respect to various angles defining our coordinate system.

\subsection{Coordinate and four momenta}

\begin{figure}[tp]
\begin{center}
\includegraphics[width=0.88\textwidth]{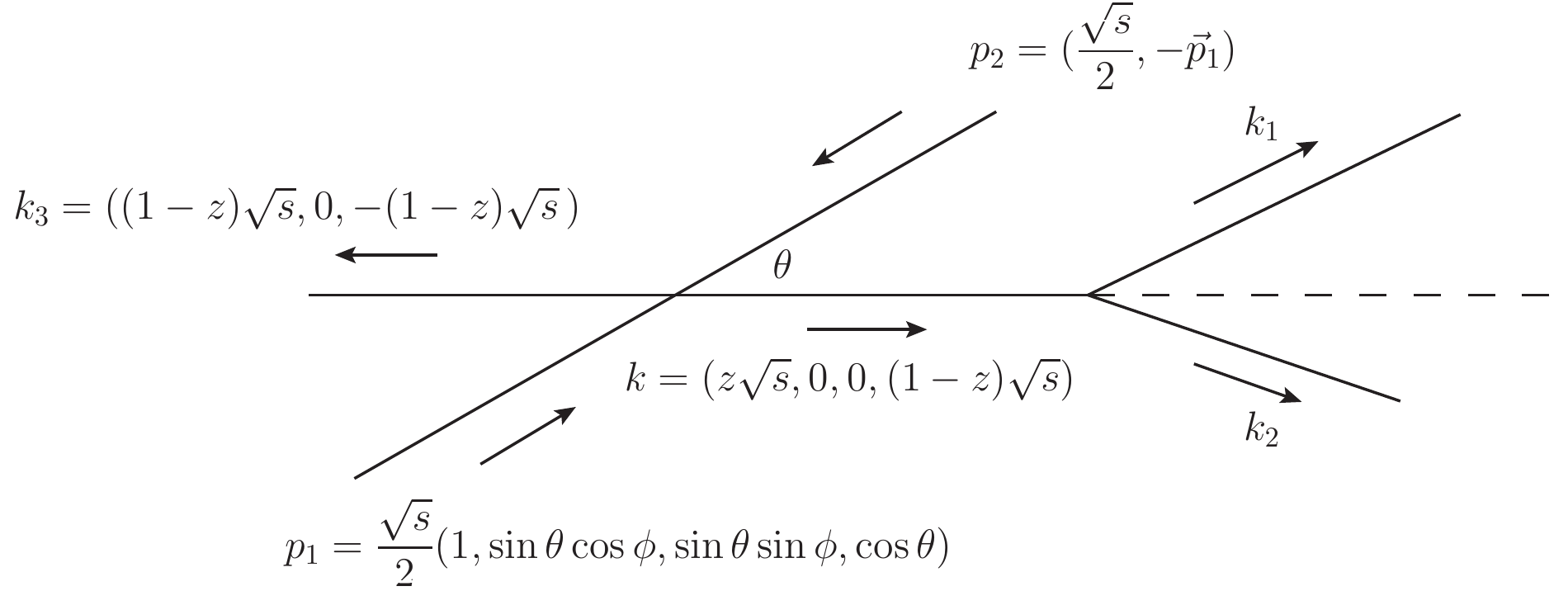}
\caption{\small Our coordinate system in $p_1p_2$ center-of-mass frame and four momenta (see also Fig.~\ref{fig:2to3:coordinate}).}
\label{fig:2to3:frame}
\end{center}
\end{figure}

The $2\rightarrow 3$ scattering process can be described in terms of 5 independent kinematic variables. Among many choices, we adopt the following coordinate system in terms of
\begin{equation}\label{eq:cooridiate1}
 \{ \sqrt{\hat{s}},\, z,\, \theta,\, \phi,\, \psi \} ~,
\end{equation}
where $\sqrt{\hat{s}}$ is total energy of the entire system, $z$ the fraction of energy flowing into the $k_1k_2$ system, namely $E_{k_1} + E_{k_2} = z \sqrt{\hat{s}}$, $\theta$ the polar angle between $p_1$ and $k_1+k_2$ directions, $\phi$ the angle between two planes made of ($p_1$, $k_3$) and ($k_1$, $k_2$) pairs, and $\psi$ the polar angle between $k_1$ and $k_1+k_2$ directions in the $k_1k_2$ rest frame. They are illustrated in Fig.~\ref{fig:2to3:frame}.

Two incoming momenta $p_1,\, p_2$ and three outgoing momenta $k_1,\, k_2,\, k_3$ in the $p_1p_2$ center-of-mass frame are parametrized in terms of variables in Eq.~(\ref{eq:cooridiate1}) as

\begin{equation}\label{app:eq:momenta:Lab}
\begin{split}
  p_1^\mu =&\ \frac{\sqrt{\hat{s}}}{2} \left ( 1,\, \sin\theta\cos\phi,\, \sin\theta\sin\phi,\, \cos\theta \right )~,
  \\[5pt] 
  p_2^\mu =&\ \frac{\sqrt{\hat{s}}}{2} \left ( 1,\, -\sin\theta\cos\phi,\, -\sin\theta\sin\phi,\, -\cos\theta \right )~,
  \\[5pt]
  k_1^\mu =&\ \frac{\sqrt{\hat{s}}}{2} \left ( z+ (1-z) \cos\psi,\,  \sqrt{(2z-1)} \sin\psi,\, 0,\, (1-z) + z \cos\psi \right )~,
  \\[5pt]
  k_2^\mu =&\ \frac{\sqrt{\hat{s}}}{2} \left ( z - (1-z) \cos\psi,\, - \sqrt{(2z-1)} \sin\psi,\, 0,\, (1-z) - z \cos\psi \right )~,
  \\[5pt]
  k_3^\mu =&\ \sqrt{\hat{s}} \left ( 1-z,\, 0,\, 0,\, -(1-z) \right )~,
  \\[5pt]
  k^\mu =&\ \sqrt{\hat{s}} \left ( z,\, 0,\, 0,\, (1-z) \right )~,
\end{split}
\end{equation}
where the momentum $k$ has the invariant mass of $m^2_k = (2z-1)\hat{s}$. Note that the $2\rightarrow 3$ process can be effectively factorized into $2\rightarrow 2$ and $1\rightarrow 2$ via an intermediate momentum $k$ (whether or not the intermediate momentum is associated with a resonance). The momenta $k_1$ and $k_2$ in Eq.~(\ref{app:eq:momenta:Lab}) are obtained by boosting those in the $k_1k_2$ rest frame,
\begin{equation}
\begin{split}
  k_1^\mu =&\ \frac{m_k}{2} \left ( 1,\, \sin\psi,\, 0,\, \cos\psi\ \right )~,
  \\[5pt]
  k_2^\mu =&\ \frac{m_k}{2} \left ( 1,\, -\sin\psi,\, 0,\, -\cos\psi \right )~,
\end{split}
\end{equation}
 along the $z$-axis with the boosting factor,
\begin{equation}\label{app:eq:boost:k1k2}
   k_z = \gamma_z\, m_k \beta_z \rightarrow \gamma_z = \frac{k^0}{m_k} = \frac{z}{\sqrt{2z-1}}~.
\end{equation}

\subsection{Helicity amplitude}
\label{app:sec:helicity}
%
%
\begin{figure}[t]
\begin{center}
\includegraphics[width=0.38\textwidth]{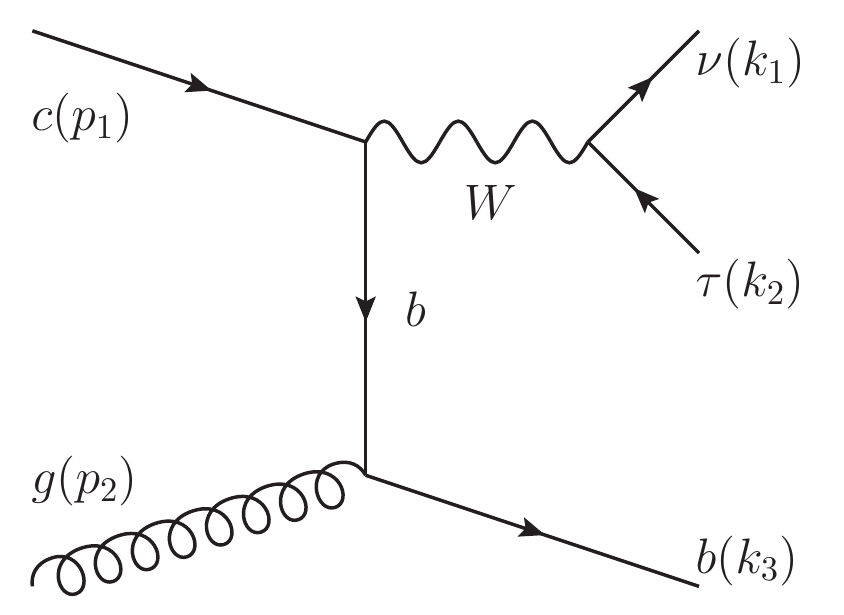}\quad
\includegraphics[width=0.38\textwidth]{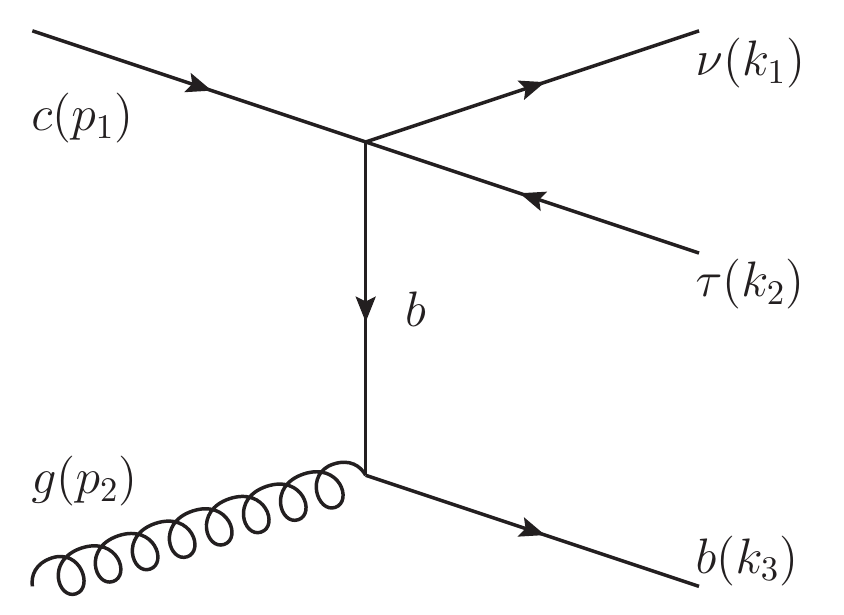}\quad
\caption{\small The $t$-channel diagrams of $cg \rightarrow \tau^+\nu \, b$ from the $W$-boson exchange in the SM (left) and four-fermion operator (right).}
\label{app:fig:tch}
\end{center}
\end{figure}
%
\begin{figure}[t]
\begin{center}
\includegraphics[width=0.38\textwidth]{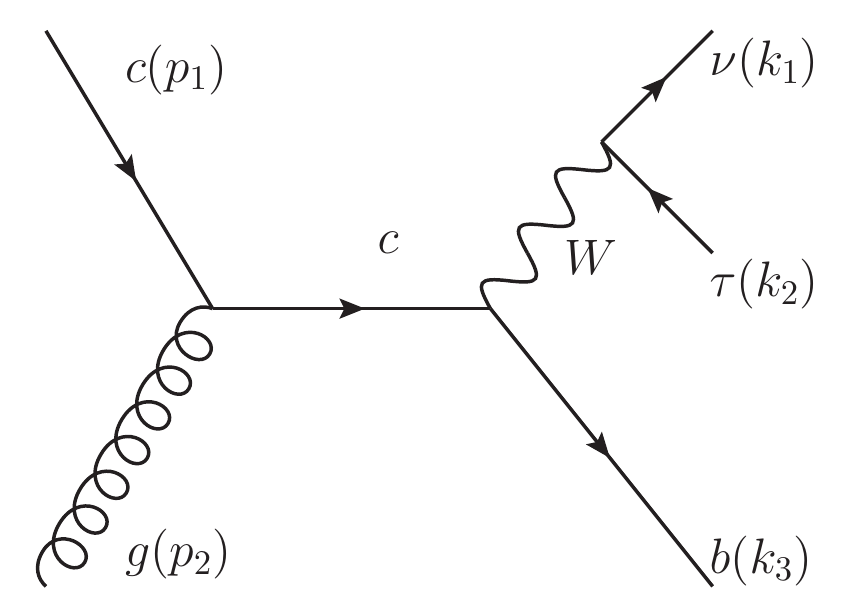}\quad
\includegraphics[width=0.38\textwidth]{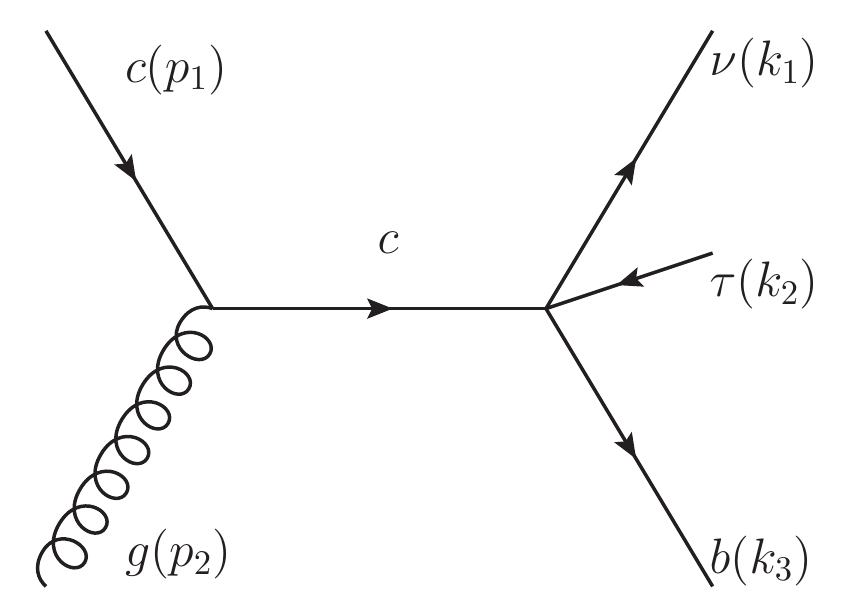}\quad
\caption{\small The $s$-channel diagrams of $cg \rightarrow \tau^+\nu \, b$ from the $W$-boson exchange in the SM (left) and four-fermion operator (right).}
\label{app:fig:sch}
\end{center}
\end{figure}
The $t$-channel amplitude in Fig.~\ref{app:fig:tch} is given by 
\begin{equation}\label{app:eq:tch}
\begin{split}
i\mathcal{M}_t= 
 i\, g_s t^a \frac{2V_{cb}}{v^2} \left ( C_{VLL}^{cb} - \frac{m^2_W}{k^2 - m^2_W + i\, m_W  \Gamma_W} \right )
 \ \bar{u}(k_3)\slashed{\epsilon}^a(p_2) \frac{\slashed{q}}{q^2}\slashed{j_L} P_L u(p_1)~,
\end{split}
\end{equation}
where $q = k_3 - p_2$ and $j^\mu_L$ is the left-handed fermion current, $j^\mu_L = \bar{u}(k_1)\gamma^\mu P_L v(k_2)$.
Similarly, the $s$-channel amplitude in Fig.~\ref{app:fig:sch} is given by
\begin{equation}\label{app:eq:sch}
\begin{split}
i \mathcal{M}_s = 
 i\, g_s t^a \frac{2 V_{cb}}{v^2} \left ( C_{VLL}^{cb} - \frac{m^2_W}{k^2 - m^2_W + i\, m_W  \Gamma_W}  \right )
 \bar{u}(k_3) \slashed{j_L} P_L \frac{\slashed{q}}{q^2} \slashed{\epsilon}^a(p_2) u(p_1)~,
\end{split}
\end{equation}
where $q=p_1 + p_2$ and $j^\mu_L = \bar{u}(k_1)\gamma^\mu P_L v(k_2)$ as before. 
The $t$-, $s$-channel momentum squared are given by
\begin{equation}
  (k_3-p_2)^2 = -(1-z)\hat{s}\, (1-\cos\theta)~,\quad (p_1+p_2)^2 = \hat{s}~.
\end{equation}

Using the expressions for the spinors in terms of our coordinates, the $t$-channel amplitudes are evaluated to be
\begin{equation}\label{eq:amp:tchannel}
\begin{split}
i\mathcal{M}^a_{t,\, L}= 
  &\, i g_s t^a \frac{2V_{cb}}{v^2} \left ( C_{VLL}^{cb} - \frac{m^2_W}{(2z-1)\hat{s} - m^2_W + i\, m_W  \Gamma_W} \right )
\\[5pt]
&\times - 2 \cos\frac{\theta}{2} \sqrt{1-z} \left ( e^{i\, \phi} \cot\frac{\theta}{2} (1+\cos\psi ) \sqrt{2z-1} + \sin\psi \right ) \sqrt{\hat{s}}~,
\\[5pt]
i\mathcal{M}^a_{t,\, R} = 
  &\, i g_s t^a \frac{2V_{cb}}{v^2} \left ( C_{VLL}^{cb} - \frac{m^2_W}{(2z-1)\hat{s} - m^2_W + i\, m_W  \Gamma_W} \right )
\\[5pt]
  &\times 2 \cos\frac{\theta}{2}\, \frac{2z-1}{2\sqrt{1-z}} \left ( e^{i\phi}  \cot\frac{\theta}{2} (1+ \cos\psi) \sqrt{2z-1} \right .
  \\[5pt]
  & \left . \hspace{2.3cm} + \frac{1}{\sqrt{2z-1}}\tan\frac{\theta}{2} (1-\cos\psi) e^{-i\phi} + 2 \sin\psi \right ) \sqrt{\hat{s}}~.
\end{split}
\end{equation}
The $s$-channel amplitudes are given by
\begin{equation}
\begin{split}
i\mathcal{M}^a_{s,\, L}= 
  &\, i g_s t^a  \frac{2V_{cb}}{v^2} \left ( C_{VLL}^{cb} - \frac{m^2_W}{(2z-1)\hat{s} - m^2_W + i\, m_W  \Gamma_W} \right )
\\[5pt]
&\times - 2 \cos\frac{\theta}{2}  \sqrt{1-z} \left ( e^{i\phi} \tan\frac{\theta}{2} (1+ \cos\psi ) \sqrt{2z-1} - \sin\psi  \right ) \sqrt{\hat{s}}~,
 \\[5pt]
  i\mathcal{M}^a_{s,\, R} =&\ 0~.
\end{split}
\end{equation}
The helicity amplitudes for other operators in Eq.~(\ref{eq:eft:banomaly:mbscale}) can be similarly obtained.
The overall amplitude grows like $\sim \sqrt{\hat{s}}$ as is expected whereas BSM amplitude grows like $\sim \hat{s}$ with respect to the SM amplitude, dictated by the Lorentz structure of the $\mathcal{O}^{cb}_{VLL}$ operator. 
As is evident in Eq.~(\ref{eq:amp:tchannel}), the $t$-channel amplitude is singular in the forward region, $\theta \sim 0$, and it leads to the logarithmic growth of the cross section, regulated by the bottom quark mass $m_b$:
\begin{equation}
 \frac{1}{-2k_b\cdot p_2} = - \frac{1}{2E_2E_b \left ( 1 - \left ( 1 - \frac{m^2_b}{E^2_b} \right )^{1/2} \cos\theta \right )}
 \rightarrow \log \frac{2E^2_b}{m^2_b}~.
\end{equation}
In practice, we need to regulate the large log by higher $p_T$ cut on $b$-jet  than $m_b$  as the coupling $\alpha_s$ is roughly 
$\alpha_s \sim 1/\log (E^2/\Lambda^2_{QCD})$.

\subsection{Differential cross section}

The cross section can be straightforwardly computed either squaring the helicity amplitudes evaluated in Section~\ref{app:sec:helicity} or evaluating the amplitudes squared directly. Since we present the full partonic differential cross section (before convoluted with PDF) with respect to four variables, $\theta,\, \psi,\, \phi,\, z$, switching from our coordinate to another choice should be straightforward. The differential cross section for each four-fermion operator, assuming all real EFT coefficients defined in Eq.~(\ref{eq:eft:banomaly:mbscale}), is given by
\begin{equation}\label{eq:ddddsigma:CVLL}
\begin{split}
  \frac{d^4\hat\sigma (cg\rightarrow \tau\nu b)}{d\cos\theta d\cos\psi\, d\phi\, dz} 
  =&\ \frac{\alpha_s}{192 \pi^3}\frac{V_{cb}^2}{v^4}\left | C_{VLL}^{cb} - \frac{m^2_W}{(2z-1)\hat{s} - m^2_W + i m_W \Gamma_W} \right |^2 \frac{\hat{s}}{1-\cos\theta} \Big [
  \\[2.5pt]
  &\sin \psi  \cos \phi  \Big \{ \sqrt{2 z-1}\,  ( 2 z^2 \cos \psi + 2 z^2 - 3 z+1 ) \sin 2 \theta
  \\[2.5pt]
  +&\, 2 (2 z-1)^{3/2}  ( (z-1) \cos\psi+z ) \sin\theta+ (2 z-1)^2 \sin^2 \theta  \sin\psi \cos\phi \Big \}
  \\[2.5pt]
  +&\, 2 \cos\theta \Big \{ -z \sqrt{2 z-1} \sin\theta \sin\psi\cos\psi \cos\phi + z(2 z^2 - 3 z + 1) \cos^2\psi 
  \\[2.5pt]
  +&\,  (4 z^3 - 6 z^2 + 4 z - 1 ) \cos\psi + z( 2 z^2 - 3 z + 1) \Big \}
  \\[2.5pt]
  +&\, (2 z-1) \Big \{ \cos^2\theta (z \cos\psi + z-1)^2
  \\[2.5pt]
  +&\, (z-1) \cos\psi \left ( 5 (z-1) \cos\psi+10 z-8 \right ) + 5 z^2 - 8z+4 \Big \} \Big ]~,
\end{split}
\end{equation}
\begin{equation}\label{eq:ddddsigma:CVRL}
\begin{split}
  \frac{d^4\hat\sigma (cg\rightarrow \tau\nu b)}{d\cos\theta d\cos\psi\, d\phi\, dz} 
  =&\ \frac{\alpha_s}{192 \pi^3}\frac{C_{VRL}^{cb\ 2} V_{cb}^2}{v^4}\frac{\hat{s}}{1-\cos\theta} \Big [
  \\[2.5pt]
  &\sin\psi \cos\phi  \Big \{  \sqrt{2 z-1}  \left(2 z^2 \cos\psi -2 z^2 + 3z-1\right) \sin 2 \theta 
  \\[2.5pt]
  +&\, 2 (2 z-1)^{3/2}  ((z-1) \cos\psi -z) \sin\theta+(2 z-1)^2 \sin^2\theta \sin\psi \cos\phi \Big \} 
  \\[2.5pt]
  +&\, 2 \cos\theta \Big \{  -z \sqrt{2 z-1} \sin\theta \sin\psi\cos\psi \cos\phi + z(2z^2 - 3z +1) \cos^2\psi 
  \\[2.5pt]
  -&\, (4 z^3 - 6 z^2 + 4 z - 1 ) \cos\psi +  z(2z^2 - 3z +1) \Big \}
  \\[2.5pt]
  +&\, (2 z-1) \Big \{ \cos^2 \theta (z \cos\psi - z+1)^2
  \\[2.5pt]
  +&\, (z-1) \cos\psi \left ( 5 (z-1) \cos\psi -10 z+8 \right ) + 5 z^2-8z+4 \Big \} \Big ]~,
\end{split}
\end{equation}
\begin{equation}\label{eq:ddddsigma:CSLSR}
\begin{split}
  \frac{d^4\hat\sigma (cg\rightarrow \tau\nu b)}{d\cos\theta d\cos\psi\, d\phi\, dz} 
  =&\ \frac{\alpha_s}{384 \pi^3}\frac{\left (C_{SL}^{cb\ 2} + C_{SR}^{cb\ 2} \right ) V_{cb}^2}{v^4}\frac{\hat{s}}{1-\cos\theta} \Big [
  \\[2.5pt]
  &(2 z-1) \left \{ 4 (z-1)^2 \cos\theta +(z-1)^2 \cos 2\theta + 11 z^2 - 14z + 5 \right \} \Big ]~,
\end{split} 
\end{equation}
\begin{equation}\label{eq:ddddsigma:CT}
\begin{split}
  \frac{d^4\hat\sigma (cg\rightarrow \tau\nu b)}{d\cos\theta d\cos\psi\, d\phi\, dz} 
  =&\ \frac{\alpha_s}{12 \pi^3}\frac{C_{T}^{cb\ 2} V_{cb}^2}{v^4}\frac{\hat{s}}{1-\cos\theta} \Big [
  \\[2.5pt]
  &\frac{1}{2} \cos^2\psi \Big \{ 4 (z-1) (z+1) (2 z-1) \cos\theta  
  \\[2.5pt]
   +&\, z (2 z^2-3z+2) \cos 2\theta + 22 z^3-57 z^2+50 z-14 \Big \}
  \\[2.5pt]
  +&\, \sin\theta \Big \{ \sin\theta \left((2 z-1) \sin^2\psi \cos^2\phi  + 2 (z-1)^2 \right)
  \\[2.5pt]
   +&\, (z-1) \sqrt{2 z-1} (3 z-1) \sin  2\psi \cos\phi  \Big \}
  \\[2.5pt]
  +&\, \sqrt{2 z-1} \left ( z^2- z+1 \right ) \sin 2\theta \sin\psi \cos\psi \cos\phi \Big ]~,
\end{split} 
\end{equation}
\begin{equation}\label{eq:ddddsigma:CSLT}
\begin{split}
  \frac{d^4\hat\sigma (cg\rightarrow \tau\nu b)}{d\cos\theta d\cos\psi\, d\phi\, dz} 
  =&\ \frac{\alpha_s}{48 \pi^3}\frac{C_{SL}^{cb} C_{T}^{cb} V_{cb}^2}{v^4}\frac{\hat{s}}{1-\cos\theta} \Big [
  \\[2.5pt]
  -&\, (2 z-1) \cos \psi  \Big \{ 2 (2 z^2 - 2 z + 1) \cos\theta
  \\[2.5pt]
   +&\,  (z-1) ((z-1) \cos 2\theta+11 z-7) \Big \}
  \\[2.5pt]
  -&\, 2 \sqrt{2 z-1} \sin\theta \sin\psi \cos\phi \Big \{ (z-1)^2 \cos\theta + 3 z^2 - 3 z + 1 \Big \} \Big ]~.
\end{split} 
\end{equation}

\section{Simulation detail}
\label{app:simulation}

\subsection{Background simulation to $pp\rightarrow \tau\nu (+ b)$}

All background samples were simulated by \textsc{\sc MadGraph}5\_aMC$@$NLO v2.3.3~\cite{Alwall:2014hca} with the default factorization and renormalization scales, interfaced with {\sc Pythia} v6.4 \cite{Sjostrand:2006za}, and they were matched at Leading order (LO) using $k_T$-jet MLM matching with appropriate {\tt xqut/QCUT}s.
The $W$+jets ($W$ not necessarily on-shell) and the Drell-Yan process $\gamma^*/Z$+jets ($Z$ not necessarily on-shell) samples were matched at LO allowing up to two extra jets in 5-flavor. In the latter, $\gamma^*/Z \rightarrow ll$+jets and $Z\rightarrow \nu\nu$+jets samples were separately simulated. The simulation of $W$+jets includes up to the order of {\tt QED=4} which covers the contribution via the vector boson fusion (VBF). We find that the contribution from VBF is not negligible when we apply $p_T$-dependent tau mistag rates in Fig.~\ref{fig:mistag:tau:CMS} (this observation, however, disappears with the $p_T$-independent tau mistag rate). We also have independently simulated $W$+jets, $Wb$+jets, and $W+2b$+jets in 4-flavor for sanity checks, among which $W$+jets is found to be dominant in the $\tau\nu$ analysis.
The $t\bar{t}$ background samples were matched allowing up to one extra parton in 4-flavor scheme whereas $t+$jets (single top) samples  were matched up to two extra partons in 5-flavor scheme. The single top samples also include contributions from $tV$ ($V=W,Z$) where $V$ decays leptonically.
The $VV$ process is simulated by five different subprocesses (categorized by the numbers of leptons and neutrinos) where a leptonic $V$ is not necessarily on-shell whereas a hadronic $V$ is produced on-shell. The subprocesses with only one leptonic $V$ were matched up to one extra parton in 5-flavor scheme whereas those with two leptonic $V$'s were matched up to two extra partons.

\subsection{Signal simulation to $pp\rightarrow \tau\nu,\, \tau\nu b$}
\label{app:signal}
The four-fermion operators were implemented in \textsc{FeynRules}\cite{Alloul:2013bka} and the resulting {\sc UFO} model file was used in \textsc{\sc MadGraph}5\_aMC$@$NLO v2.3.3~\cite{Alwall:2014hca} that we used for the generation of the signal samples.

\begin{table}[t]
\begin{center}
\scalebox{1.00}{
\begin{tabular}{c|c|c|c|c|c|c}
\multicolumn{7}{c}{$\tau\nu$ vs $\tau\nu+0,1j$}
\\[1.5pt]
\hline
Operator type &  \multicolumn{1}{c|}{$\mathcal{O}^{ij}_{SL}$} & \multicolumn{1}{c|}{$\mathcal{O}^{ij}_{SR}$} & \multicolumn{1}{c|}{$\mathcal{O}^{ij}_{T}$} & \multicolumn{2}{c|}{$\mathcal{O}^{ij}_{VLL}$}  & \multicolumn{1}{c}{$\mathcal{O}^{ij}_{VRL}$}
\\[1.5pt]
\hline
 $\sigma$ [fb] & $\sigma_{EFT^2}$ &  $\sigma_{EFT^2}$  & $\sigma_{EFT^2}$  & $\sigma_{SM-EFT}$ & $\sigma_{EFT^2}$ & $\sigma_{EFT^2}$
\\[1.5pt]
\hline
 \multicolumn{7}{c}{$i,j=1,3$}
\\[1.5pt]
\hline
$\tau\nu$&   0.1796 &  0.1802  &0.6336 & $2.255\times 10^{-3}$ & 0.2075 &  0.1916 \\[1.5pt]
\hline
$\tau\nu+0,1j$  & 0.1857 &  0.1864  & 0.7379  & $2.821\times 10^{-3}$ & 0.2572 &  0.2232 \\[1.5pt]
\hline
 \multicolumn{7}{c}{$i,j=2,3$}
\\[1.5pt]
\hline
$\tau\nu$&  12.06  &  12.09  & 39.84 & 0.1987 & 12.81 &  12.90 \\[1.5pt]
\hline
$\tau\nu+0,1j$ & 11.24 &  10.76  & 41.03 & 0.2539 & 16.33 &  16.21 \\[1.5pt]
\hline
\end{tabular}
}
\caption{The cross sections of $pp\rightarrow \tau\nu$ (without matching) and $pp\rightarrow \tau\nu+0,1j$ (with the matching up to one jet in 5-flavor scheme) at $\sqrt{s}=13$ TeV. The interference term, $\sigma_{SM-EFT}$, and quadratic term, $\sigma_{EFT^2}$, are those in Eq.~(\ref{eq:xsec:EFTcoeff}). The numbers in table are after imposing the cuts in the CMS analysis.} 
\label{tab:kfactor:signal}
\end{center}
\end{table}

\begin{table}[t]
\begin{center}
\scalebox{1.00}{
\begin{tabular}{c|c|c|c|c|c|c}
\multicolumn{7}{c}{$\tau\nu b$ vs $\tau\nu+0,1j$}
\\[1.5pt]
\hline
Operator type &  \multicolumn{1}{c|}{$\mathcal{O}^{ij}_{SL}$} & \multicolumn{1}{c|}{$\mathcal{O}^{ij}_{SR}$} & \multicolumn{1}{c|}{$\mathcal{O}^{ij}_{T}$} & \multicolumn{2}{c|}{$\mathcal{O}^{ij}_{VLL}$}  & \multicolumn{1}{c}{$\mathcal{O}^{ij}_{VRL}$}
\\[1.5pt]
\hline
  $\sigma$ [fb] &  $\sigma_{EFT^2}$ &  $\sigma_{EFT^2}$   & $\sigma_{EFT^2}$  & $\sigma_{SM-EFT}$ & $\sigma_{EFT^2}$ & $\sigma_{EFT^2}$
\\[1.5pt]
\hline
 \multicolumn{7}{c}{$i,j=1,3$}
\\[1.5pt]
\hline
$\tau\nu b$&  0.02964  & 0.02967  & 0.1315  & $0.647\times 10^{-3}$ & 0.03806 & 0.03379  \\[1.5pt]
\hline
$\tau\nu+0,1j$ & 0.06924 & 0.06978 & 0.3232 & $1.645\times 10^{-3}$ & 0.1060 & 0.08981  \\[1.5pt]
\hline
 \multicolumn{7}{c}{$i,j=2,3$}
\\[1.5pt]
\hline
$\tau\nu b$  & 1.814   & 1.809 & 8.333 & 0.05461 & 2.201 & 2.170  \\[1.5pt]
\hline
$\tau\nu+0,1j$ & 4.268 & 4.078 &  19.59 & 0.1682 & 7.264 & 7.168 \\[1.5pt]
\hline
\end{tabular}
}
\caption{The cross sections of $pp\rightarrow \tau\nu b$ (without matching) and $pp\rightarrow \tau\nu+0,1j$ (with the matching up to one jet in 5-flavor scheme) at $\sqrt{s}=13$ TeV. The interference term, $\sigma_{SM-EFT}$, and quadratic term, $\sigma_{EFT^2}$, are those in Eq.~(\ref{eq:xsec:EFTcoeff}). The numbers in table are after imposing the cuts in Eqs.~(\ref{eq:ourcut:pt}) and~(\ref{eq:cut:ours:backtoback})} 
\label{tab:taunub:signal}
\end{center}
\end{table}

The signal simulation can be generated either in 4 or 5-flavor scheme. The 4-flavor scheme suffers from the large logarithmic divergence in the $t$-channel diagram which might invalidate the perturbation. 
While 5-flavor scheme, on the other hand, correctly takes into account the resummation of large logs, it is computationally expensive to obtain sufficient statistics of $\tau\nu b$ signal events for all EFT operators. The correct simulation of $\tau\nu b$ in 5-flavor scheme requires the matching of $\tau\nu$ process allowing extra jets whose jet definition includes $b$ since the bottom quarks can come from either matrix element or parton shower. In this work, we choose 5-flavor scheme as our default for both $\tau\nu$ and $\tau\nu b$ processes~\footnote{Although the latter process should be conceptually equivalent to $\tau\nu+b$ where the $b$-quark phase space is not restricted, the cross section is overestimated due to the large logarithmic divergences, for instance, in the $t$-channel.}. The signal samples for $\tau\nu b$ process were simulated through the $pp\rightarrow \tau\nu+0,1j$ process by \textsc{\sc MadGraph}5\_aMC$@$NLO v2.3.3, interfaced with {\sc Pythia} v6.4, and they were matched at LO up to an extra jet using $k_T$-jet MLM matching. Since the $b$-jet is tagged for the $\tau\nu b$ process, we generate the signal events only for four-fermion operators with $b$-quark such as $(bu)(\tau\nu)$ and $(bc)(\tau\nu)$ with all possible Lorentz structures. Whereas the signal samples for the inclusive $\tau\nu$ analysis were generated by \textsc{MadGraph}5\_aMC$@$NLO v2.3.3, interfaced with {\sc Pythia} v8.219~\cite{Sjostrand:2007gs}, without matching.

We numerically estimate the (partial) $k$-factor of the signal cross sections by comparing the signal rates of $pp\rightarrow \tau\nu$ without matching and available matched samples of $pp\rightarrow \tau\nu + 0,1j$ described above. The comparison is presented in Table~\ref{tab:kfactor:signal} where the crude estimate of the $k$-factor is found to be roughly one. We also have checked that the differential distributions of all relevant kinematic variables agree well between two cases.

We also point out that the signal rates obtained from $pp \rightarrow \tau\nu b$ at the matrix level without the matching is not appropriate for the study of $\tau\nu b$. Not only the unmatched $\tau\nu b$ processes severely underestimate the signal rates (as is illustrated in Table~\ref{tab:taunub:signal}), also the discrepancy of differential distributions between unmatched $\tau\nu b$ and matched $\tau\nu+0,1j$ samples is not negligible.


{\small
\bibliography{lit}{}
\bibliographystyle{JHEP}}

\end{document}